\documentclass[oldversion]{aa}

\usepackage{times}
\usepackage{graphicx}
\usepackage{xspace}
\usepackage{epsfig}
\usepackage{natbib}
\usepackage{rotating}
\usepackage{dcolumn}
\usepackage{xcolor}
\usepackage[labelfont=bf]{caption}
\usepackage{lscape}
\usepackage{longtable}
\usepackage{multicol}
\usepackage{float}
\usepackage{amsmath}
\usepackage{amssymb}
\usepackage[toc,page]{appendix}

\usepackage{threeparttable}
\usepackage{threeparttablex}
\usepackage{booktabs}

\def\gsim{\;\lower4pt\hbox{${\buildrel\displaystyle >\over\sim}$}\,}
\def\lsim{\;\lower4pt\hbox{${\buildrel\displaystyle <\over\sim}$}\,}

\begin{document}

\title{Flares and rotation of M dwarfs \\with habitable zones accessible to TESS planet detections\thanks{The full Tables\,2, A1, A2 and A3 are available only in electronic form
at the CDS via anonymous ftp to cdsarc.u-strasbg.fr (130.79.128.5).}}

%\subtitle{blabla}

\author{B. Stelzer \inst{1,2} \and M. Bogner \inst{1} \and E. Magaudda \inst{1} \and St. Raetz \inst{1}}

%\offprints{B. Stelzer}

\institute{
% INST 1
 Institut f\"ur Astronomie \& Astrophysik, Eberhard-Karls-Universit\"at T\"ubingen,
 Sand 1, 72076 T\"ubingen, Germany\label{inst1} \\ \email{stelzer@astro.uni-tuebingen.de}
 \and
%% INST 2
 INAF - Osservatorio Astronomico di Palermo, Piazza del Parlamento 1, 90134 Palermo,
 Italy\label{inst2}
 % \and
%%%INST3
 % Space Telescope Science Institute....
 % USA\label{inst3}
}

\titlerunning{Flares and rotation of M dwarfs with habitable zones accessible to TESS  planet detections}

\date{Received $<$25-08-2022$>$ / Accepted $<$24-05-2022$>$}

\abstract{
\textit{Context.} {More than 4000 exoplanets have been discovered to date, providing the search for a place capable of hosting life with a large number of targets. With the Transiting Exoplanet Survey Satellite (TESS) having completed its primary mission in July 2020, the number of planets confirmed by follow-up observations is growing further. Crucial for planetary habitability is not only a suitable distance of the planet to its host star, but also the star's properties. Stellar magnetic activity, and especially flare events, expose planets to a high photon flux  and potentially erode their atmospheres. Here especially the poorly constrained high-energy UV and X-ray domain is relevant.}\\ 
\textit{Aims.} We characterize the magnetic activity of M dwarfs 
%with sufficiently long {\it TESS} observation times to spot planet transits over the entire habitable zone 
to provide the planet community with information on the energy input from the star; in particular, in addition to the frequency of optical flares directly observed with TESS, we aim at estimating the corresponding X-ray flare frequencies, making use of the small pool of known events observed  simultaneously in both wavebands. \\
\textit{Method.} We identified $112$ M dwarfs with a TESS magnitude $\leq 11.5$ for which TESS can probe the full habitable zone for transits. These $112$ stars have $1276$ two-minute cadence TESS LCs from the primary mission, which we searched for rotational modulation and flares. We study the link between rotation and flares and between flare properties, for example the flare amplitude-duration relation and  cumulative flare energy frequency distributions (FFDs).
%for our observed {\it TESS} flares. 
Assuming that each optical flare is associated with a flare in the X-ray band, and making use of published simultaneous {\it Kepler/K2} and {\it XMM-Newton} flare studies, we estimate the X-ray energy released by our detected TESS flare events. Our calibration also involves the relation between flare energies in the TESS and {\it K2} bands.\\
\textit{Results.} We detected more than $2500$ optical flare events on a fraction of about $32\%$ of our targets and found reliable rotation periods only for $12$ stars, which is a fraction of about  $11$\,\%. For these $12$ targets, we present cumulative FFDs and FFD power law fits. We construct FFDs in the X-ray band by calibrating optical flare energies to the X-rays. In the absence of directly observed X-ray FFDs for main-sequence stars, our predictions can serve for estimates of the high-energy input to the planet of a typical fast-rotating early- or mid-M dwarf.
}


\maketitle

\section{Introduction}\label{sect:intro}

With the advent of the {\it Kepler} mission \citep{2010Sci...327..977B}, M dwarfs 
are now known to be the most prolific planet hosts thanks to the abundance of such low-mass stars in our Galaxy and their favorable
  radius contrast for transit surveys. Moreover, the close-in habitable zones (HZs) of M dwarfs
  imply a high probability of finding planets that potentially host life. From {\it Kepler} detection
  statistics on average each M dwarf was estimated to be orbited by more than two small planets,
  and at least one in ten M dwarfs is expected to harbor a planet in its HZ \citep{2015ApJ...807...45D}.

  In the wake of exoplanet surveys the magnetic activity of the host stars has gained enormous interest. In
  particular the high-energy X-ray and ultraviolet (XUV) emission from the outer atmospheres of late-type stars and the
  notorious variability of their radiation are key elements to be considered in studies of the evolution
  of planet atmospheres. Various types of models for atmospheric mass loss in strongly irradiated planets
  have been developed (e.g., \citealt{2008JGRE..113.5008T}, \citealt{2008JGRE..113.7005T}, \citealt{2015ApJ...815L..12J}, \citealt{2015A&A...577L...3T}). Even if the planet
  atmospheres are not removed by the stellar irradiation, their chemistry may be affected; for example  \cite{2019AsBio..19...64T}
  have calculated that frequent flaring may destruct ozone layers.
  
  Having boosted the interest in understanding the host star's magnetic phenomena 
  because of their relevance for exoplanets, the same instruments that detect planets through their transits serve as a basis to investigate the host star's magnetic activity. In addition to {\it Kepler} and its successor, the {\it K2} mission \citep{2014PASP..126..398H},
  the Transiting Exoplanet Survey Satellite (TESS, \citealt{2015JATIS...1a4003R}) plays a major role in this field.
  At the optical wavelengths where these satellites operate,  magnetic activity is manifest most prominently in 
rotational modulation due to starspots and stochastic brightness outbursts called flares. 
Light curves (LCs)  obtained with these missions present a number of activity diagnostics, including
the period and amplitude of starspot modulations, and the energy and frequency of optical flare events (e.g.,  \citealt{Davenport14.0}, \citealt{2014ApJ...797..121H}, \citealt{Stelzer2016.0}, \citealt{Ilin19.0}, \citealt{Raetz2020.0}, \citealt{2020AJ....159...60G},
\citealt{Medina2020}). 
In contrast, no telescopes exist that are dedicated to the study of the UV and X-ray counterparts of these events,  
  which are more influential for planet evolution. 
  Since the variability of the stellar high-energy emission is difficult to constrain from observations, we
  pursue here an indirect approach, calibrating the frequency of X-ray flares on M dwarfs
  from the observations of optical flares observed with TESS. 
  
  %Photometric data from space telescopes like the Transiting Exoplanet Survey Satellite (TESS, Ricker et al. 2010) are key for exoplanet detections and at the same time serve as a basis to investigate the host star’s magnetic activity.

%TESS is a {\it NASA} mission launched in April 2018 and completed its all sky survey in July 2020. Currently, it continues operating for a second all sky survey. Targets for 2-min. cadence observations in the primary mission have been pre-selected based on the {\it TESS} Input Catalog (TIC, Stassun et al. 2018). The TIC contains astronomical and physical parameters for about 470 million point sources and 2 million extended sources. Our work is based
%  on a sample of $112$ M dwarfs with habitable zones accessible to the detection of planet transits with TESS. 

%Highly energetic X-ray radiation has a particularly strong effect on planets. Since there is only a limited number of X-ray flare observations on M dwarfs available, we propose an indirect way to estimate the X-ray flare energy of events detected in T E S S light curves.

TESS is a {\it NASA} satellite launched in April 2018, and it  completed its (nearly) all-sky survey primary mission in July 2020. It continues to operate and is currently conducting a second all-sky survey. For the primary mission, 2-minute cadence LCs and target pixel files (TPFs) are available for about $200000$ preselected targets. Target pixel files typically contain  $11 \times 11$ CCD pixels with the target located in the center of the image. Targets for 2-minute cadence observations have been preselected based on the TESS Input Catalog (TIC; \citealt{2018AJ....156..102S}). The TIC contains astronomical and physical parameters for about 470 million point sources and two million extended sources. Further, full frame images are available in 30-minute cadence for the primary mission, containing the flux of all pixels of a single CCD.

Our work is based
  on a sample of $112$ M dwarfs observed in 2-minute cadence with TESS and with their entire HZs accessible to the detection of planet transits with TESS.
We explain our sample selection and the calculation of the stellar parameters in Sect.~\ref{sect:sample}. Sect.~\ref{sect:analysis} describes our LC analysis procedure to search for rotation periods and flares. We give the results of this analysis in Sect.~\ref{sec:analysis_resulst}, including relations between flare rate and spectral type (SpT), FFDs and a discussion of detection biases. Confirmed planet host stars and TESS objects of interest (TOIs) within our sample are addressed in Sect.~\ref{subsection:flares_tois}. We finally give a conversion for flare energies in the TESS band to the {\it XMM-Newton} X-ray band in Sect.~\ref{sect:calib} and use this as a basis to construct X-ray FFDs. A summary and discussion of our results is given in Sect.~\ref{sect:discussion}.

\section{Sample}\label{sect:sample}

Our sample is drawn from the TESS Habitable Zone Star Catalog (HZCat, \citealt{Kaltenegger2019.0}). The HZCat is a subsample of the TIC. It lists 1822 stars with TESS magnitude $T<12$~mag for which TESS can detect planets out to the extent of the Earth equivalent orbital distance\footnote{meaning  the orbital distance at which the flux received by the top of the planet atmosphere is comparable to that received by Earth's atmosphere} down to a planet size of two Earth radii. The stars we selected for this work are all from the subsample that carries the ``HZflag'', meaning that TESS can probe the full HZ for transiting planets. The need to be able to detect at least two transits translates into the requirement for long continuous TESS observations. Therefore, by construction of the sample these stars are located close to the ecliptic poles where TESS provides continuous viewing over a full year \cite[see Fig.~3 of][for the distribution of the stars in ecliptic coordinates]{Kaltenegger2019.0}. This characteristic makes the sample ideally suited for a study of flares because the long baseline provides high statistics for such events for any given star. 
Apart from two exceptions, the observation times of all stars within our sample are longer than $140$ days. One star, TIC\,229586790, is only observed in one TESS sector and has an observation time of $24.9$\,d after subtracting the gaps in the LC. Thirty stars within our sample are observed in $13$ TESS sectors, corresponding to observation times of at least $296.6$\,d\footnote{The exact observation times differ from star to star because we subtracted the length of all data point gaps of each LC from its total duration.}, that is they lie within the continuous viewing zone of TESS. Table~\ref{tab:rot_act} gives an overview of the sectors in which each target was observed and the total observation durations. 
%[last sentence moved from sect.~\ref{sect:analysis}]

\begin{figure}[t]
\centering
\includegraphics[width=0.4\textwidth]{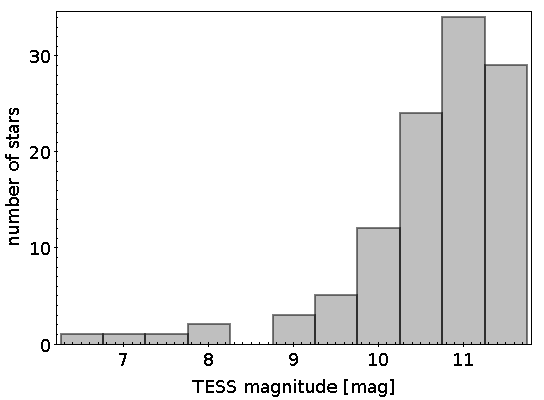}
\caption{TESS magnitude distribution of all stars from the HZCat (\citealt{Kaltenegger2019.0}) with the ``HZflag'' set and $T\leq 11.5$ mag.}
\label{fig:tmag_distr}
\end{figure}

We further introduced a more restrictive magnitude cutoff for our sample than the one applied in the HZCat, selecting only targets with $T\leq11.5$~mag. This is motivated by the fact that fainter stars have a lower signal to noise ratio, and therefore the possibility of finding flares or a rotation period in the LC analysis is lower. Fig.~\ref{fig:tmag_distr} shows the TESS magnitude distribution of our sample. Applying these selection criteria results in a sample of $112$ stars, covering a SpT range from K8 to M5 and masses from $0.14\,{\rm M_\odot}$ to $0.65\,{\rm M_\odot}$. The SpT and mass determination is described below.

\subsection{Calculation of stellar parameters}\label{sec:stellar_params}
We calculated stellar parameters from the empirical relations in \cite{Mann2015.0} (using coefficients from \citealt{Mann2016.0}), \cite{2019ApJ...871...63M}, \citet[][S16]{Stelzer2016.0} and \cite{Jao2018.0}. These relations involve the color indices $V-J$ and $J-H$ as well as the absolute magnitude in the 2MASS $K_s$ band, $M_{K_s}$. 

Therefore, we started by collecting optical and infrared photometry. We used the {\it Gaia} DR2 (\citealt{2016A&A...595A...1G}, \citealt{2018A&A...616A...1G}) and the Two Micron All Sky Survey (2MASS, \citealt{2006AJ....131.1163S}) identifiers given in the HZCat for all sample stars. {\it Gaia} and 2MASS data were  accessed via TOPCAT (\citealt{Taylor2005.0}). We extracted 2MASS photometry by using the Table Access Protocol to perform an ADQL query based on the 2MASS IDs. For the {\it Gaia} DR2 data, we performed a coordinate-based multi-cone search in TOPCAT since in that way error estimates for the {\it Gaia} magnitudes could be collected, which are not available in the {\it Gaia} archive directly. The search was based on the TIC-coordinates (epoch J2000.0). The purpose of the cone search was not to find correct {\it Gaia} counterparts since they are already listed in the HZCat, but to retrieve the {\it Gaia} photometry and distances. Therefore, we could carry out the match to {\it Gaia} J2015.5 coordinates without a proper motion correction using a large search radius and subsequently identifying the correct {\it Gaia} source among the multiple matches by the help of the HZCat. With a match radius of $20^{\prime\prime}$ we recovered 108 out of the $112$ targets.
For four targets with particularly high proper motions, >$1^{\prime\prime}$/yr (TIC\,199574208, TIC\,233193964, TIC\,392572237 and TIC\,359676790), we had to repeat the search with a larger match radius to catch the correct {\it Gaia} counterpart among the matches. The maximum match separation was $36.1^{\prime\prime}$.

\begin{figure}[t]
\centering
\includegraphics[width=0.4\textwidth,trim=1cm 0 0 0]{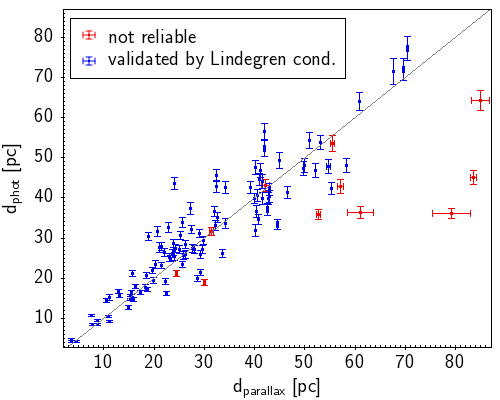}
\caption[Relation of photometric distance and {\it Gaia} parallax distance]{Relation of photometric distance and {\it Gaia} parallax distance. Stars failing the conditions of \cite{Lindegren2018.0} for a reliable {\it Gaia} parallax are marked in red. The gray line represents the \textit{1:1}-relation.}
\label{fig:distances}
\end{figure}

In order to calculate $V$ magnitudes from {\it Gaia} photometry, we applied the empirical relation in Table~2 of \cite{Jao2018.0}. This relation is valid in a range of $1.0\le G_{\rm BP}-G_{\rm RP}\le 4.0$, where $G_{\rm BP}$ and $G_{\rm RP}$ are the Gaia magnitudes in the blue and red photometer filters. All stars of our sample fulfill this requirement.
For the distances, we used the inverse {\it Gaia} parallaxes. The results were subject to a reliability check following \citet{Lindegren2018.0}, Appendix~C, Equations (C.1) and (C.2). For most targets ($101$ out of $112$), we found the {\it Gaia} distances to be reliable.
For these stars, the absolute $K_{\rm s}$-band magnitude, $M_{\rm K_s}$, was then determined using the distance modulus. Extinction can be neglected for these nearby stars.
Since $11$ targets fail the Lindegren reliability check for the {\it Gaia} parallaxes, we also  determined %the 
photometric distances, $d_{\rm phot}$. To this end, we calculated $M_{\rm K_s}$ from the empirical $M_{\rm K_s}$ versus $V-J$ relation given in Eq.~1 of S16. In order to obtain this relation, S16 performed a linear fit on the $M_{\rm K_s}$ versus $V-J$ diagram of $1078$ M dwarfs with trigonometric parallaxes given in the \textit{All-sky Catalog of Bright M Dwarfs} (\citealt{2011AJ....142..138L}).
The $M_{\rm K_s}$ values determined by applying the S16 relation were then combined with the observed 2MASS $K_{\rm s}$ magnitude to yield $d_{\rm phot}$.
Fig.~\ref{fig:distances} shows the relation of photometric distance to {\it Gaia} parallax distance for all targets. As expected, the ones with a large deviation between both distance values are found to have no reliable {\it Gaia} parallax according to the Lindegren quality conditions.
Throughout this work we use the {\em Gaia} distances except for the stars that do not fulfill the abovementioned quality criteria. For these latter ones we use the photometric distances. 
%\textcolor{orange}{Can someone please clarify if we used d\_phot when Gaia distances are "not reliable"?} \textcolor{teal}{We already talked about it, there are indeed some stars for which dphot was adopted $(8/109 \sim 7$\,\%).}

\citet[BJ18]{2018AJ....156...58B} provide distance estimates for all $1.33$ billion stars that have parallax entries in {\it Gaia}\,DR2, using a distance prior that varies depending on the Galactic longitude and latitude of each star. We extracted the distance estimates provided by BJ18 and compared them to the inverse parallax for our sample stars. The deviations are marginal (with the maximum difference being $0.2$\,pc), lending further credibility to the inverse parallaxes.
One reason for the small differences is probably the fact that all stars within our sample are nearby (they all have distances $<100$\,pc).

\subsubsection{Spectral types}\label{sec:SpT_determination}
\begin{figure}[t]
\centering
\includegraphics[width=0.4\textwidth]{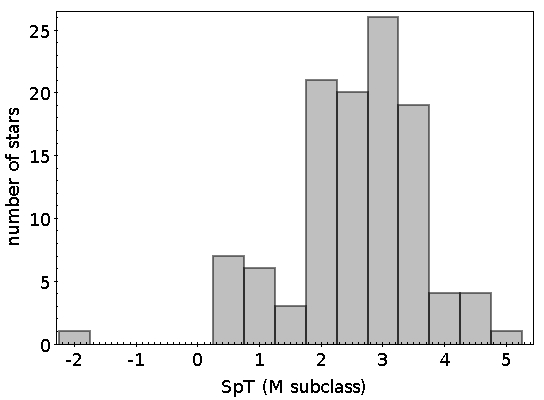}
\caption{SpT distribution of our sample. Numbers denote M subclasses. -2 stands for SpT K8.}
\label{fig:spt_distr}
\end{figure}

Spectral types were determined from $G_{\rm BP}-G_{\rm RP}$ and $G-G_{\rm RP}$ color indices using the online table {\it A Modern Mean Dwarf Stellar Color and Effective Temperature Sequence} maintained by E. Mamajek\footnote{\label{note1}The table is available at\\pas.rochester.edu/\textasciitilde emamajek/EEM\_dwarf\_UBVIJHK\_colors\_Teff.txt} as an extension of \citet{2013ApJS..208....9P}. In that table, SpTs are given in steps of half-integer subclasses. We calculated for each star the expression 
\begin{gather*}
|(G_{\rm BP}-G_{\rm RP})_*-(G_{\rm BP}-G_{\rm RP})_{\text{SpT}}|~+\\\hspace{4cm}|(G-G_{\rm RP})_*-(G-G_{\rm RP})_{\text{SpT}}|
\end{gather*}
where the subscript ``$*$'' denotes the color indices of the star and the subscript ``SpT'' the ones listed in the table for a particular SpT subclass. Then we assigned to each target the tabulated SpT that minimizes this expression. Fig.~\ref{fig:spt_distr} shows the resulting SpT distribution of our sample.

We verified that the SpTs derived in that way are consistent with the ones obtained from the SpT-color relation provided by \citet[R20]{Raetz2020.0} that was used in our previous works. R20 fitted a seventh order polynomial to the SpT versus $V-J$ relation extracted from the above-mentioned online table of E. Mamajek\footnotemark[\value{footnote}].

\subsubsection{Mass and bolometric luminosity}

Stellar masses ($M_*$) were obtained by applying the Markov chain Monte Carlo (MCMC) analysis described by S16: For each star, a set of 10000 $M_{K_s}$ values is drawn from a normal distribution with the observed value as mean and its error as standard deviation. The stellar mass is  then calculated for all $10000$ values in the data set by applying Eq.~4 in \cite{2019ApJ...871...63M}, using coefficients from their Table~6 for $n=5$, that is the case in which the relation is represented by a fifth order polynomial. \cite{2019ApJ...871...63M} denote this fifth order fit as the preferred one.

The stellar mass for each sample star is  then determined as mean of the results of all $10000$ data sets for this star. To estimate the uncertainties, the standard deviations of the MCMC analysis are added in quadrature to the error given for the empirical relation.

Bolometric corrections in the $V$ band ($BC_V$) are given in the online table by E. Mamajek\footnotemark[\value{footnote}]. We determined the bolometric corrections for our sample from this table, using SpTs as input.

The stellar parameters for all $112$ sample stars are listed in Table~\ref{tab:stellar_params_all}. It comprises the TIC identifier (col.~1),  the literature name (col.~2, obtained by searching SIMBAD for the main identifier of each star), {\it Gaia} DR2 (col.~3) and 2MASS (col.~4) identifiers, right ascension (RA, col.~5) and declination (Dec, col.~6) from the TIC, the TESS magnitude ($T$, col.~7), stellar mass ($M_*$, col.~8), distance ($d$, col.~9), SpT (col.~10) and quiescent luminosity in the TESS band ($L_{\rm qui,T}$, col. 11).
%\textcolor{orange}{Please check if the stellar radius is needed somewhere in the paper. If yes, it has to be presented in the table.} \textcolor{teal}{The stellar radius is already shown in Table~A.2}
%\textcolor{orange}{Tab A2 is only for the binary stars.}

\subsection{Common proper motion pairs}\label{sec:sample_CPMs}

Binary stars require special treatment since for many of them, the separation between the target and its companion is on the order of the TESS pixel size ($21^{\prime\prime}$), and consequently the contribution of each star to the total flux of the TESS LC cannot be determined. This implies that characteristics observed in the LCs of targets with a visual companion might not originate in the target star itself. Additionally, close binaries are known to show higher flare rates than single stars, which might be a consequence of magnetic interactions between the components (e.g., \citealt{Huang_2020}). Thus, close binaries might exhibit a behavior in terms of magnetic activity that is different from that of single stars. These magnetic interactions, however, only occur in case of small physical separation. The $12$ eclipsing M dwarf binaries studied by \cite{Huang_2020}, for example, have physical separations $\leq 0.25$\,AU. 

Our sample has been checked carefully for binary pairs, using two different approaches: 

First, we compared the {\it Gaia} DR2 proper motion (PM) vectors for all {\it Gaia} sources within the TESS target pixel file (TPF) using a $164^{\prime\prime}$ radius (see Sect.~\ref{sec:contamination}) to the target's PM. We consider objects with a deviation of less than $10$\,\% (in right ascension and declination, respectively) as companion candidates. Ten CPM pairs were identified with this procedure and verified through visual inspection in ESASky\footnote{ESASky is an application to visualize and download archived astro-
nomical data that is developed at ESAC, Madrid, Spain, by the ESAC
Science Data Centre (ESDC). It is available at https://sky.esa.int
}.

Second, the TIC coordinates of all sample stars were matched in TOPCAT with the Washington Double Star Catalog (WDS, \citealt{2020yCat....102026M}), using a match radius of 10$^{\prime\prime}$. This resulted in $19$ matches. All these $19$ targets except for two were confirmed to have a PM companion by visual inspection in ESASky. The first exception is TIC\,198187008, for which a {\it Gaia} counterpart was not present in DR2 and could only be found in eDR3, where it still has no PM data and therefore cannot be confirmed to share the target's PM. According to the WDS entry, the separation is extremely small ($0.4^{\prime\prime}$), which is probably the reason for the lack of {\it Gaia} data of the companion star. The WDS entry of TIC\,198187008 is provided with the flag ``V'', meaning ``Proper motion or [an]other technique indicates that this pair is physical''\footnote{WDS catalog description on Vizier, cdsarc.unistra.fr/viz-bin/ReadMe/B/wds?format=html, accessed 2021-10-02}. We flag TIC\,198187008 as part of a CPM pair for the further analysis. The second target with a WDS entry for which no proper motion companion was found in ESASky is TIC\,233193964. In this case, the separation of the system given in the WDS is between $230^{\prime\prime}$ and $300^{\prime\prime}$, and thus the companion is outside the TESS TPF. Therefore, the star was treated as a single star in the further analysis.

Comparing the results of the two methods, we found that $9$ of the $19$ WDS matches have also been identified as CPM pairs by our PM comparison. One target, TIC\,142086813, was found from our PM comparison to have a companion while the WDS match yielded no result.

Thus, the total number of stars found to have a PM companion is $20$. Apart from TIC\,233193964, one other binary has a very large separation of the two components: TIC\,459985740 with $450^{\prime\prime}$. Treating these two targets as single stars due to the large separation to their companions, 18 stars remain to be flagged as part of a CPM pair for further analyses.

For two CPM pairs, both components are included in our HZCat sample: TIC\,142086812 is the companion of TIC\,142086813 and TIC\,359676790 is the companion of TIC\,392572237. Therefore, the $18$ stars within the sample that have a PM companion belong to $16$ CPM pairs.

Table \ref{tab:CPMs} lists for these $16$ CPM pairs the angular separation calculated based on {\it Gaia} DR2 coordinates and taken from the WDS in case of missing {\it Gaia} DR2 coordinates for the companion star. The table also contains the distance, $G_{\rm RP}$, stellar parameters ($R_*$, $M_*$, $T_{\rm eff}$) and SpT of both components except for two cases where the {\it Gaia}\footnote{Some companion objects only appear in {\it Gaia} eDR3 that was released in December 2020, namely the companions of TIC\,198187008 (without PM) and TIC\,141025090 (with PM). For the sake of uniformity, we did not use eDR3 data since all other calculations of this work are based on {\it Gaia}\,DR2.} or 2MASS data of the companion object is incomplete. The stellar parameters of M-type companions were determined as described in Sect.~\ref{sec:stellar_params}. Small separations on the order of a few arcseconds lead to the CPM pair not being resolved in 2MASS. We caution that the empirical relations for stellar parameter calculations (cf. Sect.~\ref{sec:stellar_params}) cannot be applied to such unresolved cases. This also means that the stellar parameters of the corresponding sample stars might not be correct because the $J$, $H$ and $K_{\rm s}$ magnitudes for targets with unresolved PM companions given in the 2MASS catalog are the sum of the magnitudes of the CPM pair.
Nearly all resolved companion objects are M dwarfs and span a SpT range from K9V to M5V. The only exception is the eclipsing binary CM\,Draconis that has a White Dwarf companion (see Appendix~\ref{sec:CMDra_treatment} for the treatment of this star).
We calculated the
physical separation for all CPM pairs within our sample and found a minimum value (apart from CM\,Dra) of about $14$\,AU.
%for TIC198187008. 
This is {more than two} orders of magnitude above the maximum separation of the magnetically interacting binaries 
studied by \cite{Huang_2020}, which is, as mentioned above, $0.25$\,AU.
CM\,Dra is 
%also 
the 
only binary system within our sample that has a physical separation small enough for the components to interact magnetically 
%(angular sep.$\times$distance 
($\approx 0.02$\,AU; see Appendix~B for details). 
%All other systems have absolute separations $\gg$1000 AU (1 pc $\sim 10^5$ AU, angular separations are $\geq0.4^{\prime\prime}$) and are therefore not expected to interact magnetically.

%%%%%%%%%%%%%%%%%%%%%%%%%%%%%%%%%%%%%%%%%%%%
\section{Data analysis}\label{sect:analysis}
%%%%%%%%%%%%%%%%%%%%%%%%%%%%%%%%%%%%%%%%%%%%%

In this work, we only considered data from the primary TESS mission (that is the first $26$ sectors)  obtained in 2-minute cadence mode. In total, $1276$ 2-minute Pre-search Data Conditioning Simple Aperture Photometry (PDCSAP) TESS LCs and TPFs are available at the Mikulski Archive for Space Telescopes (MAST)\footnote{mast.stsci.edu} for our $112$ sample stars. We filtered the data using the TESS quality flags: Almost all data points with flags $\neq0$ were dismissed. We only kept ``impulsive outlier'' (value $512$, Bit $10$) as those data points might be part of a real flare and ``cosmic ray detected in collateral pixel row or column'' (value $1024$, Bit $11$).

Except for one star, TIC\,229586790, all targets were observed in multiple TESS sectors (see Table~\ref{tab:rot_act}). As explained in Sect.~\ref{sect:sample}, this is a consequence of the target selection: We chose only stars from the HZCat that were observed long enough for TESS to probe the entire HZ for transiting planets.
%(see Sect.~2). Indeed, many sample stars have observation lengths > 200 days, i. e. they lie within the continuous viewing zone of {\it TESS} near the ecliptic poles.
%Table~\ref{tab:rot_act} gives an overview of the sectors in which each target was observed and the total observation durations.

%%%%%%%%%%%%%%%%%%%%%%%%%%%%%%%%%%%% 
\subsection{Contamination analysis}\label{sec:contamination}
%%%%%%%%%%%%%%%%%%%%%%%%%%%%%%%%%%%%

TESS data are  processed by a pipeline developed at {\it NASA}'s Science Processing Operation Center (SPOC, \citealt{2016SPIE.9913E..3EJ}). Due to the coarse TESS pixels that comprise an area of $21^{\prime\prime} \times 21^{\prime\prime}$ each, the aperture chosen by the SPOC pipeline often contains other stars besides the actual target.
Even though the SPOC pipeline corrects for flux excess caused by star field crowding, these corrections cannot account for the possible variability of contaminating objects. Therefore, flares and the rotational signal in a LC with several stars inside the pipeline mask cannot be unambiguously ascribed to our target star. 
To identify such potential problematic cases, we conducted a contamination analysis based on fluxes in the Gaia $G_{RP}$ band since its wavelength range is a good approximation to the TESS bandpass (e.g.,  \citealt{2018A&A...619L..10G}). %Thus, our results do not represent the exact amount of flux contributed to the {\it TESS} LC by the contaminating source, given the corrections applied by the SPOC pipeline. %\textcolor{orange}{What exactly is meant with the phrase above? Can someone make it clearer?} %\textcolor{brown}{SR: I also do not understand exactly what is meant by that sentence. This should be rephrased. We only should make clear, that the SPOC is correcting the flux level but is not correcting for potential variability of the contaminants. We just want to understand if a contaminant could be responsible for the detected rotation signal but we do not do an own correction.} 
%However, the quiescent flux level is a benchmark for the amplitude of variability characteristics.
%\textcolor{orange}{Also here, not sure what it means. Can somebody clarify?}
%\textcolor{brown}{SR: I guess the sentence means that contaminants can affect the light curve if they are variable. The value we calculate tells us the maximum amplitude they could produce if they completely vanish.}
%Therefore, 

Our contamination factors provide a reasonable estimate to which extent 
another 
%contaminating 
source might affect the LC. To calculate them, we proceeded as follows: 
First, we listed all {\it Gaia} sources that are located within a circular area of radius $164^{\prime\prime}$ (distance from center of TESS TPF to the corners, $0.5\cdot\sqrt{2}\cdot11\cdot21^{\prime\prime}\approx 163.3^{\prime\prime}$) around the target. Then, we checked for each {\it Gaia} source if it lies within a radius of $14.85^{\prime\prime}$ (half pixel diagonal) around the center of any of the pipeline mask pixels. Finally, the flux of all sources in the mask except for the target was summed up and the ratio of this sum to the target flux is defined as a contamination factor.
%\textcolor{blue}{\bf In cases where the contamination is caused by several additional sources, the contamination factor is a benchmark for the upper limit of contamination, assuming the rather unlikely case that all contaminating sources are synchronized in their variability.}

Table~\ref{tab:rot_act} in the Appendix lists the mean, minimum and maximum contamination factors of all observation sectors for each target. There are $18$ stars with a maximum contamination $>10\%$. Ten of them are part of a CPM pair (see Sect.~\ref{sec:sample_CPMs}) with companions listed in Table~\ref{tab:CPMs}.

We discovered during the contamination analysis that three of our targets lie outside the TESS pipeline mask. Their LCs show many systematic effects that could be accidentally validated as flares in the LC analysis. These stars are TIC\,359676790, TIC\,392572237 and TIC\,471015740. We did not consider these stars any further in the LC analysis. 
One reason for targets lying outside the mask might be a nearby contaminating source that the pipeline algorithm intended to exclude and then accidentally excluded also the target. 

For the binaries in our sample listed in Table~\ref{tab:CPMs}, the maximum contamination factor of all TESS observation sectors is $>5$\% for all systems resolved with \textit{Gaia}. It is $>10$\% in all except three cases (cf. Tab.~\ref{tab:rot_act}). We note that for systems unresolved with \textit{Gaia}, contamination is underestimated since the unresolved CPM pair as a whole is treated as the target source.

%%%%%%%%%%%%%%%%%%%%%%%%%%%%%%%%%%%%%%%%%%%%%%%%%%%%%%
\subsection{Rotation period search}\label{period_search}
%%%%%%%%%%%%%%%%%%%%%%%%%%%%%%%%%%%%%%%%%%%%%%%%%%%%%%%%
\begin{figure}[t]
\centering
\includegraphics[width=0.5\textwidth,trim=0cm 0 0 0]{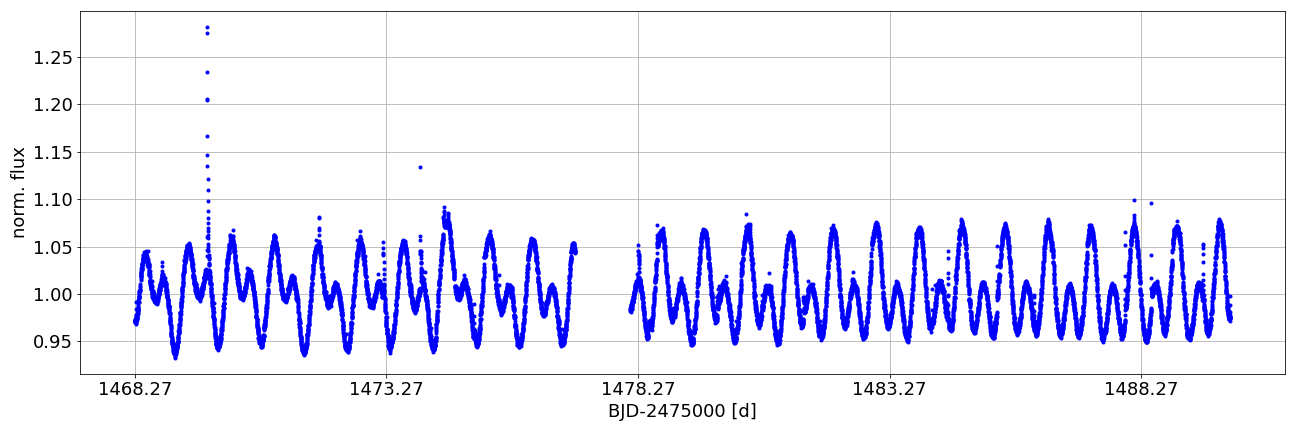}
\caption[Normalized PDCSAP LC of TIC220433364 in sector 6]{Example for a typical normalized PDCSAP LC, showing TIC220433364 in sector~6. The LC shows two humps of different height possibly resulting from two starspots on opposite sides of the star. Right before the $1478.27$\,d timestamp, there is the typical gap that can be observed in every TESS LC due to a break in observation while downloading data to Earth (cf. Sect.~\ref{period_search}).}
\label{fig:doublehump}
\end{figure}

The search for rotation periods ($P_\text{rot}$) was carried out using three different methods: autocorrelation function (ACF), Lomb-Scargle-Periodogram and sinefit, following our previous works (\citealt{Stelzer2016.0}, \citealt{Raetz2020.0}). In the version of the routine used for this work, the LC is only searched for periods up to a maximum length for which still two complete periods are covered by the duration of the LC, which is about $27$~d for individual TESS sectors. Thus, detectable periods are a priori limited to a maximum value of $13.5$~d.
To calculate the generalized Lomb-Scargle-Periodogram (GLS), we used an algorithm by \cite{2009A&A...496..577Z}\footnote{astro.physik.uni-goettingen.de/\textasciitilde zechmeister/gls.php, Fortran v2.3.02, released 2012-08-03}. The rotation period value resulting from GLS was then used as an initial guess for the sine-fit.
The third method, ACF, faces some difficulties when it comes to TESS LCs due to the data gap in the middle of each observation sector (cf. example LC in Fig.~\ref{fig:doublehump}). During this gap, data are  transferred from the satellite to Earth and therefore no measurements can be recorded (cf. \citealt{TESSobsguide}). As the ACF method needs equally spaced data points, we interpolated the gap in the middle as well as eventually occurring shorter gaps with a linear fit between the last data point before and the first one after the gap. Then, the ACF was calculated using the IDL\footnote{IDL is a product of the Exelis Visual Information Solutions, Inc.} routine A\_CORRELATE. The rotation period was obtained by determining the time lag of the first peak in the ACF.
The phase-folded LCs obtained from the three methods were plotted and evaluated by eye to verify the presence of rotational modulation.

One challenge in the search for stellar rotation periods in general is the fact that several LCs show ``double-humps'' resulting for instance from one larger and one smaller starspot on different hemispheres of the star (e.g.,  \citealt{2013MNRAS.432.1203M}). Fig.~\ref{fig:doublehump} shows an example of such a LC. For those cases, a by-eye inspection of the LC was necessary not only in order to decide if they show starspot variations but also to determine the correct value of the rotation period. The search algorithms often detected the distance between a smaller and larger hump as rotation period instead of the spacing between equal humps.

\begin{figure}[t]
\centering
\includegraphics[width=0.5\textwidth]{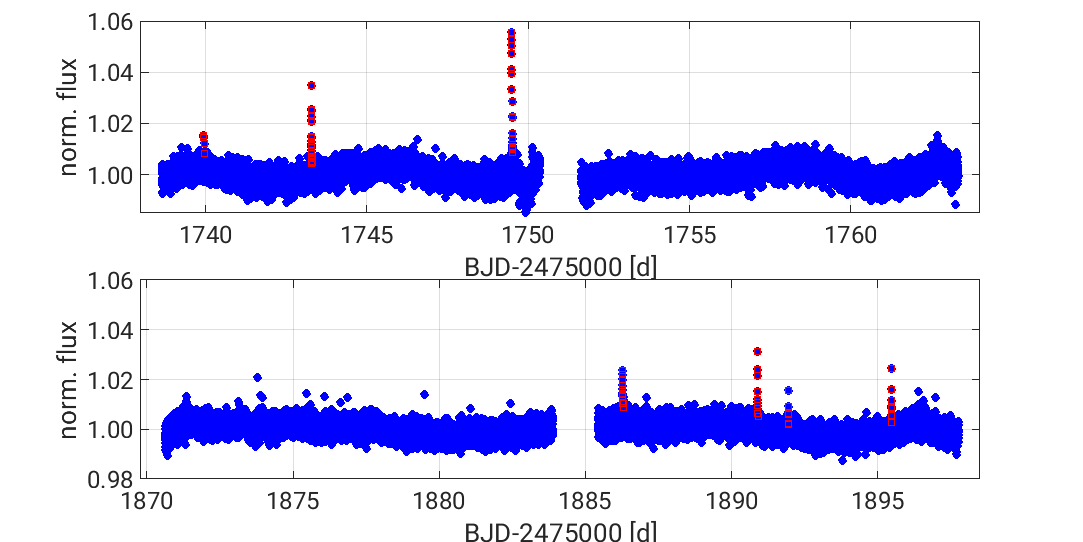}
\caption[LC of TIC\,320525520 in sector 16 and 21]{LC of TIC\,320525520 in sector~16 (top panel) and sector~21 (bottom panel). Validated flares are 
marked. 
%represented by red square markers. 
Both LCs show indications of rotational modulation, but the rotation period is not well constrained.}
\label{fig:unclear_prot}
\end{figure}

For a few targets, a rotation period was only found in some observation sectors while in others there was no rotational modulation evident or the differences between $P_{\textrm{rot}}$ values in different sectors were too big so that no clear cut result could be obtained. Fig.~\ref{fig:unclear_prot} shows as an example the LCs from two sectors of a star with not well-constrained rotation period. We disregard the periods of such cases in our work. We only consider rotation periods of stars with ``reliable'' $P_{\textrm{rot}}$ results and define ``reliable'' as follows: A rotation period is detected and is consistent in all observation sectors.

The adopted $P_{\rm rot}$ value for a given star was obtained from the results of all sectors in which it was observed as follows: First, we calculated the mean value of the $P_{\textrm{rot}}$ obtained from the three methods for each observation sector. In some cases, one method clearly failed to determine the correct rotation period. This became evident in the course of the visual inspection of the phase-folded LCs of all three methods, and the corresponding values were not taken into account. As a next step, we estimated errors for the averaged $P_{\rm rot}$ values of each sector using Eq.~2 of \cite{2004A&A...417..557L}. This equation involves the width of the peak of the window function, $\delta\nu$. According to \cite{Roberts1987}, this width can be approximated by $\delta\nu\approx 1/T$ with the length of the observation, $T$. This is possible for data sets with a sampling that is not too nonuniform. We used this approximation since all TESS LCs we analyzed have roughly the same length and the same regular two-minute cadence and are therefore uniform. 
The errors obtained from applying Eq.~2 of \cite{2004A&A...417..557L} were added in quadrature to the standard deviation of the results from the different methods to obtain the total error, $\Delta P_{\rm rot,tot}$, for each observation sector. The overall $P_{\rm rot}$ result for each target is then the weighted mean of all sectors, given by
\begin{equation}
<P_{\textrm{rot}}>=\sum_{\text{sectors}}\frac{P_{\textrm{rot}}}{\Delta P_{\textrm{rot,tot}}}\cdot\bigg(\sum_{\text{sectors}}\frac{1}{\big(\Delta P_{\textrm{rot,tot}}\big)^2}\bigg)^{-1},
\end{equation}
and the corresponding error of the weighted mean is determined as
\begin{equation}
\Delta <P_{\textrm{rot}}>=\frac{1}{\sqrt{\displaystyle\sum_{\text{sectors}}\frac{1}{\big(\Delta P_{\textrm{rot,tot}}\big)^2}}}.
\end{equation}

%%%%%%%%%%%%%%%%%%%%%%%%%%%%%%%%%%%%% 
\subsection{Flare detection and properties}\label{sec:flare_criteria}
%%%%%%%%%%%%%%%%%%%%%%%%%%%%%%%%%%%%%
\subsubsection{Flare detection method}\label{subsubsect:flare_det_method}

Our flare detection algorithm has been described by \cite{Stelzer2016.0} and \cite{Raetz2020.0} where it was applied to {\it K2} mission long and short cadence data. Here, we briefly summarize the basic steps.

First, a ``flattened and cleaned'' LC is created in order to identify outliers as data points that deviate by more than three times the standard deviation of this flattened and cleaned LC ($S_{\textrm{flat}}$). To achieve this, the original LC is subject to an iterative boxcar-smoothing and cleaning process. In this procedure, after boxcar-smoothing is applied, the smoothed LC is subtracted from the original LC in order to remove the rotational modulation caused by starspots. Then, all data points with a deviation larger than $2S_{\textrm{flat}}$ are removed. This process is repeated three times with the boxcar width being reduced in each round.
As a next step in the procedure, all parts of the LC with at least 3 consecutive data points of the original LC deviating by $\ge 3 S_{\rm flat}$ each from the mean value of the flattened and cleaned LC are flagged as potential flares. Then, there are further criteria applied in order to decide if the potential flare is validated. In short,
% \begin{center}
% \begin{tabular}{clc}
%
%&
%
(1) The flare event must not occur right before or after a gap in the LC; 
(2) 
the flux ratio between flare maximum and last flare point must be $\ge  2$;
(3) the flare maximum cannot be the last flare point;
(4) the decay time has to be longer than the rise time;
(5) a fit conducted using the flare template defined by \cite{Davenport14.0} must fit the flare better than a linear fit.
% \end{tabular}
% \end{center}
 
\subsubsection{Flare parameters}\label{subsec:flare_params}

For all flares validated with these criteria, we determined the normalized peak flare amplitude, $A_{\textrm{peak}}$, namely the difference between peak flare flux and the flux of the smoothed, interpolated LC at the time of the flare peak, normalized to the mean flux of the LC. We further determined flare start, flare end and flare duration, $\tau_F$. This latter one is the time difference between first and last flare data point, where the last flare point is the last data point after the flare peak that deviates by $\ge1S_{\textrm{flat}}$ from the mean value of the flattened and cleaned LC. Other flare parameters following from our analysis are the time of the flare maximum, the number of flare data points and the equivalent duration (ED). The ED is the integral of the LC under the flare data points. As the LC is normalized to the star's mean flux, the unit of the ED is seconds (integrating a dimensionless quantity over time). It corresponds to the time span it would take for the star to radiate away the energy of the flare at its constant, quiescent luminosity (e.g., \citealt{2012PASP..124..545H}).
 
In order to get the flare energy, $E_F$, in units of erg, the ED is multiplied by the quiescent stellar luminosity in the TESS band, $L_{\rm qui,T}$. We calculated the luminosity as $L_{\rm qui,T} = 4 \pi d^2 f_{\rm T}$, where $d$ is the distance and $f_{\rm T}$ the flux in the TESS band. The latter is obtained from the TESS magnitude ($T$) using the distance $d$ of the star, 
%$\lambda_{\textrm{eff}}=7455.64~\mathrm{\mathring{A}}$ 
the filter bandwidth $W_{\textrm{eff}}=3898.68~\mathrm{\mathring{A}}$ 
and the zeropoint  $ZP_{\lambda}=1.34\cdot10^{-9}~\mathrm{erg/cm^2/s/\mathring{A}}$. Both values were taken from the website of the Spanish Virtual Observatory.\footnote{svo2.cab.inta-csic.es/theory/fps/} 

The flare amplitude in the TESS band ($\Delta L_{\rm F,T}$) in units of erg/s is calculated by multiplying the normalized flare amplitude $A_{\rm peak}$ with $L_{\rm qui,T}$. $\Delta L_{\rm F,T}$ is the luminosity increase caused by the flare at its maximum. The total stellar luminosity in the TESS band at the flare peak is obtained by adding $\Delta L_{\rm F,T}$ and $L_{\rm qui,T}$.

%\textcolor{red}{\bf We note that the above definition of the stellar luminosity $L_{\rm qui, T}$ is lower than the one used in our previous work, e.g. \cite{Raetz2020.0}, by the volume factor four. It reflects the fact that only the projected area of the star is seen, and puts our values for the flare energies on the same scale as those works that have used the formalism by \cite{Shibayama13.0} to calculate flare energies.}

Flare rates, $\nu_{\rm F}$, were calculated by dividing the total number of validated flares (e.g., for a specific target or SpT) by the total observation time. The observation time for each individual LC is here defined as the difference between its first and last time stamp, subtracted by the summed up time spans of all data gaps. A data gap in this case is a sequence of at least two consecutive time stamps for which no flux measurements exist. This also applies to gaps at the beginning or end of the LC, in other words  data gaps between sectors. 

%%%%%%%%%%%%%%%%%%%%%%%%%%%%%%%%%%%%%%%%%%
\subsubsection{Cumulative flare energy frequency distribution}\label{sec:FFD_theory}
%%%%%%%%%%%%%%%%%%%%%%%%%%%%%%%%%%%%%%%%%%%%%%%

Cumulative FFDs are a common way to represent flare occurrence rates as a function of flare energies (e.g., \citealt{1976ApJS...30...85L}, \citealt{2014ApJ...797..121H}, \citealt{Ilin19.0}, \citealt{Raetz2020.0}). The cumulative flare frequency, $\nu (>E_{\rm F,0})$, for a given flare energy, $E_{\rm F,0}$,  is the total number of flares $N_{\textrm{flares}}(>E_{\rm F,0})$ with $E_{\rm F}\ge E_{\rm F,0}$ divided by the total observation duration of the target (see Table~\ref{tab:rot_act} for the observation durations of our sample stars). To obtain the FFD, the cumulative frequencies are calculated and  plotted against all values of $E_{\rm F,0}$ present in the flare sample. 

Usually the FFDs are represented in double logarithmic form where they can be approximated by a linear relation:
\begin{equation}
\text{log}(\nu_{\rm F,0}):=\text{log}\bigg(\frac{N_{\textrm{flares}}}{\Delta t}\bigg)=\text{log}(\beta) + \alpha\cdot \text{log}(E_{\rm F,0})
\label{eq:FFD}
\end{equation}
(cf. \citealt{1976ApJS...30...85L}), which is equivalent to the exponential power law $\nu_{\rm F,0}=\beta\cdot E_{\rm F,0}^{\alpha}$ in linear representation. It is known from observations of the Sun \cite[e.g.][]{Aschwanden00.0} and other stars \cite[e.g.][]{Audard00.0, Shibayama13.0} that $\alpha$ is a  negative value, since high-energy flares are less frequent than low-energy flares. 
In Sect.~\ref{subsect:ffd} we describe our analysis of the FFDs for the individual stars in terms of this power-law.

%%%%%%%%%%%%%%%%%%%%%%
\section{Results of rotation period and flare search}\label{sec:analysis_resulst}
%%%%%%%%%%%%%%%%%%%%%%
%%%%%%%%%%%%%%%%%%%%%%%%%%%%%%%%
%%%%%%%%%%%%%%%%%%%%%%%%%%%%%%%%
\subsection{Rotation periods}\label{sec:prot_results}
%%%%%%%%%%%%%%%%%%%%%%%%%%%%%%%%
%%%%%%%%%%%%%%%%%%%%%%%%%%%%%%%

\begin{table*}[t]
\centering
\captionof{table}[$P_{\textrm{rot}}$, SpT and {\it TESS} magnitude for targets with clearly measurable rotation periods]{
%Targets with reliable rotation periods. From left to right the columns present the identifier in the TIC, other star name, mean rotation period, SpT, {\it TESS} magnitude and parameters of FFDs (mean detection threshold energy, 
%%$\langle E_{\rm min,th} \rangle$, 
%rate of flares %from those %flares with energy 
%above the individual sector detection threshold,  
%%of each sector where the star has been detected $\nu_\text{\rm flares,>Emin}$, 
%slope $\alpha$, normalization $\beta$).
Basic stellar properties and results from the analysis of FFDs for the $12$ targets with reliable rotation period.
}
\begin{tabular}{llclrcccccc}
\hline
TIC ID & name &  $<P_{\textrm{rot}}>$ & SpT & $T$ & $\log{(\langle E_\text{min,th}\rangle})$ & {$\nu_\text{\rm flares,>Emin}$} & $\alpha$ & log($\beta$)    \\
       & & [d] & & [mag] & [erg] & {[d]$^{-1}$} &(cf. Eq. \ref{eq:FFD}) & (cf. Eq. \ref{eq:FFD})     \\
\hline
198187008\tablefootmark{**}   &          & 1.573$\pm$0.017 & M3V   & 11.44 &  32.80 &   0.18 &  $-$1.17$\pm$0.04 &  37.69$\pm$1.13\\
199574208\tablefootmark{**}   & CM Dra   & 1.268$\pm$0.015 & M4V   & 10.36 &  31.87 &   0.14 & $-$0.96$\pm$0.07 &  29.68$\pm$2.11\\
220433364\tablefootmark{**c}  & GJ2036 B & 0.854$\pm$0.005 & M4V   & 9.38  &  32.36 &   0.52 & $-$1.19$\pm$0.05 &  38.51$\pm$0.86\\
233068870                     & LP71-82  & 0.280$\pm$0.001 & M5V   & 10.25 &  31.41 &   0.37 & $-$1.005$\pm$0.05 &  31.15$\pm$1.65\\
233532220                     & G227-45  & 0.851$\pm$0.004 & M3V   & 11.41 &  32.63 &   0.10 & $-$0.84$\pm$0.06 &  26.56$\pm$1.98\\
233738219                     &          & 1.316$\pm$0.010 & M3.5V & 10.88 &  32.02 &   0.07 & $-$0.84$\pm$0.11 &  25.88$\pm$3.43\\
272232401                     & L34-26   & 2.831$\pm$0.055 & M3V   & 8.92  &  31.98 &   0.63 & $-$0.86$\pm$0.03 &  27.29$\pm$0.80\\
272785770                     &          & 3.936$\pm$0.087 & M2.5V & 9.89  &  32.62 &   0.28 & $-$1.14$\pm$0.04 &  36.86$\pm$1.03\\
287350461\tablefootmark{**c}  &          & 0.325$\pm$0.001 & M4V   & 11.29 &  32.13 &   0.07 & $-$1.08$\pm$0.26 &  33.48$\pm$8.32\\
359313701                     &          & 1.791$\pm$0.018 & M3.5V & 10.73 &  32.38 &   0.29 & $-$1.00$\pm$0.06 &  31.95$\pm$1.06\\
406857100                     & GJ4053   & 0.522$\pm$0.002 & M4.5V & 10.39 &  31.35 &   0.11 & $-$1.27$\pm$0.13 &  38.92$\pm$4.26\\
441734910 \tablefootmark{c}   &          & 1.357$\pm$0.010 & M3V   & 11.39 &  32.42 &   0.26 & $-$0.85$\pm$0.03 &  27.23$\pm$1.04\\
\hline
\end{tabular}
\tablefoot{
\tablefoottext{**}{Target with a PM companion.}                       %\\
\tablefoottext{c}{Maximum contamination factor in all observation sectors $>10$\,\% (cf. Sect.~\ref{sec:contamination} and Table~\ref{tab:rot_act})}}

\label{tab_prot}
\end{table*}

A total of $12$ stars in our sample of $109$ (that is $112$ minus the three excluded based on the contamination analysis, cf. Sect.~\ref{sec:contamination}) have a ``reliable'' rotation period according to our definition in Sect.~\ref{period_search}.

Table~\ref{tab_prot} lists the adopted $P_{\textrm{rot}}$ values for these $12$ stars together with the results from the analysis of their FFDs, which are discussed in Sect.~\ref{subsect:ffd}. The $P_{\rm rot}$ results are the weighted mean of the measurements in all observation sectors of the star, derived as explained in Sect.~\ref{period_search}. We notice that all $12$ targets are fast rotators with $P_{\rm rot}$ values ranging from $0.28$\,d to $3.93$\,d. Example LCs of one observation sector for each of them are shown in Appendix C.

Among these $12$ stars with reliable $P_{\rm rot}$ there are four that are part of a CPM pair of which two have a maximum contamination factor of all observation sectors that is $>10$\,\%. The other two systems are unresolved in \textit{Gaia}\,DR2, and thus the contamination is underestimated. One additional star has a maximum contamination factor $>10$\,\% without being part of a CPM pair (cf. Table~\ref{tab:rot_act} in the Appendix), meaning the contamination is from an unrelated neighboring {\it Gaia} source.

\begin{figure}[t]
\centering
\includegraphics[width=0.45\textwidth]{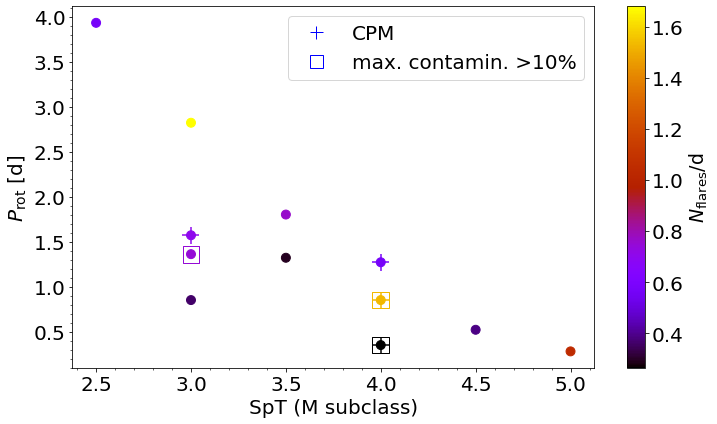}
\caption[Relation between SpT and rotation period]{Relation between SpT and rotation period for the $12$ stars with reliable $P_{\textrm{rot}}$. The flare rate of each target is color-coded. There is a clear tendency of later M-type stars having shorter $P_{\textrm{rot}}$ values.}
\label{fig:prot_SpT}
\end{figure}

As stars evolve with time, they experience changes in their rotation. During the pre-main-sequence (pre-MS) phase, the rotation period is roughly constant (e.g., \citealt{2011ApJ...727...56I}, \citealt{2021A&A...649A..96J}). The spin-up that would follow from the star's contraction is opposed by star-disk interactions that remove angular momentum (e.g., \citealt{1998A&A...333..629A}). After the circumstellar disk has disappeared, rotation is slowed down as angular momentum is carried away by magnetically driven stellar winds. The timescale over which this spin-down of rotation takes place depends on the stellar mass (e.g., \citealt{2011ApJ...727...56I}).
Fig.~\ref{fig:prot_SpT} displays $P_{\rm rot}$ versus SpT in which a strong tendency of later M-type stars having shorter rotation periods can be observed. Such a trend can be explained by the increase of the spin-down timescale for lower stellar masses (e.g., \citealt{2008ApJ...684.1390R}, \citealt{2013MmSAI..84..890S}), which implies that stars of later SpTs are more likely to still be in a relatively fast rotation state.

%%%%%%%%%%%%%%%%%%%%%%
%%%%%%%%%%%%%%%%%%%%%%
\subsection{Flares}\label{sec:flare_intro}
%%%%%%%%%%%%%%%%%%%%%%
%%%%%%%%%%%%%%%%%%%%%%

\begin{table*}[t]
\begin{center}
\caption{
%Flare parameters. Columns contain TIC identifier, time of flare start and maximum, normalized ($A_{\text{peak}}$) and total ($\Delta L_{F,T}$) flare amplitude, equivalent duration (ED) and flare energy. See Sect. \ref{subsec:flare_params} for the calculation of the flare parameters.
%\textcolor{teal}{\bf UPDATED}
Physical parameters of all validated flares calculated as described in Sect.~\ref{subsec:flare_params}.
\label{tab:flare_results}}
%\centering
%\begin{tabular}{ccccccccccccccc}
\begin{tabular}{crrrrrr}
\hline
  \multicolumn{1}{c}{TIC ID} &
  \multicolumn{1}{c}{flare start} &
  \multicolumn{1}{c}{flare max} &
  \multicolumn{1}{c}{$A_{\text{peak}}$} &
  \multicolumn{1}{c}{$\log(\Delta L_{F,T})$} &
  \multicolumn{1}{c}{ED} &
  \multicolumn{1}{c}{log($E_\text{flare}$)} \\
  &[BJD-2475000]&[BJD-2475000]&&[erg/s]&[s]&[erg]\\
\hline
233068870 & 1813.016182 & 1813.016182 & 0.008 &  28.38 & 3.40 &  31.01 \\
''        & 1816.317568 & 1816.317568 & 0.011 &  28.52 & 3.47 &  31.02 \\
''        & 1817.030065 & 1817.030065 & 0.020 &  28.77 & 8.64 &  31.41 \\
''        & 1817.062009 & 1817.062009 & 0.008 &  28.37 & 2.16 &  30.81 \\
''        & 1817.417563 & 1817.418952 & 0.014 &  28.61 & 15.88 &  31.68 \\
''        & 1818.221727 & 1818.221727 & 0.034 &  28.01 & 7.57 &  31.3 \\
''        & 1818.838391 & 1818.838391 & 0.024 &  28.86 & 7.68 &  31.36 \\
...&...&...&...&...&...&...\\
\hline
%\multicolumn{7}{l}{The full table is available in electronic form at the CDS via anonymous ftp to cdsarc.u-strasbg.fr (130.79.128.5).}\\ 
\end{tabular}
\tablefoot{The full table is available in electronic form at the CDS via anonymous ftp to  cdsarc.u-strasbg.fr (130.79.128.5).} 
\end{center}
\end{table*}

The sample contains $55$ targets for which our flare search algorithm detected and validated flares. Three of them had to be excluded from the analysis since the targets lie outside the pipeline mask, as we discovered in the contamination analysis (cf. Sect.~\ref{sec:contamination}).
All validated flares were subject to visual inspection, which resulted in rejecting the flare results of an additional $17$ targets. For a majority of those, the algorithm validated only one single event as a flare and this event was clearly an artifact. Among these $17$ stars, only TIC\,142086813 had several events validated as flares which were all dismissed as artifacts after visual inspection. After the visual inspection, a total of $2532$ validated flares remain, occurring on 35 targets, that is we detected flares on $\approx 32$\% of the total sample. %\textbf{\cite{2019ApJS..241...29Y} present a flare catalog of the {\it Kepler} mission that comprises 3420 flaring stars. Their flare incidence rate for M stars is significantly lower (9.74\%, cf. their Table 3) than our 31\%. An explanation might be that we introduced a magnitude cutoff for our sample, selecting only stars with $T\leq11.5$ mag and therefore with a higher S/N ratio (cf. Sect. \ref{sect:sample}). \cite{2020AJ....159...60G} analyzed a large sample of 24809 stars observed in the first two months of the {\it TESS} mission and find a flare occurrence rate of >40\% for mid to late M stars (SpT M4 to M6) and ~10\% for early M stars. These numbers are in better agreement with our flare occurrence rate of ~31\% although our sample does not comprise an excess of M4 to M6 stars.}\\ 

Seven of the $35$ flaring stars in our sample have a PM companion, cf. Table~\ref{tab:CPMs} in the Appendix. For four of these seven, the maximum contamination factor of all sectors is greater than $10$\,\% and two systems are unresolved by \textit{Gaia}. An additional five of the $35$ flaring stars have maximum contamination factors $>10$\,\% without being part of a CPM pair.

All flare parameters are available in an online table that comprises the TIC ID (col.~1), the time of flare start and maximum (cols.~2 and~3), normalized ($A_{\rm peak}$, col.~4) and absolute ($\Delta L_{\rm F,T}$, col.~5) flare amplitude, equivalent duration (ED, col.~6) and flare energy ($E_{\rm flare}$, col.~7). Table~\ref{tab:flare_results} shows an excerpt of this flare parameter table.

\subsection{Flares and rotation}\label{sec:flares_prot}

\begin{figure}[t]
\centering
\includegraphics[width=0.44\textwidth, trim=0.5cm 0 0 0.5cm]{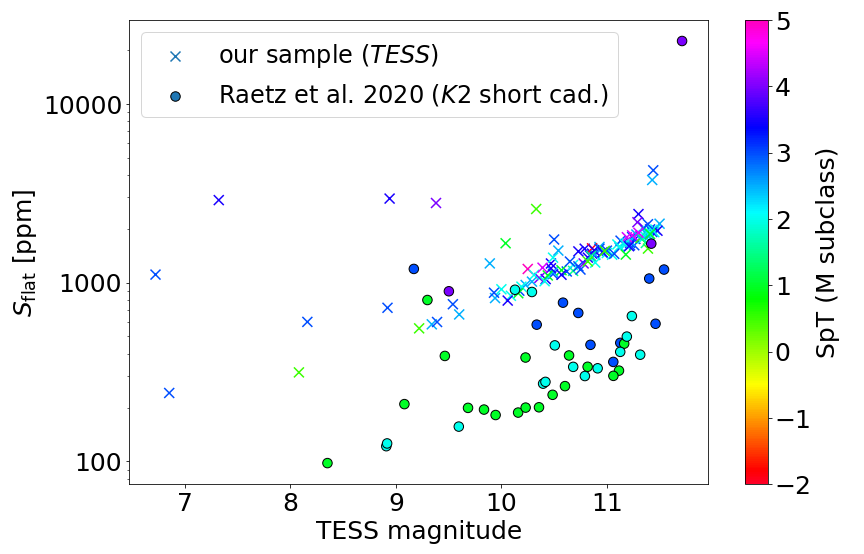}
\caption[Standard deviation of the flattened and cleaned LC vs. TESS magnitude]{Relation between the standard deviation of the flattened and cleaned LC  ($S_{\rm flat}$) and TESS magnitude. Circles represent the $S_{\rm flat}$ in {\it K2} short cadence LCs of the M dwarf sample presented by \cite{Raetz2020.0}. The conversion between {\it K2} and TESS magnitudes is described in Appendix~\ref{app:magtrans}.}
\label{S_flat}

\end{figure}
All targets with reliable $P_{\rm rot}$ also show flares, which is not surprising given that they have measurable starspots that indicate significant magnetic activity. In fact, $2138$ out of $2532$ validated flares occur on the $12$ targets with unambiguous rotation periods. The remaining $394$ flares partly occur on stars that show some indication for rotational modulation (described in Sect.~\ref{period_search}) and some occur on targets without any indication of starspot modulation. From their {\it Kepler} flare studies, \cite{2019ApJS..241...29Y}  report that for $291$ out of $3420$ flaring stars the LCs show no starspot variations. The authors give as possible explanations that the star's rotation axis might be inclined with respect to the observer's line of sight or that starspots might be located near the poles of the star and, therefore, cause much less rotational modulation in the LC as if they were near the equator. As far as our work is concerned, an additional plausible reason for observing flares but no starspot modulation in some LCs is the limited photometric precision and resulting high scatter in TESS LCs. The standard deviation of the LC after subtraction of rotational modulation and flares, $S_{\rm flat}$, (see Sect.~\ref{sec:flare_criteria}), is plotted for our sample versus TESS magnitude in Fig.~\ref{S_flat}. Almost all targets are observed in multiple TESS sectors and each sector is divided into two segments due to the data gap in the middle (cf. Sect.~\ref{period_search}), such that we obtain two $S_{\rm flat}$ values for each sector's LC. The $S_{\rm flat}$ for each target in Fig.~\ref{S_flat} is the mean of all values obtained from its TESS LCs.

For comparison, Fig.~\ref{S_flat} also shows the $S_{\textrm{flat}}$ values in {\it K2} short cadence LCs of the M dwarf sample presented by \cite{Raetz2020.0}. In order to be able to compare the $S_{\textrm{flat}}$ between the TESS and {\it K2} sample, we derived an empirical magnitude conversion between the TESS and {\it Kepler} band. This procedure is described in Appendix \ref{app:magtrans}. The {\it K2} stars from \cite{Raetz2020.0} are %thus 
plotted in Fig.~\ref{S_flat} 
at their derived TESS magnitude obtained from Eq. \ref{eq:magtrans}.

For both the TESS and the {\it K2} sample, $S_{\rm flat}$ increases for stars with fainter magnitudes, that is the signal to noise ratio worsens.
Our $S_{\rm flat}$ values are clearly higher than the ones of \cite{Raetz2020.0} for {\it K2} short cadence LCs. The difference is on average $560$~ppm. 
%For this reason, in {\it TESS} data starspot variations with low amplitudes are hidden in the noise. 

%%%%%%%%%%%%%%%%%%%%%%%%%%%%%%%%%%%%%%%%%%%%%%%%%%%%%%%%%%%%%%%%%%%%%%%%%%%%%%%%%%%%%%
\subsection{Relation between flare rate and SpT}\label{sec:flare_rate_SpT}
%%%%%%%%%%%%%%%%%%%%%%%%%%%%%%%%%%%%%%%%%%%%%%%%%%%%%%%%%%%%%%%%%%%%%%%%%%%%%%%%%%%%%%%

%\cite{2020AJ....159...60G} find in flare studies of stars observed in the first two months of the {\it TESS} mission that M4 to M6 dwarfs show the highest fractions of flaring targets. Our result is in good agreement with this finding. \textbf{\cite{2017ApJ...849...36Y} also find an increasing fraction of flaring stars from SpT M0 to M5 (cf. their Fig. 13). A rising flare frequency from M0 to M4 stars is reported by \cite{2019ApJS..243...28L}. The latter two studies are based on \textit{Kepler} and \textit{K2} long cadence data.}\\

%Table \ref{tab:spt_flares} summarizes the total number of flares and the flare rates per SpT subclass \textcolor{blue}{\bf for our sample}. 
When investigating a correlation of flare rate and SpT, detection biases have to be taken into account. Specifically, only flares with a peak amplitude in the TESS band of $\Delta L_{\rm F,T} > 3 S_{\rm flat} \cdot L_{\rm qui,T}$ 
can be detected for each target. In fact, the detection threshold is even higher than this since not only one data point is required to lie above $3\sigma$, but three consecutive data points (cf. Sect.~\ref{sec:flare_criteria}). Thus, the exact amplitude limit depends on the flare shape. A second approach is therefore to constrain the detection threshold in terms of flare energy instead of $\Delta L_{\rm F,T}$. This is discussed further in Sect.~\ref{subsect:ffd}. For the discussion of detection biases which is the topic of this section, the amplitude threshold is, however, a useful criterion. The aim of the following considerations is to assess whether the detection threshold is in any way correlated with the SpT.

\begin{figure}[t]
\centering
\includegraphics[width=0.49\textwidth]{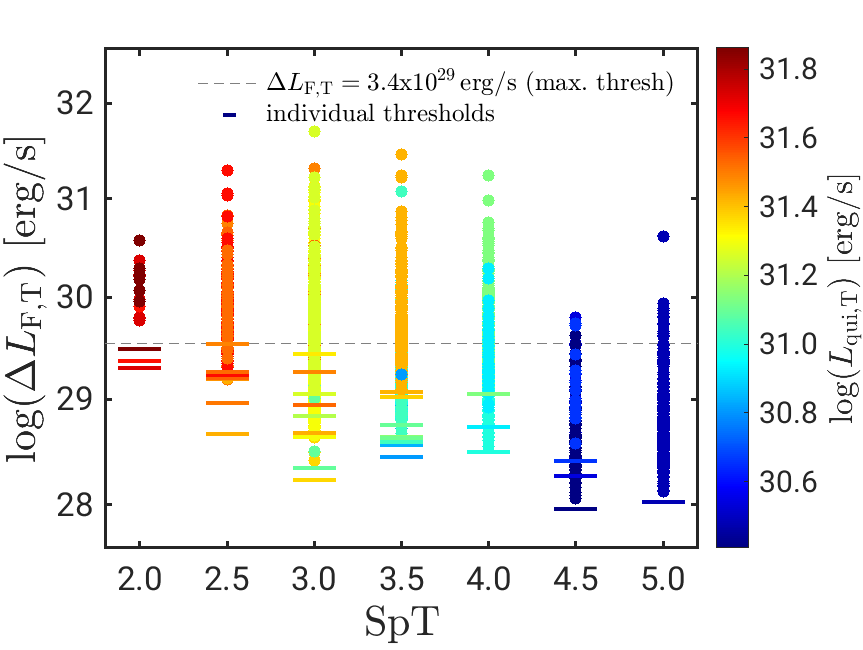}
\caption[Peak flare amplitude vs. SpT]{Relation between peak flare amplitude
in the TESS band, $\Delta L_{\rm F,T}$, and SpT. The minimum value of $\Delta L_{\rm F,T}$ that could be detected for each target is represented by a ``\_'' marker. The gray dashed line marks the highest of the flare amplitude detection thresholds for all flaring stars in this sample of 
% no factor 4
%$\boldsymbol{L_{\text{max,th}}=8.6\cdot 10^{28}}$\,erg/s 
$L_{\text{max,th}=3.4\cdot 10^{29}}$\,erg/s (see Sect.~\ref{sec:flare_rate_SpT}). The color code visualizes the expected correlation between $L_{\textrm{qui},T}$ and SpT.}
\label{fig:ampl_spt}
\end{figure}

Firstly, stars with a fainter apparent magnitude have a higher noise level, that is a larger $S_{\rm flat}$ (cf. Fig.~\ref{S_flat}). The bias regarding the SpT dependence of activity following from this $S_{\rm flat}$ versus  $T$-mag relation is probably small as there is no evident correlation between SpT and TESS magnitude within the sample (cf. color coding in Fig.~\ref{S_flat}).

Quiescent luminosity and SpT are, however, strongly correlated. Main-sequence stars of later M SpT subclasses have by definition a lower $L_{\rm qui,T}$ and, therefore, a lower flare detection threshold. Consequently, flares with lower amplitudes can be detected.
%which might also be a reason for the higher fraction of flaring stars among later M subclass stars (cf. Fig. \ref{fig:hist_spt_flares}). 
In Fig.~\ref{fig:ampl_spt}, flare amplitudes are plotted versus SpT. The flare detection threshold of each star is represented by a horizontal bar marker. We found the highest of these detection thresholds to be 
% no factor 4
% $\boldsymbol{L_{\text{max,th%}}=8.6\cdot10^{28}}$\,erg/s
$L_{\rm max,th}=3.4\cdot10^{29}$\,erg/s. Flares with higher amplitudes can be detected on all 35 flaring stars of our sample. Stars of the latest SpT class ($\geq$\,M4.5) show only a few flares above this threshold, but they have many fainter flares that are undetectable for earlier-type stars.

\begin{figure}[t]
\centering
\includegraphics[width=0.48\textwidth]{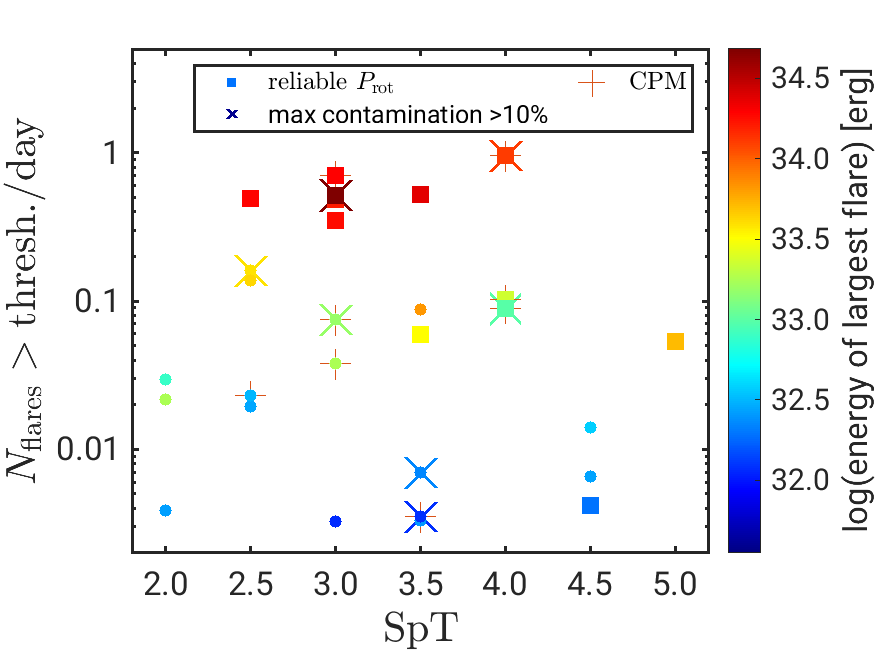}
\caption[Relation between flare rate and spectral type]{Relation between flare rate and SpT. The $12$ targets with flares and a reliable rotation period are shown with square markers. The energy of the largest flare is color-coded for each target. Only flares with amplitude 
% no factor 4
%\textcolor{teal}{$\boldsymbol{\Delta L_{F,T}> L_{\rm max,th} = 8.6\cdot10^{28}}$\,erg/s} 
$\Delta L_{F,T}> L_{\rm max,th} = 3.4\cdot10^{29}$\,erg/s are considered, %This is the highest of the flare amplitude detection thresholds within the sample. 
%Choosing it as a cutoff accounts 
to account 
for the sensitivity detection biases (cf. Sect. \ref{sec:flare_rate_SpT}).} 
\label{flare_rate_SpT}
\end{figure}

We further examine this detection bias and the result after its removal in Fig.~\ref{flare_rate_SpT}. It shows the flare rate for all flaring stars as a function of SpT considering only the flares with an amplitude greater than $L_{\rm max,th}$. 
In total, $30$ of the $35$ flaring stars within our sample show flares above the sensitivity threshold.  
The energy of the largest flare for each star is chosen for the color-coding since it is a straight-forward indicator for flare energy and free from evident detection bias (in contrast to lower flare energies). Within each SpT bin, the maximum observed  flare energy is correlated with the flare rate. Moreover, the stars with reliable rotation periods (squares) have the highest flare rates (see also Sect.~\ref{sec:flares_prot}) and higher energies of their largest flares. Further, it can be observed in Fig.~\ref{flare_rate_SpT} that the flare rate sharply drops for stars of SpT M2 and earlier and thus this is not an effect of the sensitivity bias. For stars of SpT M4.5 and M5, the amplitude cutoff at the maximum detection threshold eliminates nearly all flares. Fig.~\ref{fig:ampl_spt} already shows that these stars hardly show flares with amplitudes above the common threshold.

\begin{figure}[t]
\centering
\includegraphics[width=0.41\textwidth,trim= 0 0.8cm 0 0.4cm]{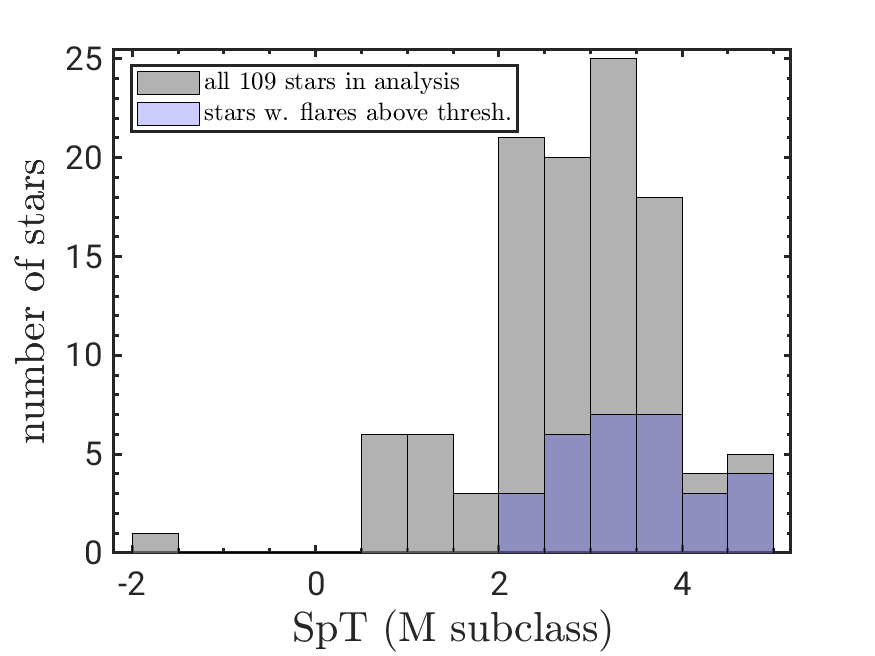}
\captionof{figure}[SpT distribution of the sample, flaring stars marked]{SpT distribution of all $109$ targets considered in the LC analysis and the $30$ stars showing flares  above the maximum amplitude threshold of 
% no factor 4
%\textcolor{red}{$8.6\cdot10^{28}$\,erg/s.
$3.4\cdot10^{29}$\,erg/s. %The amplitude threshold is determined as $L_\text{max,th}=3S_{\textrm{flat}}\cdot L_{\textrm{qui},T}$ 
(cf. Sect.~\ref{sec:flare_rate_SpT}). Numbers on the x-axis label M SpT subclasses, where ``-2'' stands for SpT K8.}
\label{fig:hist_spt_flares}
\end{figure}

The SpT distribution of the $30$ stars with flares above $L_{\rm max,th}$ is plotted in Fig.~\ref{fig:hist_spt_flares} together with that of all $109$ stars considered in the analysis. It is evident that the ratio of stars showing flares above the detection threshold  to non-flaring stars is higher for later M SpT subclasses. In particular, stars earlier than SpT M2 do not show any flares above the threshold - despite the fact that their numerical representation in the sample is comparable or even slightly higher than that of M4 to M5 stars. Overall, three groups seem to stand out, namely stars with (i) SpT $\leq$ M2,
(ii) M2 $<$ SpT $\leq$ M4, and (iii) SpT $>$ M4. In Table~\ref{tab:spt_flares} we summarize
the flare statistics for these three SpT ranges   
%In Table~\ref{tab:spt_flares} we summarize the flare statistics for the individual SpT subclasses
considering only events above the threshold $L_{\rm max,th}$.  
To obtain the total observation time for calculating the flare rates, the observation times of all targets considered in the analysis (cols.~3 and~4) or the subsample of targets exhibiting flares above $L_{\rm max,th}$  (cols.~5 and~6) for the respective SpT range are summed up. 
We caution that these  results is based on low-number statistics.

%%%%%%%%%%%%%%%%%%%%%%%%%%%%%%%%%%%%%%%%%%%%%%%%%%%%%%%%%
\subsection{Relation between flare amplitude and duration}\label{subsubsec:ampl_dur}
%%%%%%%%%%%%%%%%%%%%%%%%%%%%%%%%%%%%%%%%%%%%%%%%%%%%%%%%%%

\begin{figure*}[t]
\centering
\begin{minipage}[t]{0.48\textwidth}
\centering
\includegraphics[width=\textwidth]{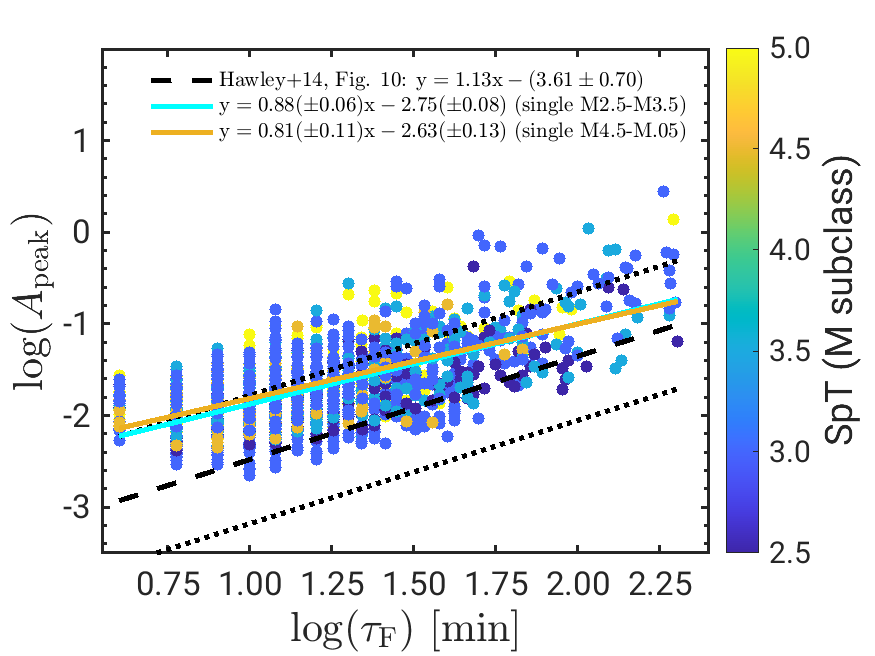}
\end{minipage}
\begin{minipage}[t]{0.48\textwidth}
\centering
\includegraphics[width=\textwidth]{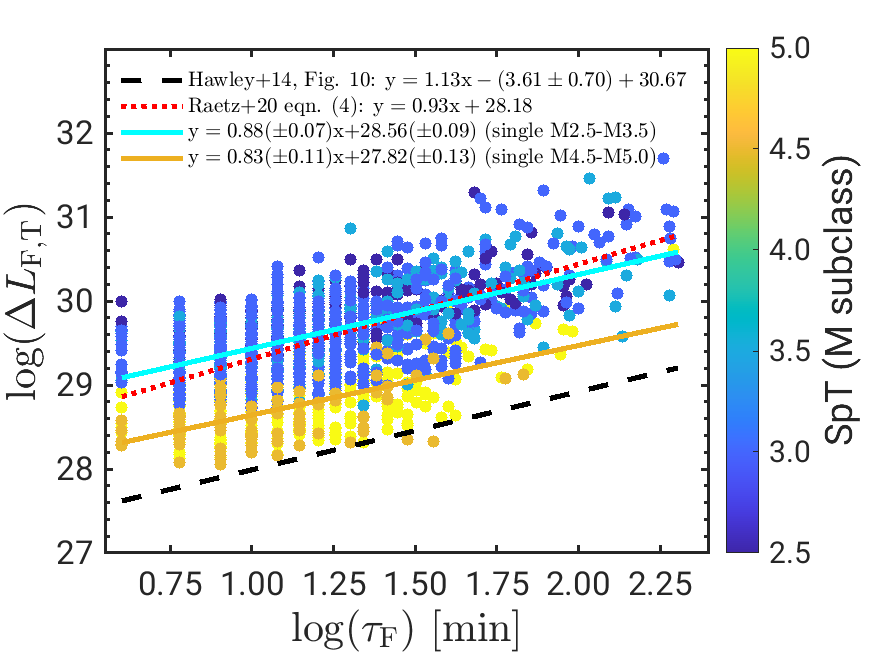}
\end{minipage}
\caption{Relative flare amplitude versus flare duration for the 8 stars with reliable $P_{\rm rot}$ that have no PM companion. The orange and turquoise lines are linear fits performed in different SpT ranges as indicated in the legend. {\it Left panel:} Normalized peak flare amplitude. {\it Right panel:} Peak flare luminosity amplitude. Relations from \cite{2014ApJ...797..121H} and \cite{Raetz2020.0} are overplotted for comparison; see Sect.~\ref{subsubsec:ampl_dur} for details.}\label{fig:flare_ampl_vs_duration}
\end{figure*}

\begin{table}[t]
\centering
\small\addtolength{\tabcolsep}{-2pt}

\caption[Flare numbers and rates for each SpT]{
%\textcolor{orange}{The flare rates without considering the threshold are not useful. Enza, please update this table considering only flares above the $L_{\rm max,th}$ threshold.} \textcolor{teal}{DONE} 
Number of flares  %($N_\text{flares}$) 
and flare rates.  %($\nu_\text{flares}$) 
%for three SpT subgroups. 
%each SpT subclass 
%considering only events above the amplitude threshold $L_{\rm max,th}$ determined in Sect.~\ref{sec:flare_rate_SpT}.  %\textcolor{orange}{Enza, to understand the caption text: What means "all targets considered in the analysis"? Is it 109? What is the sample in the rightmost column? Is it 30 stars?} %\textcolor{teal}{All targets considered in the analysis are indeed the 109 TIC stars. The rightmost sample shows only the amount of stars of that specific SpT that have flares above the threshold. So the numbers of each SpT are subsamples of those 30 flaring stars.}
}
\begin{tabular}{c|c|cc|cc}
\hline
&&\multicolumn{2}{c|}{all stars}&\multicolumn{2}{c}{flaring stars}\\
&$\Sigma$ $N_{\textrm{flares}}$&$N_{\textrm{targets}}$&$\nu_{\textrm{flares}}$ &$N_{\textrm{targets}}$&$\nu_{\textrm{flares}}$\\
&($>$ the thresh.)&&[d$^{-1}$]&&[d$^{-1}$]\\
\hline
      SpT $\leq$ M2	& 16 & 37 & 0.001 & 3 & 0.019\\
 M2 $<$ SpT $\leq$ M4 & 1240 & 67 & 0.070 & 23 & 0.207\\
      SpT $>$ M4 & 19 &	5 & 0.015 & 4 & 0.020\\
\hline
\end{tabular}
\label{tab:spt_flares}
\tablefoot{Only events above the amplitude threshold $L_{\rm max,th}$ determined in Sect.~\ref{sec:flare_rate_SpT} are considered.}
\end{table}

The left panel of Fig.~\ref{fig:flare_ampl_vs_duration} shows the normalized flare amplitude $A_{\rm peak}$ versus flare duration $\tau_{\rm F}$, the right panel shows the same plot for the absolute flare amplitude $\Delta L_{\rm F,T}$. 

The black dashed line 
%in Fig.~\ref{fig:flare_ampl_vs_duration} 
is a linear relation that we extracted from Fig.~10 in \cite{2014ApJ...797..121H}, and the black dotted lines 
%in the left panel 
represent the lower and upper envelope of this relation. These authors analyzed the relation between flare amplitude and duration for GJ\,1243, 
an 
%a particularly active 
M4 dwarf with a rotation period of $0.59$\,d in {\it Kepler} short cadence observations. This star has become a benchmark for flare studies and its applications to exoplanets \citep{2020AJ....160...36D, 2019AsBio..19...64T} because of a combination of favorable properties including its its proximity \cite[$12$\,pc;][]{GaiaeDR3}, long coverage with {\em Kepler} and elevated activity level.  The flare rate of GJ\,1243 was found by \cite{2014ApJ...797..121H} to be intermediate to high when compared to other M dwarfs. The general  result for GJ\,1243  of longer flares having a higher relative amplitude,  $A_{\textrm{peak}}$, is also present in the flare data of our HZCat sample.  However, with respect to GJ\,1243 higher amplitudes for given flare duration are measured in our sample. 

We performed linear fits using only the single stars and separating our sample into two SpT bins ranging from M2.5$-$M3.5 and from M4.5 $-$M5. No M4 stars were considered since they all have PM companions. 
According to Fig.~\ref{flare_rate_SpT}, 
the flare rate drops after SpT M4, %justifying our bin selection. 
but the linear fits performed for the two SpT subranges of our sample (orange and turquoise lines in Fig.~\ref{fig:flare_ampl_vs_duration}) are nearly indistinguishable for the normalized flare amplitude versus duration  (left panel). 
%, but yield slopes that are less steep and an offset towards higher amplitudes compared to the relation of \cite{2014ApJ...797..121H}.
The activity level of the stars, therefore, is unlikely to be responsible for the difference in our $A_{\rm peak} - \tau_{\rm F}$ relation with respect to GJ\,1243. It might, instead, be due to instrumental differences with an impact of the noise level on the measured flare shape (see below and Sect.~\ref{subsubsect:disc_flare_amplitude_duration}).

%\textcolor{red}{\bf }
After transforming the amplitudes to absolute values, $\Delta L_{\rm F,T}$, we again fitted the two selected SpT subsamples of our HZCat targets. Contrary to $A_\text{peak}$ versus $\tau_F$, the relation between flare duration and absolute flare amplitude, $\Delta L_{\rm F,T}$, is different for the two SpT subsamples, with the fit for earlier SpTs shifted upward as a result of the higher quiescent luminosities of earlier-type stars. This represents the same detection bias that is seen in Fig.~\ref{fig:ampl_spt}. 
The black dashed line %in %Fig.~\ref{fig:flare_ampl_vs_duration}, right panel also 
again refers to the results of \cite{2014ApJ...797..121H} 
for GJ\,1243 that we have  
%as in the left panel, 
converted from normalized to absolute flare amplitudes by adding the logarithm of the quiescent luminosity of GJ~1243 in the {\it Kepler} band, $\log L_{\rm Kep}$ [erg/s]=30.67 (cf. \cite{2014ApJ...797..121H}, Table 2). The relation of \cite{2014ApJ...797..121H} is shifted toward lower absolute flare amplitudes with respect to both our subsamples. Since GJ~1243 has SpT M4, in between our two bins, this cannot be an effect of SpT or $L_{\rm qui}$. 
Thus, the flares we detected for the targets of this work have higher absolute and normalized amplitudes than GJ~1243 in the \textit{Kepler} short cadence analysis of \cite{2014ApJ...797..121H}.
Hence our sample has on average a different amplitude-duration relation than GJ\,1243. 

%The slope of the amplitude versus duration fit is steeper for GJ\,1243 according to \cite{2014ApJ...797..121H}. 

The green dotted line in Fig.~\ref{fig:flare_ampl_vs_duration}, right panel is the amplitude-duration relation from \cite{Raetz2020.0} (R20, their Eq.~4), derived from {\it K2} short cadence observations of 56 bright, nearby M dwarfs (SpT K7 to M6)
and it is also shifted upward with respect to GJ\,1243. 
%The linear fits for both SpT subranges of our sample are nearly parallel to the R20 relation and they bracket the R20 relation. However, it lies closer to our late SpT bin.
In contrast to our sample, the sample of R20 comprises several slow rotators (i.e.,  $P_{\text{rot}}>10$ d) and a larger fraction of early-M-type stars (SpT K7 to M1). Our range of measured flare durations and amplitudes is comparable to that of R20.
%, but their sample shows more flares with amplitudes $\Delta L_{F} < 10^{29}$ erg/s. This might be partly due to the presence of slow rotators that tend to show flares with lower amplitudes (cf. \citealt{Raetz2020.0}, Fig. 17). The most probable explanation is, however, that flares with low amplitudes are difficult to detect with {\it TESS} due to the higher $S_{\textrm{flat}}$ in the LCs as compared to {\it K2} (cf. Fig. \ref{S_flat}). Indeed, our comparison of the {\it K2} and {\it TESS} FFD in Sect. \ref{subsubsect:calib_TESS_K2} shows several flares with energies $\lesssim 10^{32}$~erg observed with {\it K2}, but none in this energy range that are observed with {\it TESS}. 
We conclude that the flares of GJ\,1243 constitute the lower end of the amplitudes for the star's SpT and $P_{\rm rot}$ value or - equivalently - that its flares last longer for given amplitude. This might be due to a lower noise level in the data for GJ\,1243, which is from the main {\em Kepler} mission as compared to the {\em K2} data from \cite{Raetz2020.0} and the TESS mission from our work. In fact, tests with simulated flares show that for the recovered events the flare durations are more severely underestimated than the amplitudes, because part of the flare remains buried in the noise.

\subsection{Flare frequency distributions}\label{subsect:ffd}

\begin{figure}[t]
\centering
\includegraphics[width=0.49\textwidth]{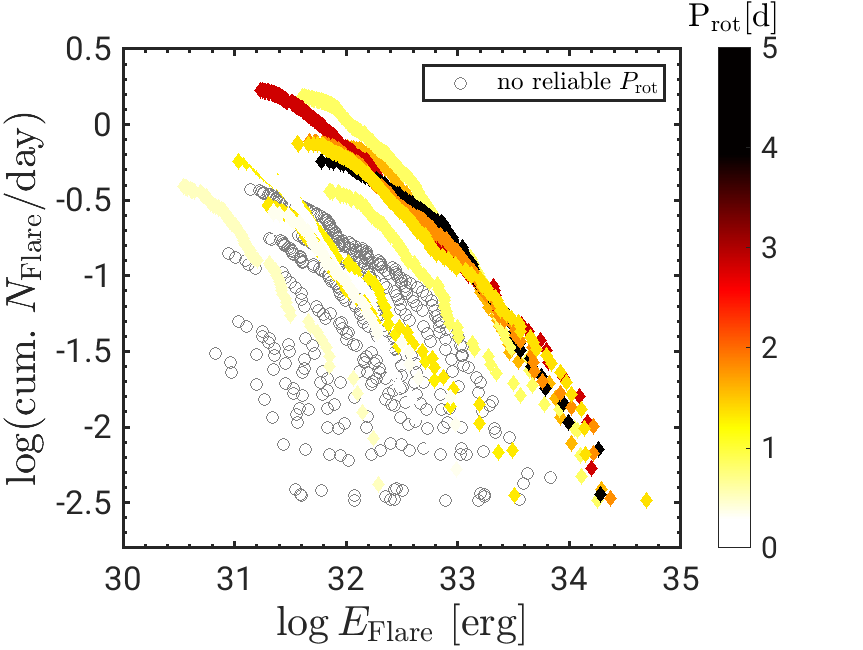}
\caption{Cumulative distribution of flare energies for all $35$ flaring stars; the $12$ stars with reliable $P_{\rm rot}$ are highlighted in color. %\textcolor{teal}{\bf UPDATED}
}
\label{fig:FFD_wo_prot}
\end{figure}

Fig.~\ref{fig:FFD_wo_prot} shows the FFDs of all $35$ flaring stars with a color-code for the rotation periods of the $12$ stars with reliable $P_{\text{rot}}$. 
In Sect.~\ref{sec:flares_prot} we have discussed that stars with reliable $P_{\text{rot}}$ show higher flare rates and this is also evident here, namely their FFDs lie above those of stars without reliable $P_{\rm  rot}$. 
In fact, the vast majority of the $>2500$ flares occur on the $12$ stars with reliable $P_{\rm rot}$. The FFD power law fit  defined in Sect.~\ref{sec:FFD_theory} was, therefore, only conducted for these $12$ targets. 

%As mentioned above, due to the condition for flare validation of at least 3 consecutive data points deviating $>3\sigma$ from the mean of the flattened, cleaned LC (cf. Sect. \ref{sec:flare_criteria}), the detection threshold for the amplitude $\Delta L_{F,T}$ is actually higher than $3S_{\textrm{flat}}\cdot L_{\textrm{qui},T}$. The limit cannot be precisely constrained because it depends on the shape of the flare. Therefore, in a second approach we determine a completeness threshold for the flare detection in terms of flare energy from the cumulative flare energy frequency distribution (FFD). For the definition of the FFD, see Sect. \ref{sec:FFD_theory}. 

A quantitative evaluation of the FFDs through a power law fit depends crucially on the determination of the completeness limit, as the detection of flares depends on the individual noise level of the LC, which is characterized by the standard deviation, $S_{\rm flat}$, defined in Sect.~\ref{subsubsect:flare_det_method}. Each FFD thus has an observation-specific cutoff flare energy, $E_{\rm min, th}$, below which not all events are detected. This usually translates into a flattening of the FFD for flare energies $< E_{\rm min, th}$. 

We determined $E_{\rm min,th}$ for each of the $12$ stars and for each of its TESS LCs  separately using the method described by \cite{Raetz2020.0}. Our technique employs the flare template of \cite{Davenport14.0} with an assumed flare duration of $360$\,s, and examines the amplitude of flares that have $3$ data points above the detection threshold of $3\,S_{\rm flat}$, namely our criterion for flare detection from Sect.~\ref{subsubsect:flare_det_method}. 
%\textcolor{orange}{DO WE HAVE TO JUSTIFY THE LENGTH 360s? DID WE CHECK HOW RESULTS CHANGE FOR DIFFERENT LENGTHS?}\textcolor{brown}{SR: The results do change with different considered flare length. It is an arbitrary choice optimized with trial and error. 360s is 1.5 times our chosen detection criteria of 3 data points above 3sigma. However, since we use the same criteria for the whole sample, the determination is homogeneous.}
The energy of this synthetic flare with an amplitude just high enough that it would be detected by our flare search algorithm is then the energy completeness threshold, $E_{\rm min,th}$.  

We note that for each star the FFD fit comprises  flares from all TESS sectors in which the star was observed, and consequently there is not a unique value of $E_{\rm min,th}$ associated with a given star but all flares from each of the sectors $s$ that have their energy fulfilling $E_{\rm f, s} > E_{\rm min,th, s}$ are considered in the power-law fit. In practice, for a given star the energy thresholds from different sectors are all within a small range. %\textcolor{orange}{Enza, what are the values?}.
 %\textcolor{teal}{The whole range goes from 30.7549\,erg to 32.3429\,erg. Specifically, I see a difference of less than 0.5\,erg at most between the $E_{\rm min,th}$ for each sector of each star.}
%For a more detailed description of the procedure, see Raetz et al. (2020).}
%We define the completeness limit following \cite{2014ApJ...797..121H}. The aim is to find the energy value where the FFD stops showing a linear behavior. This bending of the curve towards lower flare rates for low flare energies is partly ascribed to the fact that below a certain energy threshold the flare detection is incomplete, i. e. not all flares are detected. In order to find the position where the bending sets in, a linear fit is performed on the FFD, successively leaving out more and more low energy data points. The completeness limit for the flare energy is then determined as the minimum energy value where the slope of the power law fit does not change within the fit uncertainties when additional data points are left out. 

In the fitting we took account of Poisson uncertainties on the 
flare rates  as described by \cite{2014ApJ...797..121H}, that is the flare number is assumed to be Poisson-distributed as a function of flare energy, thus 
\begin{center}
\begin{equation}
\qquad\Delta \nu_\text{flares}=\frac{\sqrt{N_{\textrm{flares}}}}{t_{\textrm{obs}}},  \label{eq:nuflare_err}
\end{equation}
\end{center}
with the flare frequency $\nu_{\rm flares}$ in units of d$^{-1}$. In logarithmic representation, the error becomes
\begin{align*}
\qquad\Delta \text{log}(\nu_\text{flares})=\Delta \nu_\text{flares}\cdot\nu_\text{flares}^{-1}=\\
\qquad\bigg(\frac{\sqrt{N_{\textrm{flares}}}}{t_{\textrm{obs}}}\bigg)\cdot\bigg(\frac{N_{\textrm{flares}}}{t_{\textrm{obs}}}\bigg)^{-1}=\frac{1}{\sqrt{N_{\textrm{flares}}}}
\end{align*} 
by error propagation. This consequently gives larger errors in the cumulative rates of high flare energies due to the smaller number of flares found there.

%\textbf{The absolute values of our power law slopes range from 0.597 to 0.954. \cite{2017ApJ...849...36Y} and \cite{2019ApJS..243...28L} find power law slopes of $-$1.07 and $-$1.02, respectively, from \textit{K2/Kepler} M dwarf flare studies. (The power law slope corresponds to $-\alpha+1$ for their $\alpha$-Parameter.) \cite{Raetz2020.0} find slopes ranging from $-$0.81 to $-$0.85 for different M SpT subranges which is consistent with the result of \cite{2019ApJ...873...97L}. Most of our results are well within the range spanned by these exemplary studies.}\\

%\textcolor{red}{Die Ilin Pleijaden FFD muessen wir uns noch mal anschauen (z.B. sehe ich, dass sie steiler ist als Dein , aber eigentlich brauchen wir sie gar nicht.}

%\textcolor{red}{Wichtig waere, dass Stefanie's FFD dieselbe Steigung hat, ist das der Fall?}\textcolor{blue}{--> Fig. \ref{fig:FFD_comparison}: Für die mittlere FFD der frühen Sterne stimmt das.}\\\textcolor{red}{

\begin{figure}[t]
\centering
\includegraphics[width=0.49\textwidth]{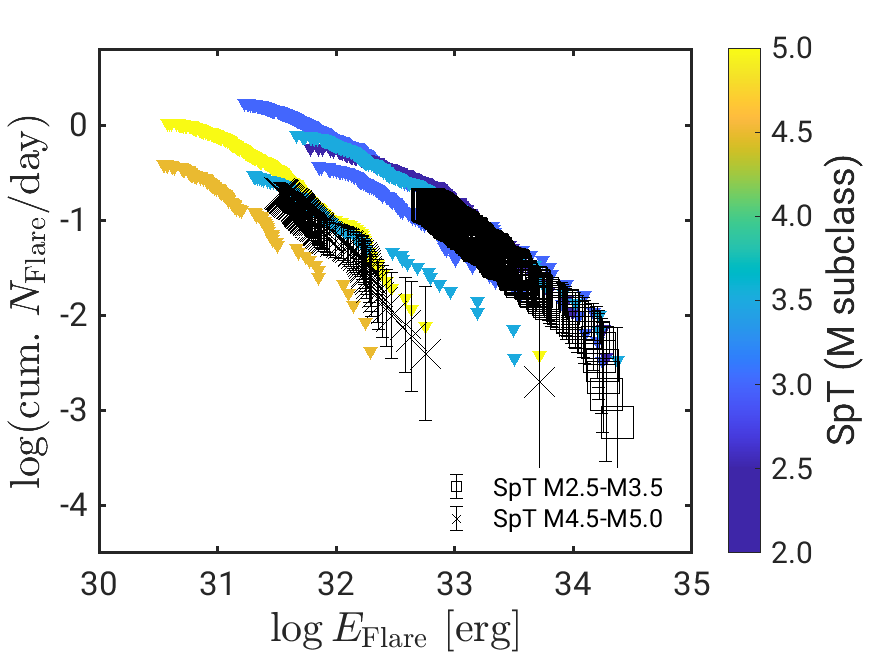}
\caption{FFDs of stars with reliable $P_{\text{rot}}$ without the five stars with PM companions or a maximum contamination in all  TESS sectors $>10$\,\%. 
%Only data points above the individual energy completeness limits defined in Sect.~\ref{subsect:ffd} are plotted.
%\textcolor{orange}{Enza: for the individual stars shown here did you plot exactly the flares that you considered for the fit of the individual FFDs, i.e. all flares that are in a given sector above the limit?} \textcolor{teal}{No, the FFD of the individual stars shows all flares, then the ``averaged'' is calculated considering only those above the discussed $E_{\rm min,th}$.}
Black square and cross symbols represent the average FFDs for the earlier and later M SpT subclass.
%stars within the sample. 
Error bars for these average FFDs 
%are depicted in grey and 
result from Eq.~\ref{eq:nuflare_err}.
Since the stars considered in the SpT-averaged FFD have a range of completeness limits, $E_{\rm min,th}$, for each of the two distributions the fits have been limited to flares with energy above the highest value of the $E_{\rm min,th}$ of all stars in the respective sample.}
\label{fig:ffd_HZsample}
\end{figure}

\begin{table*}[t]
    \centering
    \caption{FFD power law slopes and y-axis offsets for the two SpT-averaged {\it TESS} FFDs (see Sect.~\ref{subsect:ffd}) and the corresponding X-ray FFDs constructed as described in Sect.~\ref{subsection:xray_ffds}. 
    %The given errors are formal fit uncertainties. 
    %All fits are displayed in Fig.~\ref{fig:xray_ffd} in the colors given in brackets. %\textcolor{teal}{only the early SpT FFD changes to the point that the slopes and y-axis offset change too.}
    }
    \begin{tabular}{l|cc|cc}
    \hline
    &\multicolumn{2}{c|}{power law slope \textcolor{black}{($\alpha$ from Eq.3)}}&\multicolumn{2}{c}{y-axis offset \textcolor{black}{($\log{\beta}$ from Eq.3)}}\\
    SpT range&M2.5-M3.5&M4.5-M5.0&M2.5-M3.5&M4.5-M5.0\\
       \hline
    {\it TESS} (black)&\textcolor{black}{$-1.20\pm 0.04$}&\textcolor{black}{$-1.05\pm 0.06$}&\textcolor{black}{$38.68\pm 1.28$}&\textcolor{black}{$32.25\pm 1.89$} \\
    X-ray (red)&\textcolor{black}{$-0.85\pm 0.03$}&\textcolor{black}{$-0.74\pm 0.04$}&\textcolor{black}{$26.31\pm 0.89$}&\textcolor{black}{$21.48\pm 1.28$}\\
    %steepest X-ray (orange)\tablefootmark{a}&\textcolor{teal}{$\boldsymbol{-1.145\pm 0.036}$}&\textcolor{black}{$\boldsymbol{-0.997\pm 0.055}$}&\textcolor{teal}{$\boldsymbol{36.173\pm 1.204}$}&\textcolor{black}{$\boldsymbol{30.071\pm 1.764}$}\\
    \hline
    \end{tabular}
    %\tablefoot{
    %\tablefoottext{a}{Resulting FFD power law slope when {\it TESS} flare energies are converted to the X-ray band using the smallest slope within the uncertainties of our conversion derived in Sect. \ref{subsubsect:calib_multilambda}, i. e. a slope of $a=1.05$.}
    %}
    \label{tab:ffd_slopes}
    \tablefoot{The given errors are formal fit uncertainties. The colors given in brackets in col.~1 refer to Fig.~\ref{fig:xray_ffd}.}
\end{table*}

We computed average FFDs for the same two SpT ranges used in Fig.~\ref{fig:flare_ampl_vs_duration} for studying the flare amplitude$-$duration relation. Eliminating from the subsample of $12$ stars with reliable $P_{\rm rot}$ those with PM companions or a maximum contamination factor $>10$\,\% leaves a sample of five stars for the M2.5$-$M3.5 range and two stars for M4.5$-$M5.  %One ranges from M2.5 to M3.5 (five stars), the other comprises two stars of SpT M4.5 and M5. The remaining five stars have a PM companion or a maximum contamination factor of all observation sectors >10\% (cf. Sect. \ref{sec:contamination}) and are therefore not considered. 
The average FFDs for each of these two SpT subgroups have been constructed from the flares of all stars that belong to the respective group. Hereby, we considered among all flares from all sectors for all stars in the respective SpT subgroup only those above the highest $E_{\rm min,th}$ threshold, combining them into a single FFD. The use of this conservative value for the energy threshold ensures that no additional artificial substructures are introduced into the FFDs as a consequence of the different noise levels of the stars. These 
average FFDs for the two SpT subgroups are shown in Fig.~\ref{fig:ffd_HZsample} with black symbols, together with the FFDs of the seven individual stars that were taken into account. 
 
\begin{figure}[t]
\centering
\includegraphics[width=0.50\textwidth]{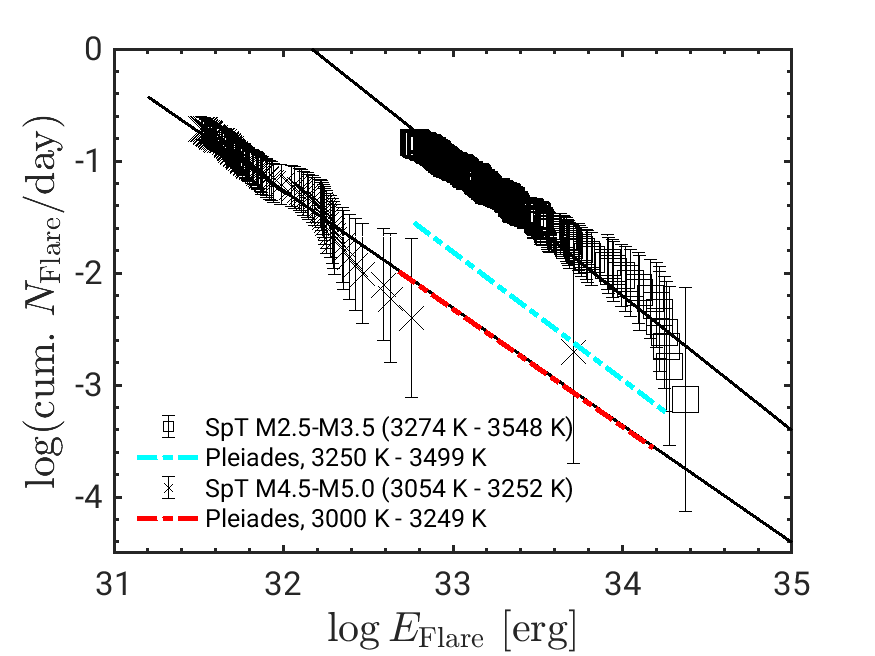}
\caption{Average FFDs for two SpT ranges as in Fig.~\ref{fig:ffd_HZsample} (black data points) but here together with power law fit. The fit parameters are listed in Table~\ref{tab:ffd_slopes}. The linear fit of the Pleiades FFDs based on {\it K2} long cadence data in two similar SpT ranges is taken from Table~4 of \cite{Ilin19.0} for energies above the completeness limits given in their work.}\label{fig:FFD_comparison}
\end{figure}

Fig.~\ref{fig:FFD_comparison} shows again the same average TESS FFDs in the SpT ranges M2.5$-$M3.5 and M4.5$-$M5. Overlaid are the power law fits, the parameters of which are listed in Table~\ref{tab:ffd_slopes}. We also display the linear fits obtained by \cite{Ilin19.0} for Pleiades stars in similar $T_{\rm eff}$ ranges. The $T_{\rm eff}$ ranges given in the legend for our two SpT bins correspond to the ranges spanned by the stars in the respective bin. 
While our slopes are similar to those for the Pleiades sample, our earlier SpT bin has a significant upward shift with respect to the corresponding Pleiades flare distribution. This 
%\textcolor{red}{\bf The slight shift of the flare frequencies of the Pleiades with respect to our field star samples 
could be due to a different rotation period distribution, and hence activity level, of the considered stars. 
%The relations from \cite{2019A&A...622A.133I} for Pleiades stars are shifted towards lower flare rates compared to the binned FFDs of our work. 
The different observing cadence to which the two data sets refer may also play a role: \cite{Ilin19.0} have used {\it K2} long cadence LCs, and 
%Another possible reason is that the analysis of these authors is based on {\it K2} long cadence LCs. 
\cite{Raetz2020.0} found that flare rates are  
lower in long-cadence as compared to short-cadence data. 
%%by a factor of
%4.6 lower in {\it K2} long cadence as compared to {\it K2} short cadence data. 
%\textcolor{gray}{Old text from Mirjam: \it Fig. \ref{fig:FFD_comparison} reveals that the slope of the Pleiades FFDs is steeper than that of our averaged FFDs: For the earlier M subtypes, our slope is $\approx -0.716\pm0.003$ versus $-1.05\pm0.01$ in \cite{2019A&A...622A.133I}, for the mid/late M subclasses it is $\approx -0.886\pm0.006$ versus $-1.14\pm0.02$ in \citep{2019A&A...622A.133I}. These authors determined the energy completeness limit for flare detection by means of synthetic flare injection.}
%\textcolor{red}{\bf While the power law slope for the cooler/later-SpT of the two samples is identical between our field stars and the  Pleiades distribution from \cite{Ilin19.0}, for the warmer/early-M stars \cite{Ilin19.0} found a slightly steeper slope ($-1.14 \pm 0.02$ vs our value of \textcolor{teal}{$-1.11 \pm 0.03$}). \textcolor{teal}{that's not entirely true. I updated the values in Table 4 after I have adjusted the $E_{\rm min}$ values from Stefanie (see email I sent you). Right now the slopes are comparable within the uncertainties..let's keep in mind that it can change if we do not correct for that factor of 4}. 
We show the Pleiades power-laws down to the energy thresholds given by \cite{Ilin19.0}. 
 As can be seen, the $E_{\rm min,th}$ values for our field M dwarfs are more sensitive than the ones for the Pleiades, especially for the later SpT bin, likely as a result of the closer distance of our sample, the different data cadence and differences in the flare detection methods.  
%They find values of $\log E_\text{min,th}\text{ [erg]}=32.77$ and $\log E_\text{min,th} \text{ [erg]}=32.68$ for the $T_\text{eff}$ ranges that correspond to our early and late SpT bin, respectively (cf. Fig. \ref{fig:FFD_comparison}). Our energy detection threshold is slightly lower for the early SpT subsample ($\log E_\text{min,th}\text{ [erg]}=\textcolor{teal}{32.16}$) and clearly lower for the late ($\text{log}(E_\text{min,th})=\textcolor{teal}{30.91}$), cf. Table \ref{tab_prot}.
%\textcolor{orange}{Enza, can you give the new thresholds Emin,th for our two average FFDs? I will then revise this part of the text.}
%These differences might affect the resulting power law slope since only data points with energies above $E_\text{min,th}$ are considered in the power law fit.  

\subsection{Flares of TOIs and confirmed planet host stars within our sample}\label{subsection:flares_tois}

\begin{table}[t]
\centering
\caption{Stars with confirmed planets or TOIs and information on flares and rotation period. 
%See Sect.~\ref{subsection:flares_tois} for details on these systems.
}
\label{tab:flares_toi}
\begin{tabular}{ccccccc}
\hline
TIC ID & other & comment & flares?&{$\mathrm{P_\text{rot}}$?}\\
\hline
198211976& TOI-2283&--&--&--\\
233193964 & GJ 687& 2 RV det. pl.& $\checkmark$ &--\\
235678745& TOI-2095&--&--&--\\
260004324& TOI-704,&conf. pl.&--&--\\
&LHS 1815 &&&\\
260708537& TOI-486&--&  $\checkmark$ &--\\
272232401 & L~34-26& direct imaging& $\checkmark$ & $\checkmark$ \\
307210830& TOI-175,& 4 conf. pl.& $\checkmark$ &--\\
&          L 98-59 &&&&\\
377293776& TOI-1450&--& $\checkmark$ &--\\
\hline
\end{tabular}
\tablefoot{A checkmark means detection.}
\end{table}

We downloaded the list of all TOIs and confirmed TESS planets from the \textit{NASA exoplanet archive}\footnote{exoplanetarchive.ipac.caltech.edu/ accessed  2021/11/26} and matched them with our sample. 
Among our $112$ stars there are eight TOIs, four of which have confirmed planets. Of these four, two have planets discovered by TESS: The M2 dwarf TIC\,260004324 alias TOI-704 (\citealt{2020AJ....159..160G}) and the M3 dwarf TIC\,307210830 (L\,98-59, TOI-175). The latter hosts a system of four  planets (e.g., \citealt{2019AJ....158...32K}, \citealt{2019A&A...629A.111C}, \citealt{Demangeon2021}).
%These two stars show no signs of starspot modulation. For TIC307210830, we detected 5 flare events, the largest having $E_{\text{flare}} \approx 2.02\cdot10^{32}$~erg. 
%Its flare rate is $\approx 0.03~\text{d}^{-1}$, thus the activity level seems rather low.\\ 
A third star, TIC\,233193964 (GJ\,687), has two planets that have been detected through radial velocity measurements. One was already known before the TESS mission (\citealt{2014ApJ...789..114B}), the other one has been discovered recently \citep{Feng2020}. 
%\textcolor{brown}{GJ687: Weiss man die inclination, also ob transiting oder nicht? In der Antwort schreibst Du, dass einer der Planeten in der HZ ist, dann sollte ihn TESS doch sehen.}}. \textcolor{green}{Soweit ich weiß, ist die inclination nicht bekannt. Ich sehe keine Transits in den Lichtkurven, aber das muss nichts heißen.}
%For this star, we found 8 flare events, but no rotational modulation.
TIC\,272232401 was found to host a planet with a wide orbital separation of $7506^{+5205}_{-2060}$\,AU \citep{Zhang2021} which was discovered by direct imaging. 
%\textcolor{brown}{Dieser letzte Satz passt hier nicht so richtig, weil Du erst im nächsten Absatz sagst, welche der TOIs "aktiv" sind. Ich würde den Satz hier rausnehmen und eventuell in die Discussion/Conclusions am Ende vom Paper.} 

Table~\ref{tab:flares_toi} gives an overview for which of the eight TOIs and planet hosts we detected flares and rotation periods. 
The detailed results of our analysis, such as the $P_\text{rot}$ values and the flare rates with 
%and without 
a cutoff at the maximum amplitude threshold of  $3.4\cdot 10^{29}$~erg/s  
%\textcolor{orange}{Enza: Table A3 has to be updated: (1) it makes no sense to present flare rates without any detection limit, therefore remove column 4. (2) Numbers in col.5 which are for the L\_{max,th} have to be updated. (3) I wonder if we should provide also flare rates above the cutoff in the FFD. Finally, here in the text has to be revised.} \textcolor{teal}{Let me know first if I have to include the flare rate above the cutoff too and then I will update it :)}
%\textcolor{orange}{Enza, I forgot to mention, Table A3 should have the following columns (in this order): TIC ID, observation sectors, obs. duration, contamination factor (3 columns), Prot, Nflares\_tot, nu\_fl above the Lmax\_th value, nu\_fl above the Emin threshold, Emin threshold. For Emin I think you should use the mean of all Emin for each star. Does it make sense?} \textcolor{teal}{There is indeed one thing that does not make sense to me: with nu\_fl above $E_{\rm min}$ you mean considering all flares that are above the $E_{\rm min}$ of a specific sector of each flaring star? So for those $30$. If it is so I do not have this info for all of them, but only for those with reliable Prot. Indeed, I remind you I already gave the $<E_{\rm min}>$ for these 12 stars for which we performed the fit in Table~1. I'm not so sure what you want to show here, sorry :)}
(cf. Sect.~\ref{sec:flare_rate_SpT}), can be found in Table~\ref{tab:rot_act}.
Flare rates of all TOIs with two-minute cadence TESS LCs have been presented by \cite{Howard22.0}; this list should include the TOIs from our Table~\ref{tab:flares_toi} but their full target list is not available yet. 
%For this star, we detected flares at a rate of $1.68\,\text{d}^{-1}$ and a rotation period of 2.83\,d. 
%Table \ref{tab:flares_toi} summarizes the information on all \textcolor{blue}{\bf 8} confirmed planet host stars and TOIs within our sample.     

\section{Calibrating X-ray flare energies}\label{sect:calib}

The part of the stellar radiation that has the most important effects on planets is
the high-energy UV and X-ray emission. Especially crucial is the variability of the
stellar irradiation, both the amount
of the brightness changes (flares) and the frequency of the events. 
However, only a very limited number of UV and X-ray observations of M dwarfs during
flare events are available. Therefore, an indirect
way must be found to estimate the energy of XUV flares that
reaches the planet. 

Combining the extensive data base of optical flares detected with  TESS in our sample
with a collection of simultaneous optical and X-ray observations of flares on M dwarfs 
we propose here a calibration from the radiative output of flares in the optical TESS band to {\em XMM-Newton}'s X-ray band. Arriving from the observed TESS flare properties
at an estimate of the properties of the same events in the X-ray band is a 
multistep procedure. Since -- to the best of our knowledge - there are no simultaneous 
TESS and X-ray observations of M dwarf flares available we take a detour involving {\it K2} data. 
We first calibrate the TESS flare properties to their {\it K2} equivalents making use of an early-M 
dwarf observed with both TESS and {\it K2}  
(Sect.~\ref{subsubsect:calib_TESS_K2}), then we estimate the X-ray output of the flares in Sect.~\ref{subsubsect:calib_K2_xrays}
making use of a simultaneous {\it K2} and {\em XMM-Newton} observation of the 
Pleiades (\citealt{Guarcello19.0}) and a second simultaneous {\it Kepler} and {\it XMM-Newton} 
flare study by \cite{Kuznetsov21.0}. 
%\cite{2021ApJ...912...81K}. 
In Sect.~\ref{subsubsect:calib_multilambda}, we combine the results to estimate the X-ray energy of our observed TESS flares. Finally, we construct X-ray FFDs based on these estimates in Sect.~\ref{subsection:xray_ffds}.

\subsection{Simultaneous multiwavelength observations of M dwarf flares}

\subsubsection{From TESS to K2}\label{subsubsect:calib_TESS_K2}

In the course of a comparative study of TESS and {\it K2} observations in a sample of M dwarfs observed with both missions 
(Raetz et al., in prep.), wecomputed the frequency distributions of the flare
energies derived from both instruments. When comparing the TESS and {\it K2} 
flare rates one must be aware that there is a time-lapse of  about $4$\,yrs between the {\it K2}
and the TESS observations. In such a long time interval the stellar activity level can
undergo drastic changes. In fact, only one star in the sample from Raetz et al. (in prep.)
provides evidence for a stable activity level, by displaying only very moderate differences
in the spot variability pattern of the {\it K2} and the TESS LCs. This star, TYC~1330-879-1,\footnote{alias TIC\,372611670 and EPIC\,202059229.} with SpT M1 is used here to examine the 
difference in the flare energies measured with the two photometric space missions.

\begin{figure}[t!]
\centering
\includegraphics[width=9cm,angle=0]{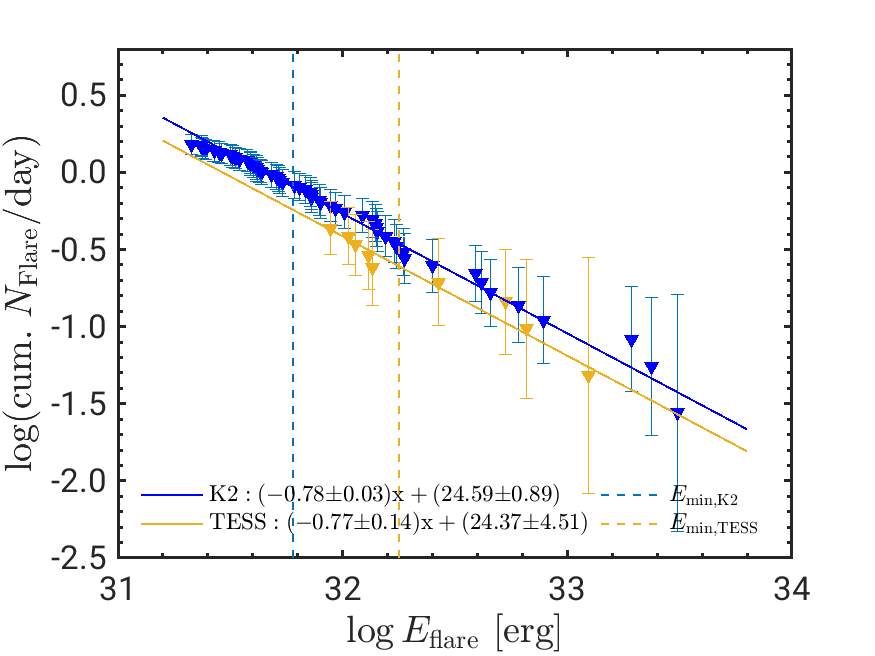}
\caption{TESS and {\it K2} FFD of TYC~1330-879-1 yielding the same flare rates in both instruments. Only data points above the sensitivity limits $E_{\rm min,th}$, marked by the vertical dashed lines, are considered in both fits. See Sect.~\ref{subsect:ffd} for the determination of $E_{\rm min,th}$.
%The determination of the detection limit is described in Sect.~\ref{subsubsect:calib_TESS_K2}.
}
\label{fig:EPIC202059229_ffd}
\end{figure}

In Fig.~\ref{fig:EPIC202059229_ffd} we show the FFDs observed with {\it K2} and TESS for TYC~1330-879-1. 
For the power law fits, we took into account only data points above the minimum detectable flare energy, 
$E_{\rm min,th}$, which was determined as described 
in Sect.~\ref{subsect:ffd}. 
%by \cite{Raetz2020.0}.
%\footnote{For TYC 1330-879-1, for the determination of the flare energy completeness threshold it was not possible to apply the procedure described in Sect.~\ref{sec:FFD} to the {\it TESS} FFD due to the low number of flares. No bending of the FFD is visible in this case. Hence, we applied the method of \cite{Raetz2020.0} for both the {\it K2} and {\it TESS} FFD of our calibration star for the sake of uniformity. \cite{Raetz2020.0} used the flare template of \cite{2014ApJ...797..122D} and varied the flare amplitude for a given flare duration of 360~s until (in the case of {\it TESS} short cadence data) 3 flare points lie above the detection threshold of $3S_\text{flat}$ (cf. Sect. \ref{sec:flare_criteria}). The energy of this synthetic flare with an amplitude just high enough that it would be detected by our flare search algorithm is then the energy completeness threshold, $E_\text{min,th}$. For a complete description of the procedure, see \cite{Raetz2020.0}.}
The values of $E_{\rm min,th}$ are shown as dashed vertical lines for both FFDs in the respective color. For the TESS FFD, only four data points lie above the limit and are considered in the fit. 
%The fits were conducted using the curve\_fit function of the scipy.optimize package. 
%\textcolor{orange}{Enza, please perform the fit with your tool also on these two FFDs.}
%\textcolor{teal}{I have re-done the fit after I corrected for the factor of 4. I considered the $E_{\rm min}$ Stefanie gave me, one value for each data (K2 and  TESS).}

The FFDs derived with the two instruments and the power law fit results are consistent with each other 
within the uncertainties. We conclude that the sensitivity to flares is similar for {\it Kepler} and TESS, 
and there is no conversion necessary between flare energies in the {\it K2} and TESS band.
In agreement with this result, \cite{2020AJ....160...36D} found similar TESS and {\it Kepler} FFDs of 
the active M4 star GJ~1243. They also examined the flux response of the TESS and {\it Kepler} filter 
to a $10000$\,K blackbody curve representing a flare and concluded that the flare energies derived with 
the two instruments are very similar.

\subsubsection{From K2 to X-rays}\label{subsubsect:calib_K2_xrays}

Opportunities to study the relation between flares in white-light and in X-rays have been rare, and the corresponding literature is not abundant. 
We make here use of two studies   with flares observed simultaneously in both wavebands, 
\cite{Guarcello19.0} and \cite{Kuznetsov21.0}, 
in order to establish a calibration between optical and X-ray flare energies. 

The work of \cite{Guarcello19.0} is based on 
the simultaneous {\it K2} and {\em XMM-Newton} (PI Drake, obs. IDs 0761920-101, -201, -301, -401, -501, -601) observations of the Pleiades cluster.  
%have constituted a rare opportunity to study the relation between flares in white-light and in X-rays.
The $12$ brightest flares in this data set have been analyzed by \cite{Guarcello19.0} and
a relation between the flare energy in the {\it K2} band and in soft X-rays was presented. 
%\textcolor{blue}{\bf The authors fitted their $E_{K2}$ vs. $E_X$ relation with six different methods, obtaining slopes between 0.64 and 1.64. This corresponds to a slope range of 0.61 to 1.56 for the inverse relation, $E_X$ vs. $E_{K2}$.} 
%\textcolor{red}{Text hier oben in blau wuerde ich weglassen. Tut nichts zur Sache; wir haben anderes Sample also gutes Recht auf andere slopes.}
We used the energy values given in Table~4 of \cite{Guarcello19.0}.
Their flare sample comprises some events on F and G stars. We kept only the $10$ flare events on 
M-type stars to adapt the sample of stars used for the {\it K2} to X-ray calibration to our M dwarf sample. 

We complemented these data with the results of 
%a second study
the study of
\cite{Kuznetsov21.0} who also studied flares observed simultaneously with {\it Kepler} and {\it XMM-Newton}. 
They give flare energies in both bands for nine flare events in their Table~2. For our energy calibration, 
we excluded one event with a much larger X-ray flare energy than the remaining ones.
The three flaring stars analyzed by \cite{Kuznetsov21.0} are of SpT K7, M2, and M3 and have rotation 
periods of $8.55$\,d, $1.50$\,d and $6.04$\,d, respectively. 
The SpT of the ten M-type stars in the \cite{Guarcello19.0} sample ranges from K8 to M5 and the $P_{\rm rot}$ 
ranges from $0.22$\,d to $4.49$\,d.
%The 10 M-type stars in the \cite{Guarcello19.0} sample comprise a SpT range from K8 to M5 and a $P_{\text{rot}}$ range from 0.22~d to 4.49~d. 
These values are in good agreement with the SpT and $P_{\rm rot}$ range of our sample 
(cf. Fig.~\ref{fig:prot_SpT}).

%\textcolor{blue}{\bf We combined the results of these two studies mentioned above to better constrain the $E_X$ vs. $E_{K2}$ relation and kept only the events on M-type stars to make sure that the calibration sample represents our targets well.}

\begin{figure}[t]
\begin{center}
\includegraphics[width=0.44\textwidth,trim=1cm 0 0 0]{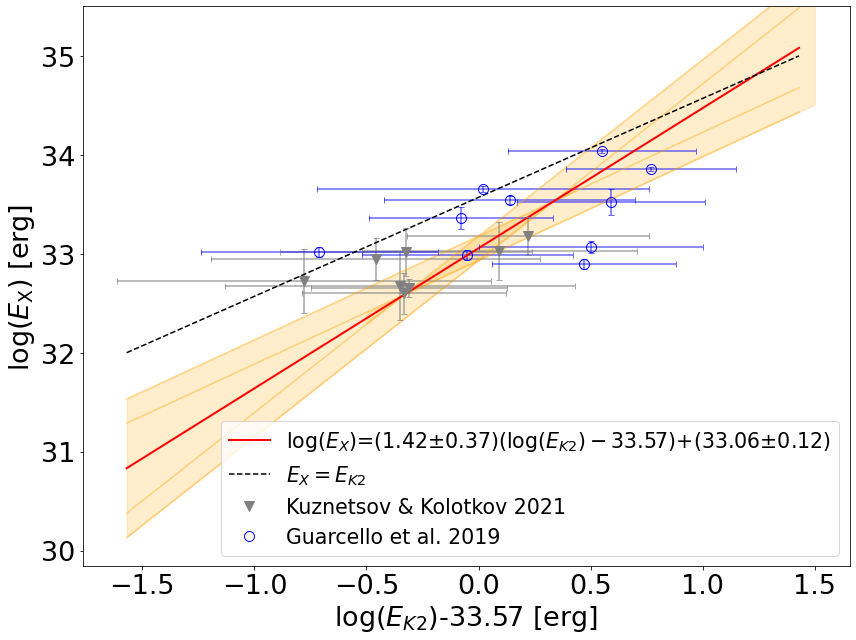}
\caption{{\it XMM-Newton} versus {\it K2} flare energy of the M stars from \cite{Guarcello19.0}, 
Table~4 and \cite{Kuznetsov21.0}, Table~2. The red line displays a linear fit taking into account 
uncertainties in both $E_{\rm X}$ and $E_{\rm K2}$. $E_{\rm K2}$ was shifted to minimize the error 
of the y-axis offset. The orange area comprises fit uncertainties in y-axis offset and slope. 
See Sect.~\ref{subsubsect:calib_K2_xrays} for details.}
\label{fig:guarcello_Ex_Ewlf_shift}
\end{center}
\end{figure}

Fig.~\ref{fig:guarcello_Ex_Ewlf_shift} shows the X-ray flare energy versus optical flare energy in the {\em Kepler/K2} band for the flare events of the ten M dwarfs in the \cite{Guarcello19.0} sample, together with the eight flare events of \cite{Kuznetsov21.0}. We fitted a linear relation of the type $y=ax+b$ to these data, where $x=E_{\rm K2}$ is the flare energy in the {\em Kepler/K2} band and $y = E_{\rm X}$ the flare energy in the {\it XMM-Newton} X-ray band.
We note that \cite{Guarcello19.0} and \cite{Kuznetsov21.0} use different {\it XMM-Newton} energy bands 
($0.3-7.9$\,keV versus $0.2-12$\,keV). However, given the difference of only $0.1$\,keV at the low energy 
end and the fact that most of the stellar X-rays are emitted at soft energies ($<5$\,keV), this is unlikely 
to affect our analysis.

To minimize the error of the y-axis offset, $b$, we shifted $E_{\rm K2}$ such that the zeropoint of 
the x-axis lies in the middle between the highest and lowest {\it K2} flare energy of the flares 
shown in Fig.~\ref{fig:guarcello_Ex_Ewlf_shift}. This 
way smaller errors of the y-axis offset are achieved. 
%is justified since our goal is to derive a relation for converting optical flare energies to the X-ray band, hence the relation is only applied to \textcolor{blue}{\bf an energy range of} $\log{E_{\rm K2}}\,\text{[erg]}\textcolor{blue}{\mathbf{=}}31...35$, not \textcolor{blue}{\bf around} $\log{E_{\rm K2}}\,{\rm [erg]}=0$. 
%\textcolor{orange}{Enza, Stefanie: do you understand the above sentence?}
%\textcolor{brown}{SR: Hm, I don't understand the sentence either. My guess would be, that Mirjam meant that she removed a constant offset from the x-axis before doing the linear fit as shown in Fig. 16. But since this offset is clearly seen in the axis can't this sentence be removed here and in the figure caption?} 
%\textcolor{teal}{First, I wonder: how did Guarcello+19 calculate the energy? Did they consider $4 \pi d^2$ or $\pi d^2$? In the paper seems to be not explicitly specified. Same thing regarding Kuznetsov \& Kolotkov+21. 
%Second: where Mirjam wrote $\log E_{K2}~\rm [erg] =0$ I think she meant 33.57, according to the shift she applied to the x axis..but still I do not understand how the relation is applied only for $\log E_{K2}=31..35$.}
In order to be able to consider uncertainties in both $E_{\rm X}$ and $E_{\rm K2}$ in the fit, 
we used orthogonal distance regression (ODR){\bf \footnote{The fit was performed using the scipy.odr package.}}. 
Symmetric uncertainties are required for the fitting, and we took the larger values of the asymmetric 
errors presented in the literature accepting that this leads to a slight overestimate of the errors.  
%However, we prefer this solution as compared to taking the minimum or average of the lower and upper error since this causes an underestimation of the uncertainty.}
The fit result is also displayed %as a red line 
in Fig.~\ref{fig:guarcello_Ex_Ewlf_shift}. The black dashed line marks the 1:1 relation 
(i.e., $E_{\rm X} = E_{\rm K2}$). 
%Fig. \ref{fig:guarcello_Ex_Ewlf_shift} shows the X-ray flare energy versus optical flare energy in the {\it Kepler/K2} band for the flare events of the 10 M dwarfs in the \cite{Guarcello19.0} sample, together with the 8 flare events of \cite{2021ApJ...912...81K}. We fitted a linear relation of the type $y=ax+b$ to these data, where $x=E_{\rm K2}$ is the flare energy in the {\it Kepler/K2} band and $y=E_{\rm X}$ the flare energy in the {\it XMM-Newton} X-ray band. 
%The fit result is displayed as a red line in Fig. \ref{fig:guarcello_Ex_Ewlf_shift}. The black dashed line marks the 1:1 relation (i. e. $E_{\rm X}=E_{\rm K2}$). In order to be able to consider uncertainties in both $E_{\rm X}$ and $E_{\rm K2}$ in the fit, we used orthogonal distance regression (ODR). To obtain symmetric uncertainties, we took the larger values where asymmetric errors were given.
%To minimize the error of the y-axis offset, $b$, we shifted $E_{\rm K2}$ such that the zeropoint of the x-axis lies in the middle between the highest and lowest {\it K2} flare energy of the flares shown in Fig.~\ref{fig:guarcello_Ex_Ewlf_shift}. This is justified since our goal is to derive a relation for converting common optical flare energies to the X-ray band, hence the relation is only applied to energies $\log E_{\rm K2}\text{ [erg]}\sim31...35$,  not at $\log E_{\rm K2}\text{ [erg]}\sim0$.
As a result of the fit, we obtained 
\begin{equation}
\text{log}(E_{\rm X})=(1.42\pm0.37)\cdot(\text{log}(E_{\rm K2})-33.57)+(33.06\pm0.12),
\label{eq:K2Xray}
\end{equation}

\noindent where $E_{\rm X}$ and $E_{\rm K2}$ are given in erg.

%We note that \cite{Guarcello19.0} and \cite{2021ApJ...912...81K} use different {\it XMM-Newton} energy bands (0.3-7.9\,keV versus 0.2-12\,keV). However, given the difference of only 0.1\,keV at the low energy end and the fact that most of the stellar X-rays are emitted at soft energies, this is unlikely to affect our analysis.

\subsubsection{Calibration based on the multiwavelength data}\label{subsubsect:calib_multilambda}

The calibrations presented in Sects.~\ref{subsubsect:calib_TESS_K2} 
and~\ref{subsubsect:calib_K2_xrays} can be combined to estimate for a flare observed
with TESS its equivalent energy output in the {\em XMM-Newton} X-ray band. 
As discussed in Sect.~\ref{subsubsect:calib_TESS_K2}, no conversion is necessary between flare energies in the TESS and {\it K2} band. Thus, we obtain the flare
energy in the X-ray band by applying the relation obtained from the fit to the Pleiades M-type stars of 
\cite{Guarcello19.0} and the field M-type stars of \cite{Kuznetsov21.0} 
%\cite{2021ApJ...912...81K} 
directly to our TESS flare energies:
\begin{equation} 
\text{log}(E_{\rm X}) = a \cdot (\text{log}(E_{TESS})-33.57) + b, \label{eq:tess_xray}
\end{equation}
with the values of $a = 1.42\pm0.37$ and $b=33.06\pm0.12$ derived in Sect.~\ref{subsubsect:calib_K2_xrays} and $E_{\rm X}$ and $E_{\rm K2}$ in units of erg. 

The optical/X-ray flares on which we calibrated the relation between $E_{\rm K2}$ and $E_{\rm X}$ comprise flare energies between roughly  $10^{32.5}$\,erg and $10^{34.5}$\,erg. Our calibration between the TESS and {\it K2} FFDs has been derived for slightly lower energies ($\lesssim 10^{33.5}$\,erg, cf. Fig.~\ref{fig:EPIC202059229_ffd}). However, no change in
  power law slope is expected at higher energies. 
%  \textcolor{orange}{I would also think that the PL does not change (again) at high energies, but I remember having seen a paper where they see exactly that. I do not remember which paper. Stefanie?}
  Flare energy frequency distributions with similar slopes to that found by us for TYC~1330-879-1 have been observed
  on superflare stars with energies up to $10^{36}$\,erg \citep{Shibayama13.0}.
  %\textcolor{orange}{I do not understand this sentence: This TYCHO star seems not to make part of our sample of 12 stars with FFD. Moreover what is the meaning of the sentence. You have an idea? I would remove it.}
  \cite{Raetz2020.0} studied the FFDs of M dwarfs 
  covering a range of activity levels, which translate into a vertical offset of the
  FFDs but do not affect the slope. A similar result is seen in the literature compilation of FFDs by \cite{Ilin19.0}. 
  Therefore, the one-to-one
correspondence between TESS and {\it K2} flare frequencies can safely be 
extrapolated to higher flare energies.

%The {\it K2} and {\it Kepler} M star flares with X-ray counterparts that we used to calibrate the $E_{\rm X}$ versus $E_{\rm K2}$ relation (cf. Fig. \ref{fig:guarcello_Ex_Ewlf_shift}) cover roughly the same energy range as the flares we detected in our sample of HZCat stars. 

\subsection{X-ray FFDs}\label{subsection:xray_ffds}

\begin{figure}[t]
\centering
\includegraphics[width = 0.5\textwidth]{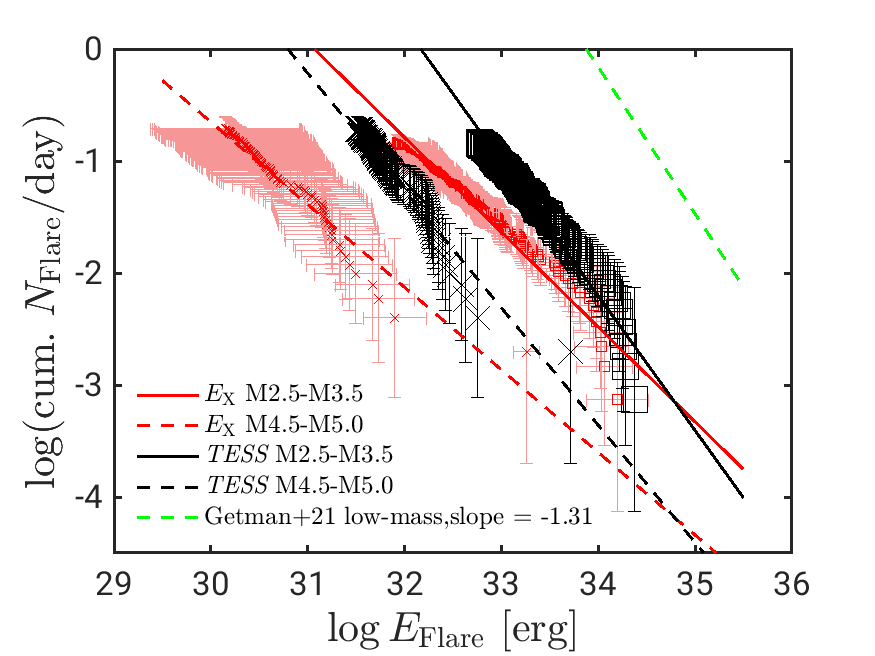}
\caption{Average TESS FFDs for stars with reliable $P_{\rm rot}$ in the early (M2.5$-$M3.5) and late (M4.5$-$M5) SpT range derived in Sect.~\ref{subsect:ffd}  (black) and constructed X-ray FFDs based on the energy conversion derived in Sect.~\ref{subsubsect:calib_multilambda} (red). Error bars on $E_{\rm X}$ represent the uncertainty caused by this conversion. Error bars on the flare rate are the same for the observed TESS (black) and constructed {\it XMM-Newton} (red) FFDs and result from Eq.~\ref{eq:nuflare_err}. 
%The orange dashed and solid lines show the power law fit when flare energies are transformed to the X-ray band using the smallest slope within the uncertainty in the relation in Eq.~\ref{eq:tess_xray} (that is $a = 1.05$). 
The green line represents the  extrapolation to small flare energies for the observed X-ray FFD that \cite{Getman2021.0} found for PMS stars with masses $\leq 1M_\odot$. The slopes of our power law fits are given in Table~\ref{tab:ffd_slopes}.}
\label{fig:xray_ffd}
\end{figure}

We can now use the relation between optical and X-ray flare energies derived in Sect.~\ref{subsubsect:calib_multilambda} to construct X-ray FFDs for our HZCat sample directly from the observed TESS flares.
We do this separately for the two average TESS FFDs of our early (M2.5$-$M3.5) and late
(M4.5$-$M5) samples with reliable rotation period. The X-ray FFDs are
%This is
obtained by shifting each energy value of the average TESS FFDs from Fig.~\ref{fig:ffd_HZsample} into the {\it XMM-Newton} X-ray band with the relation in Eq.~\ref{eq:tess_xray}. The result is shown in Fig.~\ref{fig:xray_ffd} (red) together with the observed TESS FFDs (black). For comparison, we also show the observed X-ray FFD in the low-mass regime ($\leq 1\,M_\odot$) from the pre-MS star study of \cite{Getman2021.0}. 
%as a green dashed line. 
We subtracted a constant value of $\log{365.25}$ from the published relation of \cite{Getman2021.0} since they give flare rates in units of ${\rm yr^{-1}}$.

We performed a linear fit on our constructed X-ray FFDs. In order to be able to consider both the errors in $E_{\rm X}$ resulting from the energy conversion and the flare rate uncertainties, we again used ODR. The resulting FFD power law slopes and their uncertainties, which are the formal errors of the fit, are given in Table~\ref{tab:ffd_slopes} together with the power law slopes of our two SpT-averaged  TESS FFDs. %\textcolor{gray}{\it Old text by Mirjam:
%They are less steep than our results for the constructed X-ray FFDs since the slope of our $E_{\rm K2}$ to $E_{\rm X}$ calibration (Eq.~\ref{eq:K2Xray}) is steeper than the 1:1 relation.
%\textcolor{orange}{\bf Enza, Stefanie: can you confirm that the above text in grey is wrong? Below is my next text.} \textcolor{teal}{I would say that your text is correct, by looking at the results and eqn~5.}\textcolor{brown}{SR: Yes, Mirjams sentence is wrong and the one in red below is correct.}
The synthesized X-ray FFDs are less steep than the observed TESS FFDs because the slope of our $E_{\rm K2}$ to $E_{\rm X}$ calibration (Eq.~\ref{eq:K2Xray}) is steeper than the 1:1 relation. 
Further, the X-ray FFDs are shifted toward lower flare energies with respect to the optical ones. This directly follows from the fact that almost all flare events used for the calibration show higher energies in the {\it K2} band than in X-rays (cf. Fig.~\ref{fig:guarcello_Ex_Ewlf_shift}). 
%To illustrate the effect of the \textcolor{red}{\bf  uncertainty on the slope} in the $E_{\rm X}$ versus \textcolor{red}{\bf $E_{\rm TESS}$} relation on the X-ray FFDs, the orange solid and dashed lines in Fig.~\ref{fig:xray_ffd} display the FFD power law fit when flare energies are transformed with $\log{E_{\rm X}} = 1.05 (\log{E_{\rm K2}} - 33.57) + 33.06$, i. e. with the smallest slope that is within the uncertainty of our $E_{\rm X}$ versus $E_{\rm K2}$ calibration. The corresponding slopes 
%%of both X-ray FFDs 
%are also given in Table~\ref{tab:ffd_slopes}. \textcolor{red}{\bf As expected} they are steeper %with relation to the X-ray FFDs obtained using an energy conversion slope of $a=1.42$ 
%and nearly parallel to the observed {\it TESS} FFDs since the slope of the energy conversion in this case is close to 1.

We note that our transformation from optical to X-ray FFDs involves the assumption of a 1:1 correspondence of the flare occurrence in both calibration steps: from TESS to {\it K2} as well as from {\it K2} to X-rays. There are two aspects about this assumption: first, the actual physical 1:1 correspondence in flare occurrence, and second, the 1:1 correspondence in observations which might be impeded by different detection biases for the two instruments. That each TESS flare is associated with a flare in the {\it K2} band should be fulfilled as they both represent white-light emission. Moreover, we have demonstrated that TESS and {\it K2} yield identical FFDs (see Sect.~\ref{subsubsect:calib_TESS_K2}). 

The optical to X-ray calibration is more complex.   
Regarding the physical 1:1 correspondence in flare occurrence in this case, it is instructive to have a look at results for our Sun.
In the standard solar flare picture, optical flares are the result of bombardment of the lower atmosphere with particles accelerated in the corona; the ensuing density enhancement and heating of the corona produces the soft X-ray flare (e.g., \citealt{1982SoPh...78..107A}, \citealt{1983ApJ...266..383C}), which is detected in the case of stellar flares with {\it XMM-Newton}.
For the case of the Sun there seems not to be an unexpectedly high number of soft X-ray flares without an optical counterpart or vice versa: {\cite{Milligan2018.0}} systematically studied the number of solar flares greater than GOES C1 class in solar Cycle\,24 that were observed simultaneously by instruments operating at different wavelengths. For various combinations of instruments, they compared the actual number of simultaneously observed flares to the number that can be expected theoretically, assuming that the instruments operate independently and considering for each instrument its rate of success to observe a flare. For the Solar Optical Telescope (SOT, \citealt{SOT}), the number of flares captured simultaneously with the X-Ray Telescope of the {\it Hinode} mission (XRT, \citealt{XRT}) was higher than the expected value. Since \cite{Milligan2018.0} took into account the detection sensitivity of each instrument,
  %their result can be interpreted such
this means that they did not find a discrepancy that could be attributed to an actual lack of counterparts in either of the two bands.

The second caveat in our transformation of optical to X-ray FFDs regards detection biases. Even if each soft X-ray flare comes with an optical counterpart and vice versa, it might happen that one of the two events is not captured by the respective instrument due to sensitivity biases or other 
%preventing 
circumstances. This means that we are translating any potential bias in TESS flare observations while we are not considering the observation biases in the X-ray band. Apart from the sensitivity of each individual instrument toward flare detection, here the viewing geometry of a flare event may also play a role for its visibility in different bands: For instance, \cite{Flaccomio2018.0} explained X-ray flares without optical counterpart in the star forming region NGC\,2264 by events occurring behind the limb such that the lower-lying optical emission remains hidden to our view while the more extended coronal loops are still partly visible. We note that the majority of their X-ray flares had an optical counterpart,
%and in the absence of a clear observational answer
strengthening
our hypothesis of each optical flare being associated with an X-ray flare and the other way round.
%can be considered as a reasonable guess.

To summarize, we have constructed X-ray FFDs by translating TESS flare rates from the optical energy at which they have been observed to the calibrated X-ray flare energy. 
These X-ray FFDs represent a prediction for the FFDs that hypothetical observations of our stars in the {\em XMM-Newton} energy band would yield. They are based on strong assumptions, (1) that each optical flare is associated with an X-ray flare and vice versa as discussed in detail above, and (2) that the sensitivity and biases for flare detection are the same for TESS and {\em XMM-Newton}. As long as no observational constraints for X-ray FFDs are available our prediction may serve as a useful input to planet irradiation models for active M dwarfs. 
%\textcolor{blue}{\bf This implies that our X-ray FFDs include }the sensitivity and biases towards flare detection \textcolor{blue}{\bf from the optical {\it TESS} observations, transferred to the X-ray wavelength range.}

%Based on that assumption, when we translated our {\it TESS} flare rates from the optical energy at which they have been observed to the calibrated X-ray flare energy we have constructed X-ray FFDs that would be observed by a hypothetical X-ray instrument covering the energy range of {\it XMM-Newton} and having the same sensitivity and biases towards flare detection as {\it TESS}. 

\section{Discussion}\label{sect:discussion}

% - FOLLOWING IS TEXT ON FACTOR 4
%\textcolor{red}{\bf We note that in the flare literature there are different ways to calculate the stellar luminosity that constitutes a normalization factor of the photometric LC. \cite{Shibayama13.0} base their calculation on surface fluxes and obtain the luminosity (of flare and star) considering only the part of the area projected onto the line-of-sight, and several authors have followed this approach. Others use the formalism we applied \cite[e.g.][]{2014ApJ...797..121H, Howard19.0} to obtain the quiescent luminosity from magnitude and distance. The results differ by a factor of four. Therefore, care is needed in a direct comparison of flare amplitudes and energies computed with these different methods.... Moreover, the flare amplitudes and energies calculated by the different studies are valid for the energy band probed by the respective instrument. Our study of the M dwarf TIC... observed both with K2 and TESS and the study of GJ... by DavenportXX shows that the FFDs of both missions are nearly identical but this may not be the case for other studies involving e.g. ground-based instruments.... The method of \cite{Shibayama13.0} yields the bolometric flare energy....} 

\subsection{Optical flares and rotation on HZCat stars}

We analyzed the activity and rotation properties of $109$ M dwarfs using TESS LCs.
The stars were selected to have the characteristic that
planets in their entire HZ can be detected with TESS through their transits. For two of our targets, in fact, TESS has discovered planets that were confirmed by follow-up observations. 
Two targets have planets discoverd with other methods,  
%One additional target was a known planet host already before the {\it TESS} mission 
and four additional stars within our sample are classified as TOI.

\subsubsection{Statistics of rotation period detections}\label{subsubsect:disc_prot_statistics}

We detected rotation periods on $12$ stars, only $11.0$\,\% of the sample. This fraction
is low compared to numbers found for M dwarfs in {\it Kepler} studies, for example \citealt{2013MNRAS.432.1203M} ($63.2$\,\%), \citealt{Stelzer2016.0} ($72.4$\,\%) or \citealt{Raetz2020.0} ($82.1$\,\%).
%Weascribe this to the 
One possible reason for the low period detection fraction in our sample is the 
lower photometric precision of TESS with respect to {\it Kepler} and {\it K2}
LCs (cf. Fig.~\ref{S_flat}), which makes the notoriously small amplitudes of spot modulations in M dwarfs
difficult to distinguish from the noise. \cite{2020ApJS..250...20C} examined rotation periods of $1000$ TOIs using 2-minute cadence LCs of the first $20$ months of the TESS mission. They found only $131$ targets with unambiguous rotation periods, that is a fraction of $13.1$\,\%, which is in good agreement with the $11.0$\,\% for our smaller sample.
However, when comparing our results to literature studies, it also has to be considered that our period search is a priori limited to a maximum period of $13.5$\,d (cf. Sect.~\ref{period_search}). \cite{2013MNRAS.432.1203M} find rotation periods below this limit for a fraction of only $7.5$\,\% of their sample, and thus our value of $11.0$\,\% is even slightly higher. In the sample of \cite{Stelzer2016.0}, $17.9$\,\% of the analyzed stars have $P_{\rm rot}<13.5$\,d. \cite{Medina2020} searched for rotation periods with TESS photometry and ground based photometry from MEarth using as subsample the volume-complete $15$\,pc mid- to late-M dwarf sample of \cite{Winters2021}. 
They included  previously published rotation periods from the literature for about $1/4$ of their sample and they found that in their total sample of $125$ stars $27.2$\,\% have  $P_{\rm rot}<13.5$\,d. 
%For 35 of their stars, they adopted previously published $P_\text{rot}$ values from the literature. 34 of the total sample of 125 stars have P\_rot<13.5 d, which corresponds to a fraction of 27.2\%. 
Thus, our fraction of $11$\,\% of stars with measured $P_{\rm rot}$ from TESS photometry is 
%on average not much 
by at most a factor of two
lower than 
the results from Kepler, K2 or ground-based studies. 
%the results presented in the studies cited above that determined rotation periods from {\it Kepler/K2} or ground based photometry.}

\subsubsection{Statistics of flare detections}\label{subsubsect:disc_flare_statistics}

Flares are present on a much larger number of stars than rotation periods ($35$ out of $109$ stars, that is $32.1$\,\% of the sample). 
However, $84\,\%$ of the flares we detect are found on the $12$ stars with reliable $P_{\rm rot}$ that are all fast rotators. \cite{Raetz2020.0} established a  bimodality in the flare rate with frequent events on stars with $P_{\rm rot} \lesssim 10$\,d and low flare rate for slower rotators. An enhanced flare frequency on fast rotators was also found in other studies, for example  \cite{Zeldes22.0}.  

The observed flare rate depends also on the observing cadence and SpT of the sample. 
\cite{2019ApJS..241...29Y} have presented a flare catalog of the {\em Kepler} mission that comprises 
$3420$ flaring stars. Their flare incidence rate for M stars is significantly lower 
($9.74$\,\%) than our result. We note that their work is based on {\it Kepler} long cadence LCs, which is likely the main reason for the lower flare incidence rate.  \cite{2018ApJ...859...87Y} compared flare properties in long and short cadence LCs of the entire {\it Kepler} mission data set and found that a majority ( about $60$\,\%) of the short cadence flares have no long cadence counterpart.  \cite{Raetz2020.0} found an even lower fraction of $31$\,\% of short cadence flares recovered in long cadence for their M dwarf sample. 
%This discrepancy certainly also affects the flare incidence rate. 
%, cf. their Table 3) than our 31\%.  
%\textcolor{red}{Hatten wir nicht (ueberraschend) schon irgendwo anders Hinweise gefunden,
%dass {\it TESS} mehr flares sieht als Kepler?} \textcolor{blue}{Ich weiß nicht mehr - wahrscheinlich schwer zu belegen durch den zeitlichen Abstand. Der niedrige Wert hier liegt aber (mal wieder) an der long cadence, die Info hat gefehlt.}
\cite{2020AJ....159...60G} analyzed a large sample of $24809$ stars observed in the first two months of 
the TESS mission in 2-minute cadence. They found a flare occurrence rate of $>40$\% for mid- to late-M stars (SpT M4 to M6) 
and $10$\,\% for early-M stars. These numbers are in better agreement with our flare occurrence rate of 
$32$\,\% although our sample does not only comprise mid to late-M stars. 
% an excess of M4 to M6 stars.
\cite{2020AJ....159...60G} 
%find in flare studies of stars observed in the first two months of the {\it TESS} mission 
also found 
that M4 to M6 dwarfs show the highest fractions of flaring targets. This is in good agreement with the SpT distribution of the flaring stars in our sample. An increase of the fraction of flaring stars 
from SpT M0 to M5 was also seen by \cite{2017ApJ...849...36Y} and by \cite{2019ApJS..243...28L}, both
based on Kepler and {\it K2} long-cadence data. \cite{Rodriguez2020.0} found the same for LCs of $1376$ nearby M dwarfs from the All-Sky Automated Survey for Supernovae (ASAS-SN, \citealt{Shappee2014.0}, \citealt{Kochanek2017.0}).

%\textcolor{orange}{I think we should put and discuss here the flare rate of the two average samples, i.e. the number $\nu$ at the Emin,th threshold for the two black distributions of Fig.13.} 
To account for detection biases %related to the fact that only flares with a total amplitude $>3S_\text{flat}\cdot L_\text{qui}$ can be detected for each star (cf. Sect. \ref{sec:flare_rate_SpT}) and to make the flare rates of our 35 flaring stars comparable with each other, 
we calculated 
%different 
flare rates for our sample 
%\textcolor{gray}{\it Old text from Mirjam: One where all flares for the respective target are considered and another one for only flares with luminosity amplitudes $> 6.6\cdot 10^{29}$\,erg/s. This threshold is the highest value of $3S_\text{flat}\cdot L_\text{qui}$ found among our 35 flaring stars.}
considering two different detection completeness limits regarding (i) the flare luminosity amplitude (described in Sect.~\ref{sec:flare_rate_SpT}) and (ii) the flare energy (described in Sect.~\ref{subsect:ffd}). As a universal limit for the minimum flare amplitude required for detection within all stars of our sample we had determined a luminosity amplitude $L_{\rm max,th}$ with a value of 
% no factor 4
%\textcolor{teal}{$ 8.6\cdot 10^{28}$}\,erg/s.}
$ 3.4\cdot 10^{29}$\,erg/s. 
The average flare rate over all $109$ stares and considering only flares above $L_{\rm max,th}$ is $0.043\,\rm d^{-1}$ ($1275$ flares on $30$ stars). 
%\textcolor{gray}{\it Old text from Mirjam: 
%The average flare rate of these 35 stars regardless of the threshold is $\overline{\nu}_F\approx 0.302~\textrm{d}^{-1}$ and for flares with amplitudes above the threshold it is $\overline{\nu}_{F,\text{th}}\approx 0.144~\textrm{d}^{-1}$.}
Considering only the $30$ stars that present events above this critical luminosity amplitude
%, $L_{\rm qui,T, th}$, 
we find $\overline{\nu}_{\rm F,L_{\rm max,th}}\approx 
%flare rate is 0.169 1/d for all 35 flaring stars
0.168~\textrm{d}^{-1}$. %\textcolor{teal}{No, 0.144 is the mean calculated including all 35 flaring stars. If you want only those with flares above the thresold than the mean rate is 0.168.}
%\textcolor{orange}{Enza, why is your flare rate for the 35 stars different from the value that Mirjam has given?} \textcolor{teal}{Done, but if I take ONLY those flares above the universal threshold the stars are 30 and not 35.}\textcolor{orange}{\bf Do you get Mirjam's value if you include 35 instead of 30 stars?}
However, we found a 
strong SpT dependence of the flare rate, with no flares at all detected on stars with SpT earlier than M2.0 and a peak in the flare rate around M3$-$M4. %The flare rates above $L_{\rm max,th}$ for the two SpT subgroups we examined are \textcolor{teal}{$\boldsymbol{\bar{\nu}_{\rm F,earlyM,L_{\rm max,th}}=0.375\,\rm d^{-1}}$} (M2.5$-$M3.5 \textcolor{red}{with \textcolor{teal}{$509$} flares, see Table~\ref{tab:spt_flares}}) and \textcolor{teal}{$\boldsymbol{\bar{\nu}_{\rm F,midM,L_{\rm max,th}}=0.029}\,\rm d^{-1}$} (M4.5$-$M5, \textcolor{red}{with only \textcolor{teal}{$15$} flares in total}).

Similar to $L_{\rm max,th}$ we can define a conservative threshold for the completeness in terms of flare energy from the FFDs. The FFD analysis was carried out only for the $12$ stars with measured rotation period,
 seven of which have contamination factors $<10$\,\% and are single stars. This latter subsample comes up for %which come up for 
{$\sim 50$\,\%} of all detected flares. 
%\textcolor{orange}{Enza, confirm this percentage of flares for the 7 stars.}
%\textcolor{orange}{No, the text wants to state the fraction of all flares that are on the 12 stars with Prot. Actually, this number is simply: 2138/2532 = 84\%. I guess I asked you just to confirm these numbers. Can you confirm them?} 
We take as energy threshold the largest value of the  $E_{\rm min,th}$ among all those present in the sample.
%%stars that define each of the two above M subclasses. 
%We find for this parameter, 
%% no factor 4
%%${(\langle E_{\rm min,th} %\rangle)_{\rm max} = %32.1639}$\,erg 
%${(\langle E_{\rm min,th} %\rangle)_{\rm max} = %32.776}$\,erg for the %M2.5$-$M3-5 range and % no %factor 4
%%${(\langle E_{\rm min,th} %\rangle)_{\rm max} = %30.9079}$\,erg 
%${(\langle E_{\rm min,th} %\rangle)_{\rm max} = %31.510}$\,erg for the %M4.5$-$M5 range.
  %\textcolor{teal}{This is still not entirely true. We did not calculate the ``averaged'' (I hate call it this way :) FFD considering all $12$ TIC stars with reliable period. We neglected those with contamination factor < 10\% and that are CPM. So they are in total $7$ for this analysis.} %\textcolor{gray}{\bf We take as energy threshold the largest value of the median $E_{\rm min,th}$ measured in the individual sector light curves for a given star. We find for this parameter, $(\langle E_{\rm min,th} \rangle)_{\rm max} =...$} 
The %corresponding 
rate of all events above this  $(\langle E_{\rm min,th} \rangle)_{\rm max}$ averaged over the $7$ stars is $\bar{\nu}_{\rm F,L_{\rm max,th}}=0.682\,\rm d^{-1}$ %($1337$ flares)
.
%Separately, for the SpT-averaged FFDs} {\bf of sub-group M2.5$-$M3.5 we found {$\boldsymbol{\bar{\nu}_{\rm F,midM,L_{\rm max,th}}=0.663}\,\rm d^{-1}$} ($5$ stars with a total of {$900$} flares) and $\boldsymbol{\bar{\nu}_{\rm F,midM,L_{\rm max,th}}=0.734}\,\rm d^{-1}$ ($2$ stars with a total of $369$ flares) for SpT of M4.5$-$5. Note that these values for the flare rates are not directly comparable with each other because the energy threshold is lower for later-type stars.

%\textcolor{orange}{Enza, can you please calculate all these numbers?}

%\textcolor{blue}{\bf With the exception of TIC272232401 (L34-26),} the flare rates of the \textcolor{blue}{\bf eight} planet hosts and TOIs within our sample (cf. Table \ref{tab:flares_toi}) are $\leq 0.062~\text{d}^{-1}$ (without amplitude threshold) and thus way below the average flare rate of the whole sample given above. 

Among the eight known planet host stars and TOIs, only TOI-1450 and L34-26 show flares above the amplitude  threshold. The latter is the most active star among the planet hosts and the only one for which we measured a rotation period. It is fast-rotating ($P_{\rm rot}=2.83$\,d) and shows flares at a frequency of %1.68\,d$^{-1}$ without detection threshold and 
$0.48$\,d$^{-1}$ above the detection threshold, $L_{\rm max,th}$. The planet, however, should not be influenced by the host star's activity, given its large orbital separation (cf. Sect.~\ref{subsection:flares_tois}). 

\subsubsection{Relation between flare amplitude and duration}\label{subsubsect:disc_flare_amplitude_duration}

We found a linear relation between flare amplitude and duration for our sample of 2532 flares 
that is similar to the one determined by \cite{Raetz2020.0} based on {\it K2} short cadence data (cf. Fig.~\ref{fig:flare_ampl_vs_duration}). The slope found by \cite{2014ApJ...797..121H} for {\it Kepler} short cadence observations of the highly active M dwarf GJ\,1243 is  slightly steeper than what we found for our sample. More drastic is the shift between our amplitude-duration distributions and that of GJ\,1243.

%\textcolor{blue}{Hawley misst die flare duration auch als Differenz zw. erstem und letztem Flare Punkt. Allerdings legen sie die Flare Punkte durch visuelle Inspektion fest (mindestens 2 Leute stimmen zu, dass ein Punkt zum Flare gehört).}\\
%Thus, flare amplitudes of {\it Kepler} short cadence flares on GJ1243 grow faster with flare duration than it is the case for our sample. 

Since GJ\,1243 has been considered a prototypical flare star it is relevant to examine the characteristics of its events in the context of larger samples of M dwarfs.  
There are three possible sources for the differences with respect to our amplitude-duration distributions: 
%First, it might be due to differences of \textcolor{red}{\bf flare parameters obtained from} $K2$ short cadence 
%%flare 
%observations with respect to {\it TESS}. \textcolor{red}{\bf However,}
%the fact that the slopes in the amplitude-duration relation for both our SpT subsamples is nearly parallel to that of \cite{Raetz2020.0} stands against a systematic {\it Kepler/K2} versus {\it TESS} difference.  %%being the reason for the steeper slope that \cite{2014ApJ...797..121H} found for GJ1243. 
%Another 
A possible 
explanation could be a systematic difference in flare duration measurements between our analysis and that of \cite{2014ApJ...797..121H}.
Analogously to our method, \cite{2014ApJ...797..121H} also measure the flare duration as the time difference between the first and last flare point. In doing so, they determine flare points by visual inspection while we rely on the output of our flare search algorithm. This makes it difficult to draw a conclusion on whether systematic differences in the flare duration measurement are responsible for the different slopes in the amplitude-duration relation. 
%In favour of this explanation is the fact that we use the same flare detection algorithm as \cite{Raetz2020.0} which results in nearly parallel slopes. 
 %The ranges of flare duration and amplitude are roughly consistent with each other for our HZCat sample and GJ1243. 
%The fact that the slopes in the amplitude-duration relation for both our SpT subsamples is nearly parallel to that of \cite{Raetz2020.0} stands against a systematic {\it Kepler/K2} versus {\it TESS} difference being the reason for the steeper slope that \cite{2014ApJ...797..121H} found for GJ1243. 
Secondly, the different instruments used could play a role.  
The higher standard deviation in TESS LCs compared to {\it K2} (cf. Fig.~\ref{S_flat}) might lead to underestimated flare durations with  TESS inasmuch as flare points at the end of the decay phase disappear in the noise. We expect this to have a larger effect on longer flares, simply because they have a longer decay phase. Flare amplitudes should be affected much less by a higher noise level since they are determined as the difference between the peak flare flux and the flux of the smoothed, interpolated LC at the time of the flare peak. 
A potential understimation of the flare duration would shift the TESS amplitude-duration relation to the left, as observed when we compare our sample to GJ\,1243. 
Finally, GJ\,1243 might simply exhibit flares of a certain shape such that the flare amplitude grows faster with the duration. 
%The potential underestimation of the flare duration should therefore rather lead to a steeper slope for our {\it TESS} sample compared to the {\it Kepler} study of \cite{2014ApJ...797..121H} as our longer flares might be increasingly underestimated in duration. 
%Analogously to our method, \cite{2014ApJ...797..121H} also measure the flare duration as the time difference between the first and last flare point. In doing so, they determine flare points by visual inspection while we rely on the output of our flare search algorithm. This makes it difficult to draw a conclusion on whether systematic differences in the flare duration measurement are responsible for the different slopes in the amplitude-duration relation. In favour of this explanation is the fact that we use the same flare detection algorithm as \cite{Raetz2020.0} which results in nearly parallel slopes.

We further found for our HZCat stars that the relative flare amplitude $A_{\rm peak}$ as a function of the flare duration is indistinguishable for the early and mid/late SpT subsample, whereas the absolute flare amplitude $\Delta L_{\rm F,T}$ is shifted upward for the earlier SpTs (cf. Fig.~\ref{fig:flare_ampl_vs_duration}). This is a consequence of their higher quiescent luminosity and ensuing higher detection threshold.

\subsubsection{Slope of flare frequency distributions}\label{subsubsect:disc_ffd}

We have presented a detailed study of cumulative FFDs, which yielded
for the $12$ stars with reliable rotation period values 
%The absolute values of our power law slopes range 
from $-1.27$ to $-0.84$
%0.597 to 0.954 
for the power law slopes $\alpha$ from Eq.~3. 
Converting the differential flare distributions given by 
\cite{2017ApJ...849...36Y} to the cumulative form yields power law slopes of $-1.19\pm0.33$ for the $15$ M stars within their sample of 540 that have {\it Kepler} short cadence data.  Similarly, the cumulative power law slope found by \cite{2019ApJS..243...28L} is  $-1.02\pm0.11$ for $K2/Kepler$ long cadence data. 
%The power law slopes we cite from the latter two studies correspond to $-\alpha + 1$ for their $\alpha$-Parameter.
%\textcolor{orange}{Stefanie, could you please check this statement? i.e. how are signs and values of alpha and beta defined in Raetz+20 and Lu+19?} \textcolor{brown}{SR: my power-law fit is defined as:
%$\rm log(\nu)=\beta\,\rm log(E_{\rm F})+C$ \\ The derivative of the cumulative flare number distribution is given by $\dfrac{\mathrm{d}\nu}{\mathrm{d}E_{\rm F}}\sim E_{\rm F}^{-\alpha}$, where $\alpha=1+|\,\beta\,|$ \\ Mirjam is citing my $\beta$ here. Yang et al. (2017) and Lu et al. (2019) are giving the $\alpha$. Mirjam recalculated the  $\beta$ from their $\alpha$ to compare with her values.}
\cite{Raetz2020.0} found slopes 
ranging from $-0.81$ to $-0.85$ for different M SpT subranges, which is consistent with the result of \cite{2019ApJ...873...97L}. These latter authors found a value of $-0.86$ for the $548$ M dwarfs within their sample and a slope of $-0.78$ for a subsample of $375$ fast rotating M dwarfs ($P_{\rm rot}<10$\,d) based on {\it K2} long cadence data.
%We note that for some of our stars, the determined completeness limit might be too low, leading to an underestimation of the power law slope since too many low-energy events are considered in the fit and the shape of the FFD flattens at the low-energy end. Moreover, 
We note that a steepening of the FFD at the high-energy end, and therefore a deviation from the single power law shape, has been observed in previous flare studies, e.g., \cite{Lurie2015.0} and \cite{Silverberg2016.0}. Both these studies are based on data from the {\it Kepler} mission. The authors suggest that the steepening of the FFD toward higher flare energies might not be a real feature, but either a result of CCD saturation effects or due to low-number statistics.
For the $12$ stars with reliable $P_{\rm rot}$ in our sample, the single power law fits are consistent with the shape of the FFDs at the high-energy end under consideration of flare rate uncertainties. These uncertainties are increasing for high-energy events as only a small number of them is observed for each star. This leads to a blurred shape of the FFD, which makes it difficult to assess 
%both the accuracy of the completeness limit and the question whether all FFDs of our targets show a single power law shape. 
the shape of the power law at the highest observed flare energies. 
With this caveat in mind, our power law slopes for the FFDs are in good agreement with the range spanned by previous studies.

In summary, the consistency of our results with previous literature studies on larger
samples shows that the rotation and flare properties of this subsample
with HZs accessible for TESS transit detections is representative
for the class of M dwarfs. 

\subsection{From optical to X-ray flares}

Given the importance of these stars for planet detections, we 
went one step further and made some inferences on the expected properties of their 
X-ray flares. Such events quite inevitably take place on active M dwarfs, and the
ionizing high-energy radiation has likely a stronger impact on the evolution of the
planet than the lower-energy optical flare photons. However, a similarly detailed study of 
X-ray flare energy distributions is impossible in the foreseeable future because no
X-ray space-mission dedicated to long-term monitoring of a significant number of stars
is at the horizon. Therefore, combining the few available data sets in which flares were observed at more than one wavelength is a useful alternative for estimating the X-ray FFD. 
To this end, we have combined the simultaneous {\it XMM-Newton} and {\it K2/Kepler} observations of 
10 M-type stars in the Pleiades presented by \cite{Guarcello19.0} and three field M-type stars from 
%\cite{2021ApJ...912...81K} 
\cite{Kuznetsov21.0} with the simultaneous
TESS and {\it K2} observation of a single M dwarf, TYC~1330-879-1, analyzed by us specifically
for the scope of transforming observed TESS flare properties to the {\it K2/Kepler} band. 
We found that the FFDs of TESS and the {\it K2} mission map one-to-one to each other, namely 
the energies of flares seen in the TESS band are equal to the corresponding {\it K2} flare energies. \cite{2020AJ....160...36D} also found equivalent TESS and {\it K2} FFDs for the active M4 star GJ1243. Therefore, we can directly use the simultaneous {\it K2/XMM-Newton} sample to calibrate optical TESS flare energies to the X-ray band.

Under the assumption of a 1:1 correspondence between the occurrence of optical and X-ray flares, we converted the observed TESS FFDs to synthetic X-ray FFDs. This involves the propagation of eventual biases in TESS flare detection while those in the X-ray band are not considered.      
Our constructed X-ray FFDs are shifted toward lower flare energies with respect to the observed TESS FFDs in the respective SpT subset (cf. Fig.~\ref{fig:xray_ffd}). This is explained by the fact that in our calibration flare sample the energy released in the optical band is larger than that released in the X-ray band. Due to the shortage of simultaneous flare observations for M dwarfs in the optical and X-ray band, our calibration relies on the relatively low number of 18 events. The uncertainty in our $E_{\rm X}$ versus $E_{\rm K2}$ relation allows for a slope of $1$, which results in X-ray FFDs with slopes very similar to those of the observed TESS FFDs. \cite{Flaccomio2018.0} found for flares of pre-MS stars observed with CoRoT and {\it Chandra} a relation between $E_{\rm opt}$ and $E_{\rm X}$ of $E_{\rm opt}=6.37\cdot E_{\rm X}^{0.8}$ (for their reduced major axis fit). This translates to a slope of $1.25$ in the $\log E_{\rm X}$ versus $\log E_{\rm opt}$ relation. Their result is therefore consistent with the slope of our calibration based on the \cite{Guarcello19.0} Pleiades M-type stars and the \cite{2021ApJ...912...81K} field M stars.  

Observational constraints on X-ray flare statistics are not available for M dwarfs to the best of our knowledge. X-ray FFDs have  been studied for pre-MS stars (see \citealt{Getman2021.0} and references therein). The completeness limits for the flare energies
of these observations are extremely shallow (e.g., $\log{E_{\rm F,min}} \,{\rm [erg] }= 36.2$
for the MYStiX/SFiNC sample from \cite{Getman2021.0} and
log $\log{E_{\rm F,min}} \,{\rm [erg] }= 35.3$ for the \cite{Stelzer2007.0} sample in the Taurus and Orion star forming regions.) The slopes 
of these pre-MS X-ray FFDs are, however, similar to those found from optical FFDs by us and other studies discussed above. If the \cite{Getman2021.0} results are extrapolated to smaller flare energies, the observed X-ray FFDs lie at $E_F\approx10^{35}$\,erg about $1-2$ dex above our predictions (see Fig.~\ref{fig:xray_ffd}). Toward lower energies, the difference increases up to  about 5 dex at $E_F\approx10^{29}$\,erg.
 %\textcolor{red}{Hier waere es wirklich gut noch 1 Figur einzufuegen mit den beiden Geraden (ohne
           %   Daten) fuer Deine X-ray FFDs und der gruenen Getman Verteilung. In meinem 2007 paper habe ich
           %   leider den y-offset nicht angegeben. Da sieht man mal wie schlecht es ist, wenn man nicht alle
            %  Info veroeffentlicht. Sonst haettest Du so ein aehnliches Diagramm wie Ilin machen koennen aber fuer
            %X-ray FFDs. Aber ich denke Vergleich mit Getman-gruen reicht.}
One reason might be that we underestimate X-ray flare rates by the direct translation of optical flare rates. This could be caused by a detection bias reducing the number of optical flares observed with respect to their X-ray counterparts. On the Sun, for instance, white-light flares are detected more rarely than soft X-ray flares because of their shorter duration (e.g., \citealt{Namekata2017.0}). In any case, flare rates do not come without detection biases and therefore strongly depend on the instrument in use. It is thus likely that different properties of our hypothetical X-ray instrument with relation to {\it Chandra}, which is the basis of the \cite{Getman2021.0} study, influence flare frequencies.
            
Another possibility to explain this excess in the flare frequency of the pre-MS sample might be the youth of the stars. The X-ray activity level (irrespective of flares) is well known to decrease with stellar age  \citep[e.g.][]{Preibisch2005,Magaudda2020}. 
The origin of this ``quiescent'' X-ray emission is not understood, and unresolved nano-flares are a possible explanation  \cite[e.g.][]{Aschwanden00.0}. 
\cite{Coffaro22.0} inferred from modeling the X-ray spectra of solar-type stars with observations of our Sun that at an age of $\sim 200$\,Myr the whole corona of the star is filled with magnetic structures, and the enhanced X-ray flux of younger stars must be due to increased (unresolved)  flaring. 
 %Magaudda et al. (A\&A subm.)
        % \textcolor{orange}{Enza, please insert correct and updated reference for your paper.}
         %   found that field M dwarfs with saturated X-ray emission (defined by their position in the $L_{\rm X}$ vs $P_{\rm rot}$ plot) display between $0.8-2$ dex lower $L_{\rm X}$ values than pre-MS stars in the same mass range. 
Our FFDs apply to the most active of the field M dwarfs (as only fast rotators with clear spot modulation signal were considered) and are certainly in the saturated regime  where from inference from solar-type stars, as explained above,  part of the X-ray emission should be due to unresolved flaring. 
%In such an interpretation 
Given the negative power law of the FFDs one can expect that resolved flares decay in a similar way throughout stellar evolution as the quiescent (nano-flaring) X-ray luminosity. 
The discrepancy in
the flare rates between our predicted X-ray FFD for active M dwarfs and the observations for
pre-MS stars are therefore plausibly explained by the decrease of activity with age.
Finally we caution that the ``low-mass'' sample of \cite{Getman2021.0} is defined as
stars with mass $<1$ $M_\odot$, and thus does not represent the young analogs of our stars. 
% \textcolor{orange}{Enza, please insert correct and updated reference for your paper, and insert/update reference for Preibisch and Coffaro...}. \textcolor{teal}{DONE}
 
As \cite{Ilin2021.0} show in their comprehensive summary
of published optical stellar FFDs for M dwarfs, different studies have come up with distributions
that have vertical offsets of up to two orders of magnitude. Our TESS FFDs would lie
at the upper end of these distributions consistent with the high activity of the $12$ stars with reliable $P_\text{rot}$ within our sample. 
            %\textcolor{orange}{Beate, check if still true after Enza revises Fig.11, plus comment on the factor 4.}
One may,
therefore, conjecture that the X-ray FFDs of less active M dwarfs are analogously shifted
downward to lower frequencies. Future studies of the relation between quiescent X-ray luminosity
and X-ray variability, for example with the eROSITA all-sky survey \citep{Predehl2021.0}, may help to constrain the
dependence of the flare rate on the overall level of magnetic activity. 
%For X-ray flares on M dwarfs no FFDs are available in the literature, however, we combined the XX flares observed  simultaneously with {\it XMM-Newton} and K2 observations by Guarcello + Kusnetsov... to derive the relation between K2 and {\it XMM-Newton} flare energies. Together with the assumption that every X-ray flare has a corresponding {\it TESS} event (and viceversa), this has allowed us to construct hypothetical X-ray FFDs \textbf{(see Sect. \ref{subsection:xray_ffds}) that indicate the frequency of flare events above a certain X-ray Energy and effects on hypothetical planet atmospheres.}

\begin{acknowledgements}
This paper is based on data collected with the TESS mission, obtained from the MAST data archive at the Space Telescope Science Institute (STScI). Funding for the  mission is provided by the {\it NASA} Explorer Program. STScI is operated by the Association of Universities for Research in Astronomy, Inc., under {\it NASA} contract NAS 5–26555.\\
This work has made use of data from the European Space Agency (ESA) mission
{\it Gaia} (\url{https://www.cosmos.esa.int/Gaia}), processed by the {\it Gaia}
Data Processing and Analysis Consortium (DPAC,
\url{https://www.cosmos.esa.int/web/Gaia/dpac/consortium}). Funding for the DPAC
has been provided by national institutions, in particular the institutions
participating in the {\it Gaia} Multilateral Agreement.\\
This publication makes use of data products from the Two Micron All Sky Survey, which is a joint project of the University of Massachusetts and the Infrared Processing and Analysis Center/California Institute of Technology, funded by the National Aeronautics and Space Administration and the National Science Foundation.\\
This research has made use of ESASky, developed by the ESAC Science Data Centre (ESDC) team and maintained alongside other ESA science mission's archives at ESA's European Space Astronomy Centre (ESAC, Madrid, Spain).\\
The research described in this paper makes use of Filtergraph, an online data visualization tool developed at Vanderbilt University through the Vanderbilt Initiative in Data-intensive Astrophysics (VIDA).\\
This work made use of the gaia-kepler.fun crossmatch database created by Megan Bedell.

\end{acknowledgements}

%\nocite{Mann2016.0}
\bibliographystyle{aa} %aa.bst
%\bibliography{references}
\bibliography{tessHZstars_short2}

\newenvironment{localsize}[1]
{%
  \let\orignewcommand\newcommand
  \let\newcommand\renewcommand
  \makeatletter
  \input{bk#1.clo}%
  \makeatother
  \let\newcommand\orignewcommand
}
{%
  \clearpage
}

 \begin{appendix}
% %\sectionfont{\Large}

% \newgeometry{left=1cm, right=1cm, top=2cm, bottom=1.8cm}
 \renewcommand*\thetable{A.\arabic{table}}
 \setcounter{table}{0}
 \renewcommand*\thefigure{A.\arabic{figure}}
    \setcounter{figure}{0}

% %\input{./Tables/TableA1_Factor4.tex}
% %\begin{landscape}

\begin{table}
\section{Data Tables.}

%\\\normalfont{
Stellar parameters, TESS contamination factors, rotation periods, flare parameters and  observation times for the M~dwarf sample.  The full tables are available in electronic form at the CDS via anonymous ftp to  cdsarc.u-strasbg.fr (130.79.128.5).
%}}

\onecolumn
\small\addtolength{\tabcolsep}{-4pt}

\begin{TableNotes}
\item[(**)] denotes stars with PM companion.
\end{TableNotes}

%\begin{table}
\begin{longtable}{ccccccccccccc}
\caption{Stellar parameters for our $112$ sample stars; see Sect.~\ref{sec:stellar_params} for details.}\label{tab:stellar_params_all}\\
\hline
 TIC ID&Other name &Gaia DR2 ID &2MASS ID &RA & Dec&$T$&$M_*$&$d$&SpT&log($L_{\textrm{qui},T}$)\\
            &  &&&[deg]&[deg]&[mag]&[$M_{\odot}$]&[pc]&&[erg/s]\\
\hline
\endfirsthead

 TIC ID&Other name &Gaia DR2 ID &2MASS ID &RA & Dec&$T$&$M_*$&$d$&SpT&log($L_{\textrm{qui},T}$)\\
            &  &&&[deg]&[deg]&[mag]&[$M_{\odot}$]&[pc]&&[erg/s]\\
\endlastfoot
  38759628 &L 130-37& 4677345043203760896 & 04280568-6209254 & 67.024 & -62.157 & 10.42 & 0.331 $\pm$ 0.007 & 18.241 $\pm$ 0.012 & 3.5 & 31.15 \\
  55745883 &UCAC4 141-004923& 4665366619931917824 & 04593230-6153042 & 74.885 & -61.884 & 10.17 & 0.389 $\pm$ 0.008 & 19.899 $\pm$ 0.01 & 3.0 & 31.33 \\
  140635646 &UCAC4 064-003335& 4624560719769978368 & 04471915-7719483 & 71.83 & -77.33 & 10.68 & 0.435 $\pm$ 0.009 & 32.497 $\pm$ 0.023 & 1.5 & 31.55 \\
  140998116 &UCAC4 066-003833          & 4624829103685457024 & 05130604-7653221 & 78.275 & -76.89 & 10.62 & 0.341 $\pm$ 0.007 & 22.157 $\pm$ 0.013 & 3.0 & 31.24 \\
  $..$&$..$&$..$&$..$&$..$&$..$&$..$&$..$&$..$&$..$&$..$\\
\hline
  \end{longtable}
\end{table}

\normalsize

% % %\newgeometry{left=1.7cm, right=1.5cm, top=1.5cm, bottom=3cm}

% % %\begin{landscape}

% % % \onecolumn
% % % \setcounter{table}{0}
\small\addtolength{\tabcolsep}{-4pt}

\begin{table}[htbp]

\caption[Stellar parameters for CPM pairs]{Stellar parameters for CPM pairs.}\label{tab:CPMs}
\raggedright
\begin{tabular}{ccccccccccccc}
\\
\hline
TIC ID                 &                  {\it Gaia} DR2 ID                 &                  2MASS ID                 &                  dist                 &                  $G_{RP}$                 &                  $R_*$                 &                  $T_{\textrm{eff}}$                 &                  $M_*$                 &                  SpT                 &                  sep.                            \\
                 &                                   &                                   &                  [pc]                 &                  [mag]                 &                  $[R_\odot]$                 &                  [K]                 &                  $[M_\odot]$                 &                                   &                  [$^{\prime\prime}$]                                       \\
                 \hline
142086812                 &                  5260759175761501696                 &                  06334337-7537482                 &                  8.840$\pm$0.002                 &                  8.26                 &                  0.425$\pm$0.013                 &                  3442$\pm$80                 &                  0.443$\pm$0.009                 &                  M3V                 &                  21.74      \\
142086813                 &                  5260759175761502976                 &                  06334690-7537301                 &                  8.838$\pm$0.003                 &                  9.04                 &                  0.332$\pm$0.010                 &                  3330$\pm$82                 &                  0.338$\pm$0.006                 &                  M3.5V                 &                           
%\vspace{0.2cm}  
\\                                                                                  149988104                 &                  5482312349304478976                 &                  06011889-6047092                 &                  20.292$\pm$0.020                 &                  11.19                 &                  0.286$\pm$0.009                 &                  3282$\pm$82                 &                  0.286$\pm$0.006                 &                  M3.5V                 &                  6.85   \\
--                 &                  5482312418025591808                 &                  06011898-6047023                 &                  20.371$\pm$0.030                 &                  11.78                 &                  0.236$\pm$0.007                 &                  3194$\pm$81                 &                  0.232$\pm$0.005                 &                  M4V                 &                                 
%\vspace{0.2cm} 
\\ 
  $..$&$..$&$..$&$..$&$..$&$..$&$..$&$..$&$..$&$..$\\
\hline                                                                                              \end{tabular}
\tablefoot{Some companion stars have no or incomplete {\it Gaia}/2MASS photometry. Therefore, stellar parameters cannot be determined.\newline \tablefoottext{*}{White Dwarf accompanying the eclipsing binary CM~Draconis (TIC\,199574208)}.\newline\tablefoottext{**}{{\it Gaia} parallax distance  not reliable according to Lindegren conditions (cf. sec. \ref{sec:stellar_params}). $d_{\textrm{phot}}\approx42.9$ pc}\newline\tablefoottext{+}{separation taken from WDS. All other separations are calculated using {\it Gaia} coordinates.}
}
\end{table}

% % %\end{landscape}

% % %\input{./Tables/TablesA3.tex}

% % %\newgeometry{left=1.7cm, right=1cm, top=1.5cm, bottom=2cm}
% % %\input{./Tables/TableA3.tex}
\small\addtolength{\tabcolsep}{10pt}

\begin{longtable}{lccrrrccccccccc}

\caption[Contamination factors for all sample stars]{Results of the rotation and flare  search and contamination analysis for all stars. Gaps in each LC were subtracted from the observation length  (cf. Sect. \ref{sec:flare_rate_SpT}).}\label{tab:rot_act}\\%Listed are the {\it TESS} sectors in which the star was observed and the total observation length (cols. 2\&3), Columns 4-6 list the mean, minimum and maximum contamination factors of all \textit{TESS} sectors (see Sect. \ref{sec:contamination} for details). The last three columns are: the rotation period, the total number of validated flares, and the rate of flares with $\Delta L_{F,T}\geq3.4\cdot10^{29}$\,erg/s  which is the maximum detection threshold of all flaring stars 

\hline
      &&&\multicolumn{3}{c}{contamination factor}&&&\\
    TIC ID&observation sectors&obs. duration&mean&min&max&$P_\text{rot}$&$N_\text{flares,tot}$&$\nu_\text{flares,thresh}$\\
    &&[d]&[\%]&[\%]&[\%]&[d]&[d]$^{-1}$&[d]$^{-1}$\\
\hline
\endfirsthead

\multicolumn{10}{l}%
{\tablename\ \thetable\ -- Continued from previous page} \\
\hline
     &&&\multicolumn{3}{c}{contamination factor}&&&\\
    TIC ID&observation sectors&obs. duration&mean&min&max&$P_\text{rot}$&$N_\text{flares,tot}$&$\nu_\text{flares,thresh}$\\
    &&[d]&[\%]&[\%]&[\%]&[d]&[d]$^{-1}$&[d]$^{-1}$\\
\hline
\endhead
\hline
\multicolumn{4}{l}{Continued on following page.}
\endfoot
\endlastfoot

38759628&1-13&302.32&0.6&0.5&0.7&$-$&5&0.003\\
55745883&1-6;8-13&278.64&2.5&0.0&5.2& $-$&$-$&$-$\\
140635646&1-9;11-13&281.35&0.4&0.1&0.5& $-$&$-$&$-$\\
140998116&1-13&304.2&3.8&0.2&9.6& $-$&$-$&$-$\\
$..$&$..$&$..$&$..$&$..$&$..$&$..$&$..$&$..$\\
\hline
\end{longtable}
\tablefoot{\tablefoottext{a}{``$\infty$'' is set when the target is outside the mask. These stars have been excluded from the analysis.}
\tablefoottext{**}{denotes CPM pairs. There are $18$ targets with a maximum contamination $>10\%$, meaning the flux in the aperture mask coming from objects other than the target makes for more than $10$\,\% of the target's flux in at least one observation sector. Ten of these $18$ stars are part of CPM pairs where contamination is mainly caused by the companions listed in Table \ref{tab:CPMs}.}}

% % %\clearpage
%  \restoregeometry
\twocolumn
% %%%%%%%%%%%%%%%%%%%%%%%%%%%%%%%%%
 \section{Special treatment of the eclipsing binary CM Draconis}\label{sec:CMDra_treatment}
% %%%%%%%%%%%%%%%%%%%%%%%%%%%%%%%%%%%

% \renewcommand*\thetable{B.\arabic{table}}
% \setcounter{table}{0}
\renewcommand*\thefigure{B.\arabic{figure}}
\setcounter{figure}{0}
\begin{figure}[h]
\centering
\includegraphics[width=0.5\textwidth]{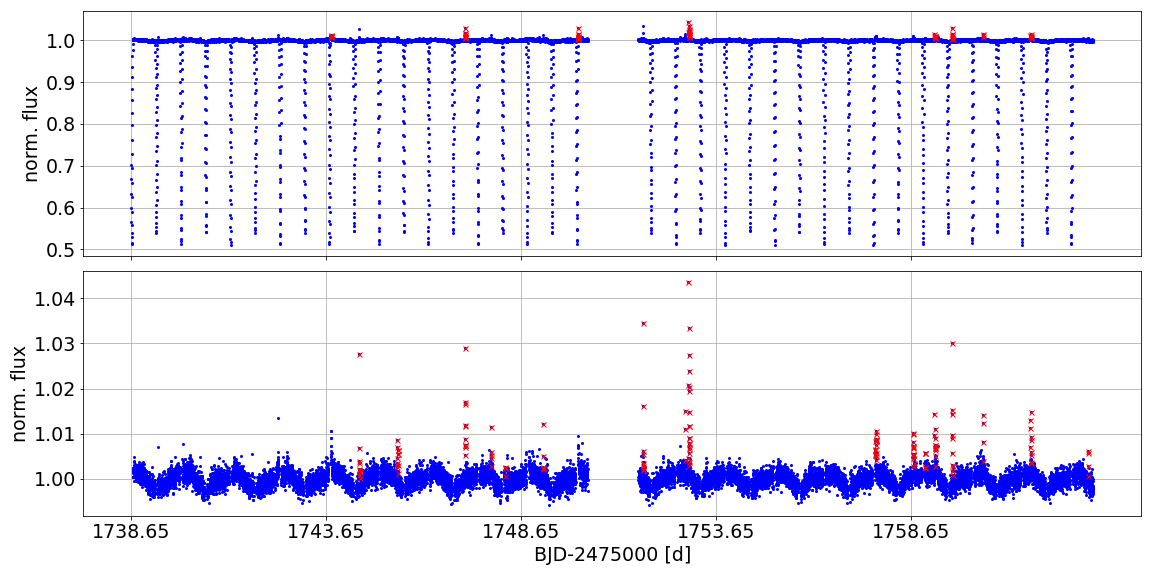}
\caption[LC of eclipsing binary CM Dra]{\textit{Top:} original LC of the eclipsing binary CM Draconis (TIC\,199574208) in sector~$16$. \textit{Bottom:} LC of the same sector after removal of the eclipses as it was used for flare and rotation period search. Validated flares in both panels are marked in red. The rotation period found in the analysis is identical to the orbital period of $1.27$\,d, meaning the spin is synchronized to the orbital motion by tidal interaction between the two components. This tidal locking of CM Dra is also reported in the literature (e. g \citealt{1998A&A...338..479D}, \citealt{2009ApJ...691.1400M}).}
\label{fig:cm_dra}
\end{figure} 

\normalsize
\noindent
TIC\,199574208 is better known as the eclipsing binary (EB) CM\,Draconis. The binary system itself comprises two very similar red dwarfs  ($M_1=0.23M_\odot$, $M_2=0.21M_\odot$) orbiting each other with a period of $T=1.27$\,d (e.g., \citealt{2012AN....333..754P}). Using the {\it Gaia} parallax we estimate a separation of 1.17 mas, and thus the two sources are not resolved by {\it Gaia}. The unresolved EB has another companion object: A White Dwarf at a separation of 26.7$^{\prime\prime}$ sharing the same proper motion that might contaminate the photometry is listed in Table \ref{tab:CPMs}.

The eclipses dominate the LC and make it difficult to analyze other characteristics such as flares or starspot variations. For instance, the eclipses cause a significantly higher standard deviation of the LC and therefore raise the flare detection threshold. Since this work focuses on flare and rotation period search, all data points of the eclipses were removed from all LCs of this star before the final analysis. Fig.~\ref{fig:cm_dra} shows as an example in the top panel the original LC of sector~$16$ with eclipses and in the bottom panel the LC after removal of the eclipses as it was used for the final analysis. In this sector, the flare search algorithm validated 18 flares in the LC without eclipses while only 8 were found in the original LC.

% \clearpage
% \newgeometry{left=2.3cm, right=1.7cm, top=1.5cm, bottom=1.9cm}
 \section{Example LCs of stars with reliable rotation periods}\label{sec:LCs_prot}
 \renewcommand*\thefigure{C.\arabic{figure}}
 \setcounter{figure}{0}

\begin{figure*}[h!]
\centering
\parbox{18cm}{\includegraphics[width=\textwidth]{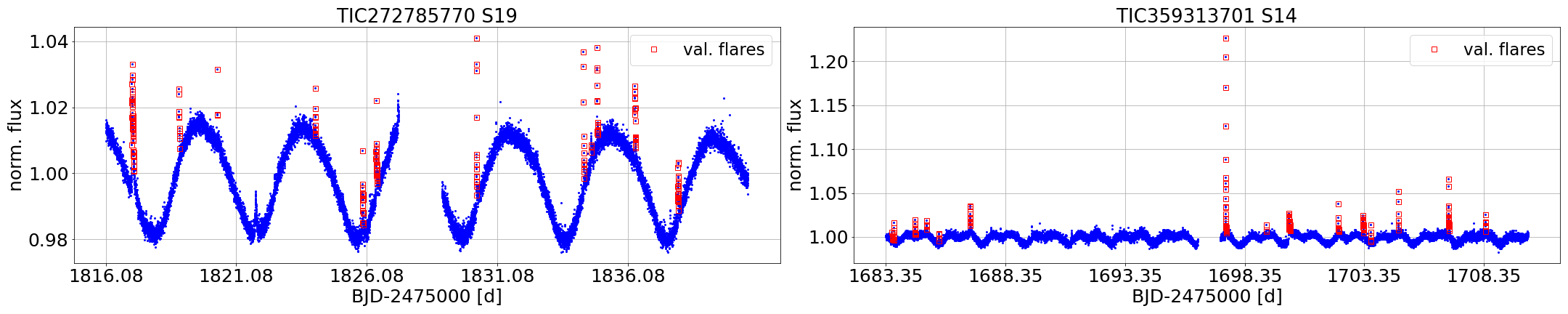}}
\parbox{18cm}{\includegraphics[width=\textwidth]{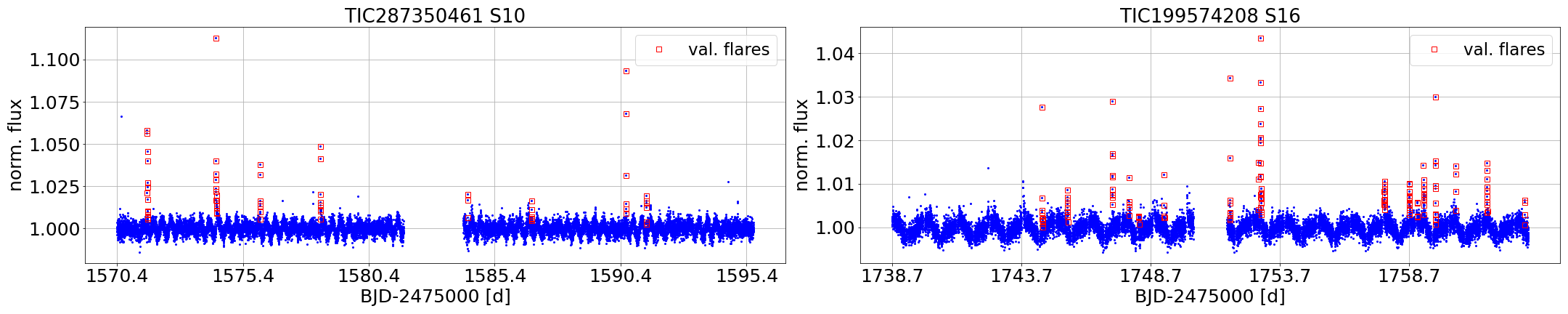}}
\parbox{18cm}{\includegraphics[width=\textwidth]{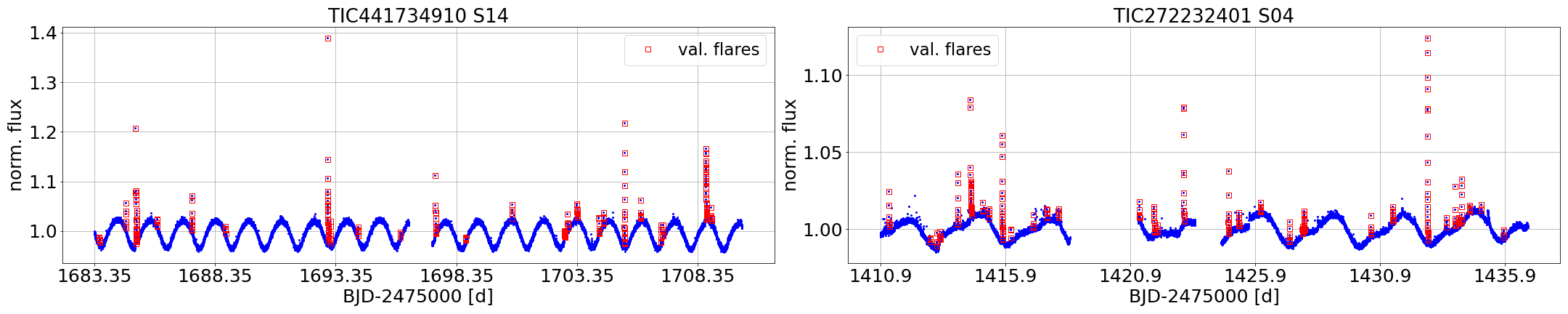}}
\parbox{18cm}{\includegraphics[width=\textwidth]{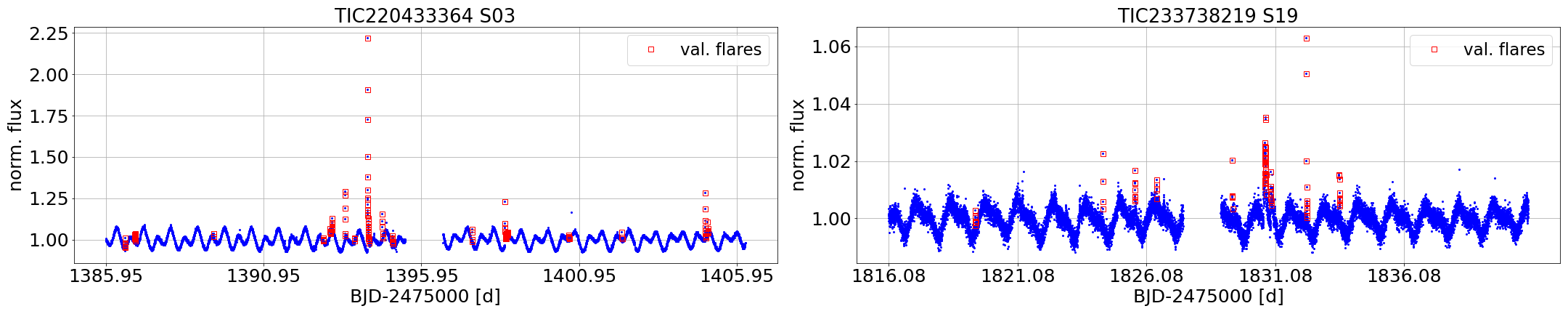}}
\parbox{18cm}{\includegraphics[width=\textwidth]{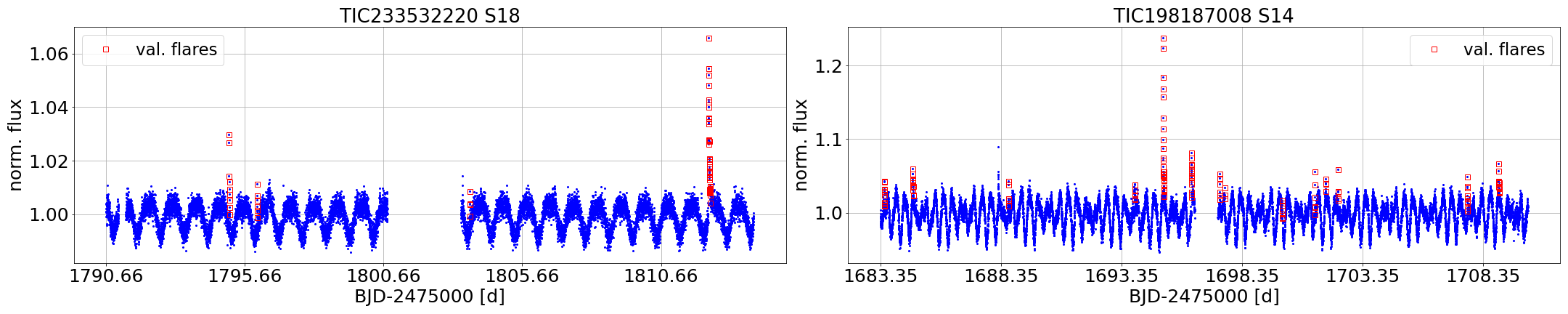}}
\parbox{18cm}{\includegraphics[width=\textwidth]{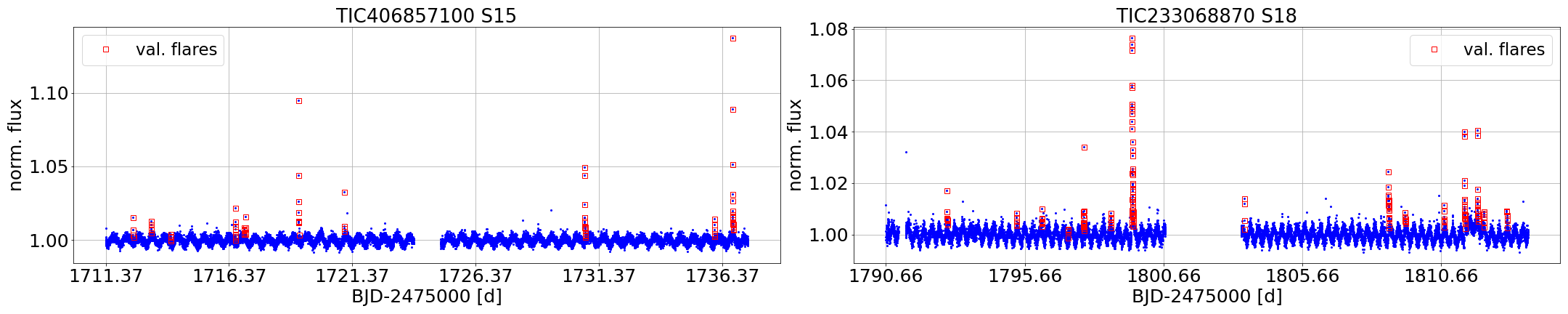}}

\caption{Examples of TESS LCs; one TESS sector is displayed for each of the $12$ stars with reliable rotation period.}
%}
\end{figure*}

% \restoregeometry
 \section{Estimating Kepler Magnitudes using Gaia and TESS photometry}\label{app:magtrans}
\renewcommand*\thefigure{D.\arabic{figure}}
\renewcommand*\thetable{D.\arabic{figure}}
\setcounter{figure}{0}

\begin{figure}[h!]
\includegraphics[width = 0.5\textwidth]{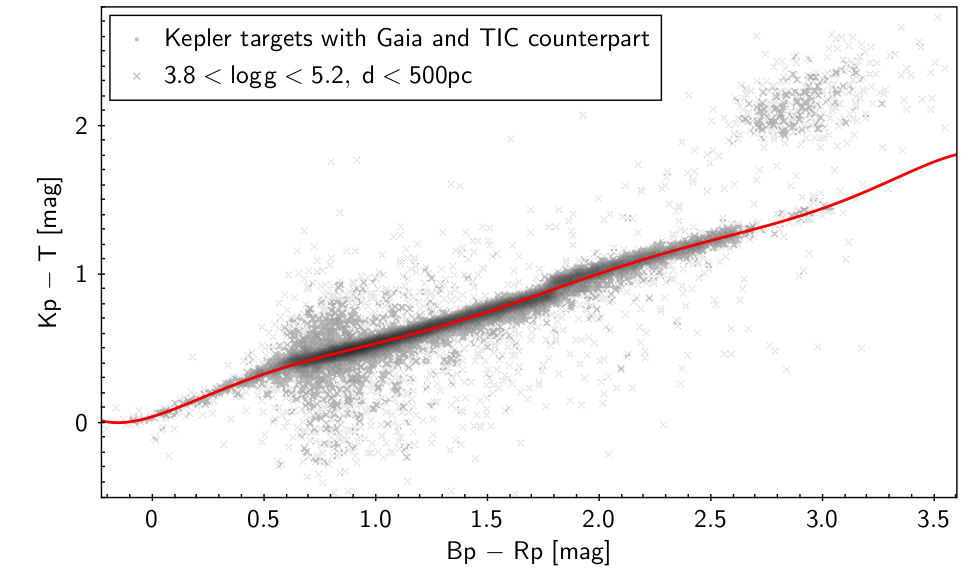}
\caption{Fit of $K_p-T$ vs. $BP-RP$ for about $19000$ nearby main sequence stars.}\label{fig:Kp_vs_T}
\end{figure}

In order to compare the standard deviations of the flattened LCs ($S_\text{flat}$) of the short cadence {\it K2} data from \cite{Raetz2020.0} to our TESS results (see Sect.~\ref{sec:flares_prot}) we had to convert the {\it K2} magnitudes to TESS magnitudes. To this end, we made use of the gaia-kepler.fun crossmatch database created by Megan Bedell\footnote{https://gaia-kepler.fun/}. This website provides cross-matched catalogs of {\it Gaia} data for stars observed by {\it Kepler/K2}. In particular, we downloaded the 1" cross-match of {\it Gaia} DR2 \citep{2018A&A...616A...1G} with the {\it NASA} Exoplanet Archive {\it Kepler} Stellar Properties Catalog (for Q1-Q17, DR25) consisting of all stellar targets with long-cadence observations from {\it Kepler} \citep{2016ksci.rept....8M}. This catalog was further matched with \cite{2018AJ....156...58B}. The resulting downloaded catalog contains 201312 entries. 
For our purpose we selected the subsample of 20379 nearby (<\,500\,pc) main-sequence stars (3.8\,<\,log\,g\,<\,5.2). By using the provided {\it Gaia} IDs we cross-matched the catalog with the TESS Input Catalog (TIC v8, \citealt{2018AJ....156..102S}) which results in 19009 entries. This final catalog lists {\it Gaia}, {\it Kepler} as well as TESS magnitudes.

Fig.~\ref{fig:Kp_vs_T} shows the color-color plot using {\it Gaia} $BP-RP$ and the difference between {\it Kepler} and TESS magnitudes. To estimate the TESS magnitudes of the sample presented by \cite{Raetz2020.0}, we fitted a 7th order polynomial to these data. The result of the fitting is
\begin{equation}
Kp-T=A+Bx+Cx^{2}+Dx^{3}+Ex^{4}+Fx^{5}+Gx^{6}+Hx^{7}
\label{eq:magtrans}
\end{equation}
with $x\,=\,BP-RP$ and the coefficients given in Table~\ref{tab:TKpmag}.
By inserting the {\em Gaia} color and the {\it K2} magnitude of the R20 sample, we can estimate the TESS magnitude.

\begin{table}
\centering
\caption{Coefficients for the magnitude calibration given in Eq.~\ref{eq:magtrans}.}
\begin{tabular}{lr}
\hline \hline
Coefficient & Value \\ \hline
$A$ & 0.0449377231 \\
$B$ & 0.4603859028 \\
$C$ & 0.8820614439 \\
$D$ & -2.0683447014 \\
$E$ & 1.8798211314 \\
$F$ & -0.8203383843 \\
$G$ & 0.1721529852\\
$H$ & -0.0139446163\\
\hline
\end{tabular}
\label{tab:TKpmag}
\end{table}

%\newgeometry{left=2.3cm, right=1.7cm, top=1.5cm, bottom=1.9cm}
\section{FFDs with energy completeness limits for the $12$ stars with reliable rotation periods %\textcolor{teal}{30.03.2022: New $E_{\rm Flare}$ and $E_{\rm min}\rightarrow$ Fig to be updated with my office computer (after we are sure about the factor 4)}
}\label{app:ffds}

\renewcommand*\thetable{E.\arabic{table}}
\setcounter{table}{0}
\renewcommand*\thefigure{E.\arabic{figure}}
\setcounter{figure}{0}

\begin{figure*}[t]
\centering
\includegraphics[width=\textwidth]{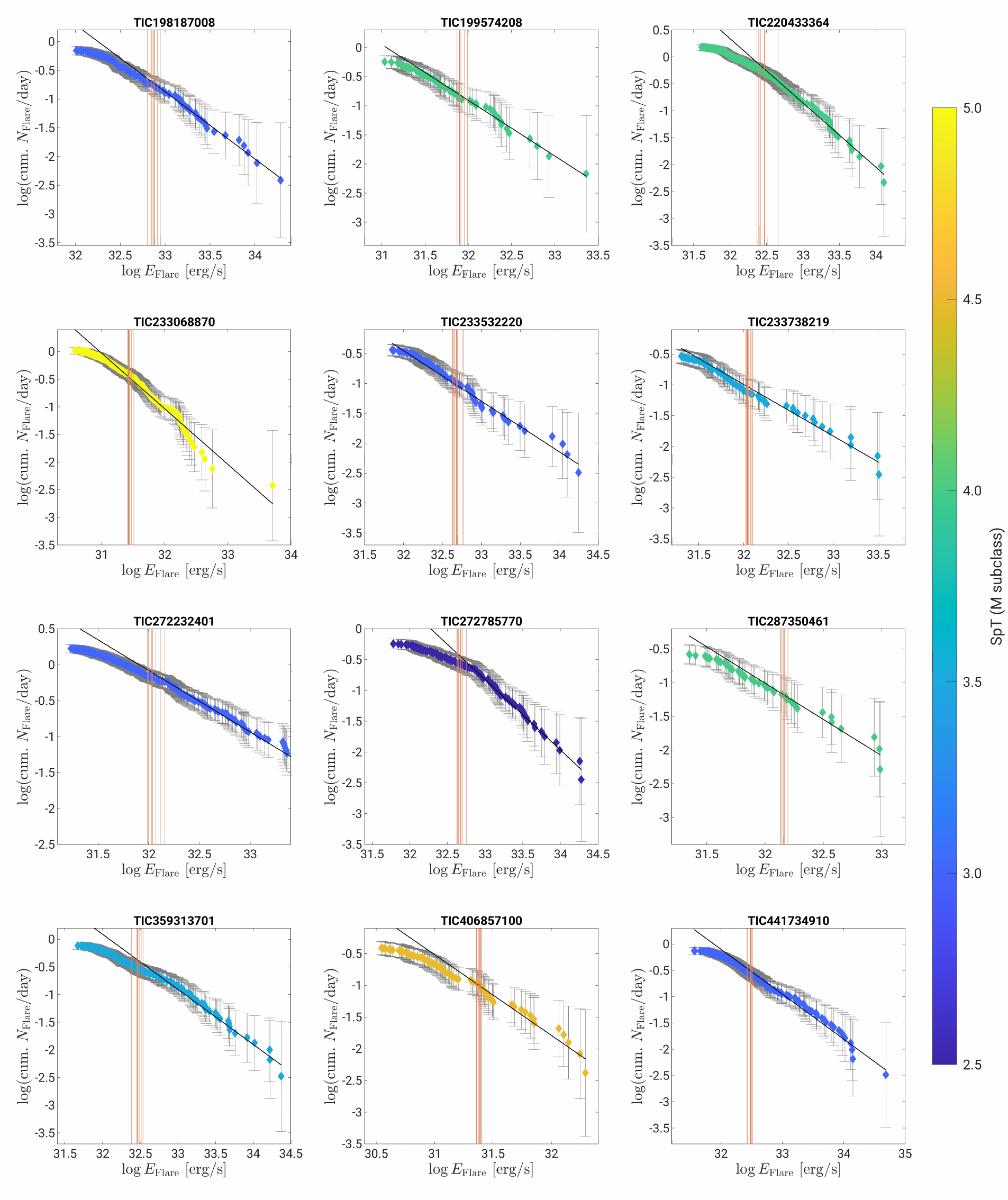}
\caption{Cumulative distribution of the flare energies for the $12$ stars with reliable rotation period determined from TESS light curves. The red solid line shows the energy threshold of each TESS sector that we adopted to select the flare events considered in the  power law fit.}
\end{figure*}

\end{appendix}
\end{document}